\documentclass[]{aa}
\usepackage{graphicx}
\usepackage{ulem}
\usepackage{color}
\usepackage{amsmath,amssymb}
\usepackage[varg]{txfonts}
\usepackage{hyperref}

\def\simle{\mathrel{\hbox{\rlap{\hbox{\lower4pt\hbox{$\sim$}}}\hbox{$<$}}}}
\def\simgr{\mathrel{\hbox{\rlap{\hbox{\lower4pt\hbox{$\sim$}}}\hbox{$>$}}}}

\begin{document}

\title{Ultra-luminous X-ray sources and neutron-star--black-hole mergers from
   very massive close binaries
at low metallicity} \titlerunning{ULXs and NSBH mergers in very massive binaries with CHE}

\author{Pablo Marchant\inst{1,2}\thanks{Email: pablo@astro.uni-bonn.de}
\and Norbert Langer\inst{1} \and Philipp Podsiadlowski\inst{3,1}
\and Thomas M. Tauris\inst{4,1} \and Selma de Mink\inst{5} \and Ilya
Mandel\inst{6}\and Takashi J. Moriya\inst{7}
}

\institute{Argelander-Institut f\"ur Astronomie, Universit\"at Bonn, Auf dem H\"ugel 71, 53121 Bonn, Germany
\and
Center for Interdisciplinary Exploration and Research in Astrophysics (CIERA)
and Department of Physics and Astronomy, Northwestern University, 2145 Sheridan
Road, Evanston, IL 60208, USA
\and
Department of Astrophysics, University of Oxford, Oxford OX1 3RH, UK
\and
Max-Planck-Institut f\"ur Radioastronomie, Auf dem H\"ugel 69, 53121 Bonn, Germany
\and
Anton Pannenkoek Institute for Astronomy, University of Amsterdam, NL-1090 GE
Amsterdam, the Netherlands
\and
School of Physics and Astronomy, University of Birmingham, Birmingham B15 2TT, UK
\and
Division of Theoretical Astronomy, National Astronomical Observatory of Japan,
National Institues of Natural Sciences, 2-21-1 Osawa, Mitaka, Tokyo 181-8588,
Japan
}

\date{}

\abstract {The detection of gravitational waves from the binary black
   hole (BH) merger GW150914 may enlighten our understanding
of ultra-luminous X-ray sources (ULXs), as BHs of masses
$>30M_\odot$ can reach luminosities $>4\times10^{39}~{\rm erg~s^{-1}}$ without
exceeding their Eddington luminosities.
It is then important to study variations of evolutionary channels for merging
BHs, which might instead form accreting BHs and become ULXs.
It was recently shown that very massive binaries with
mass ratios close to unity and tight orbits can undergo efficient rotational mixing and
evolve chemically homogeneously, resulting in a compact BH binary. We study similar systems by
computing $\sim 120\,000$ detailed binary models with the
\texttt{MESA} code covering a wide range
of masses, orbital periods, mass ratios and metallicities. For
initial mass ratios
$q\equiv M_2/M_1\simeq0.1-0.4$, primaries with masses above $40M_\odot$ can evolve chemically homogeneously, remaining
compact and forming a BH without experiencing Roche-lobe
overflow. The secondary then
expands and transfers mass to the BH, initiating a ULX
phase. At a given metallicity this channel is expected to produce the most massive accreting stellar BHs and the brightest ULXs.
We predict that $\sim 1$ out of $10^{4}$ massive stars evolves this way,
and that in the local universe $0.13$ ULXs per $M_\odot~\rm yr^{-1}$ of
star-formation rate are observable, with a
strong preference for low-metallicities. An additional channel is
still required to explain the less luminous ULXs and the full population of
high-mass X-ray binaries. At metallicities
$\log Z>-3$, BH masses in ULXs
are limited to $60M_\odot$ due to the occurrence of pair-instability supernovae
which leave no remnant,
resulting in an X-ray luminosity cut-off for accreting BHs. At lower
metallicities, very massive stars can avoid exploding as pair-instability supernovae
and instead form BHs with masses above $130M_\odot$, producing a gap
in the ULX luminosity distribution. After the ULX phase,
neutron-star-BH binaries that merge in less than a Hubble time are
produced with a low formation rate $<0.2~\rm Gpc^{-3}yr^{-1}$.
We expect that upcoming X-ray observatories will test these predictions,
which together with additional gravitational wave
detections will provide strict constraints on the origin of the most massive
BHs that can be produced by stars.
}


\maketitle

\section{Introduction}
\label{sect:intro}

One of the most puzzling discoveries made by the Einstein Observatory are the
off-nucleus X-ray point sources with luminosities above
$10^{39}~{\rm erg~s^{-1}}$ \citep{LongvanSpeybroeck1983}, which owing to their
extreme luminosities were termed ultra-luminous X-ray sources (ULXs).
Compared to the typical properties of high-mass X-ray binaries (HMXBs),
such high luminosities are difficult to
explain in terms of accreting compact objects, as the Eddington limit for
neutron stars ($\sim 10^{38}{\rm erg~s^{-1}}$, hereafter NS) and stellar mass black holes
($\sim 2\times 10^{39}{\rm erg~s^{-1}}$ for a $10M_\odot$ black hole, hereafter BH)
is well below the luminosities of some of the observed sources.
One possibility to explain these high luminosities is
to consider the existence of a
population of intermediate-mass BHs (IMBHs), with masses
between $\sim 10^2-10^5M_\odot$, possibly arising from the collapse of
primordial stars (eg. \citealt{MadauRees2001}) or formed in dense globular clusters
(eg. \citealt{MillerHamilton2002}).

As the Chandra X-ray Observatory and other facilities opened up the possibility of studying
populations of X-ray point sources in galaxies down to much lower luminosities,
\citet{Grimm+2003} showed that ULXs generally correspond to the tail of the HMXB
population, and their number is strongly correlated with star-formation rate
(SFR). \citet{Swartz+2011} estimated a typical number
of $\sim 2$ ULXs per $M_\odot~{\rm yr^{-1}}$ of SFR for a local
sample of galaxies, while \citet{Luangtip+2015} observed that
there is a scarcity of ULXs in luminous
infrared galaxies, with an estimated number of $0.2$ ULXs per $M_\odot~{\rm
   yr^{-1}}$ of SFR.
All of this points towards both a
stellar origin for ULXs and possibly a strong metallicity dependence, which
disfavors the IMBH scenario.
Using a
sample of 64 galaxies at various metallicities down to
$\log Z\,{\sim}\,{-}3$, \citet{Mapelli+2010} found that
the number of ULXs per $M_\odot~{\rm yr^{-1}}$ of SFR scales with metallicity as
$Z^{-0.55\pm 0.23}$ (see also \citealt{Prestwich+2013}).
Although this appears to rule out IMBHs as the central engines of most ULXs,
there remain a handful of particularly bright sources (in excess of
$3\times10^{41}~{\rm erg~s^{-1}}$) which appear to form an
independent population \citep{Sutton+2012,Swartz+2011}. The term
hyper-luminous-X-ray-source has been coined for these objects,
with ESO243-49 HLX-1 being the best current candidate for an IMBH
\citep{Farrell+2009}. However, some of these have been confirmed to
be background AGN \citep{Sutton+2015}, reducing the number of known objects of this class.

If ULXs have a stellar origin, there are various potential explanations for
their high luminosities. For instance,
beaming of the radiation emitted would imply that the actual
full-sky luminosity of these sources is much lower, so that ULXs could consist
of BHs with masses below $10M_\odot$ accreting at or below the Eddington rate.
This could be a purely geometrical effect \citep{King+2001}
or the result of relativistic beaming \citep{Kording+2002}, but
measurements from ionization nebulae around some ULXs
appear to confirm the isotropic estimate of their luminosities
\citep{PakullMironi2003}. On the other hand, \citet{Begelman2002} and
\citet{RuszkowskiBegelman2003} proposed that the photon-bubble
instability, which acts on radiation-dominated accretion disks and produces
clumping, would cause photons to be radiated away through low
density regions allowing for accretion rates up to 10 times the
Eddington rate.
The presence of a corona supported by strong magnetic fields could
also help to counter the radiation pressure and allow super-Eddington accretion
\citep{SocratesDavis2006}.

A clear case of super-Eddington accretion is the
recently observed NS-ULX \citep{Bachetti+2014}, for which accretion rates above
a hundred times the Eddington rate are required to explain its luminosity in
excess of $10^{40}~{\rm erg~s^{-1}}$. The flux observed from this object has
both a pulsed component with a period of $1.37$ days, and a non-pulsed component,
so beaming alone appears insufficient
to explain its nature.Very recently, two more ULXs powered by NSs
have been discovered \citep{Israel+2016,Israel+2017}, and some argue that a significant
fraction of ULXs could contain a NS accretor \citep{KingLasota2016}.
The very high luminosities of these accreting NSs do not
necessarily imply that
accreting BHs can also radiate well above their Eddington luminosities, since the
accretion flows around NSs and BHs should differ substantially.

It should also be considered that, although Galactic BHs
are limited to masses below $\sim20M_\odot$ due to strong wind mass
loss
\citep{FryerKalogera2001,Spera+2015,Sukhbold+2016}, in lower metallicity environments BH masses could
reach up to $45M_\odot$ \citep{HegerWoosley2002,Belczynski+2016b}, with the
mass being limited by the effects of
pair-instability supernovae (PISNe) and pulsational-pair-instability
supernovae (PPISNe). Such massive BHs can easily account for
the luminosity of ULXs, requiring accretion rates only slightly above the
Eddington limit to explain some of the brightest sources \citep{Zampieri&Roberts2009}.
For massive enough progenitors, it is expected
that PISNe can be avoided, resulting in BHs with masses above $\sim 130M_\odot$
and possibly causing a gap in the BH mass distribution \citep{HegerWoosley2002}. However, for single
stars this requires extremely high zero-age main-sequence masses and low metallicity.

\subsection{Formation channels for ULXs and merging binary BHs}

The commonly assumed model for ULX formation involves the occurrence of a
CE phase in an initially very wide binary \citep{Rappaport+2005}.
In these models,
the envelope of the primary is stripped in a common-envelope (CE) phase, which significantly
reduces the orbital period. The primary then collapses to a BH, and when the
secondary expands and initiates Roche-lobe overflow (RLOF) the system becomes an active X-ray source. Whether a
standard HMXB or a ULX is produced depends on the mass of the BH formed,
providing a simple explanation for the continuous luminosity
distribution function from HMXBs to ULXs, although accretion rates
$\sim10$ times Eddington are still required to explain the brightest sources.
A different possibility is the formation of ULXs containing a BH through
dynamical interactions in star clusters \citep{MapelliZampieri2014,MacLeod+2016}, which could
potentially produce more massive BHs at a given metallicity, as the progenitor
of the BH can evolve as a single star and avoid envelope stripping in a binary.

\begin{figure}%[!ht]
   \begin{center}
   \includegraphics[width=0.85\columnwidth]{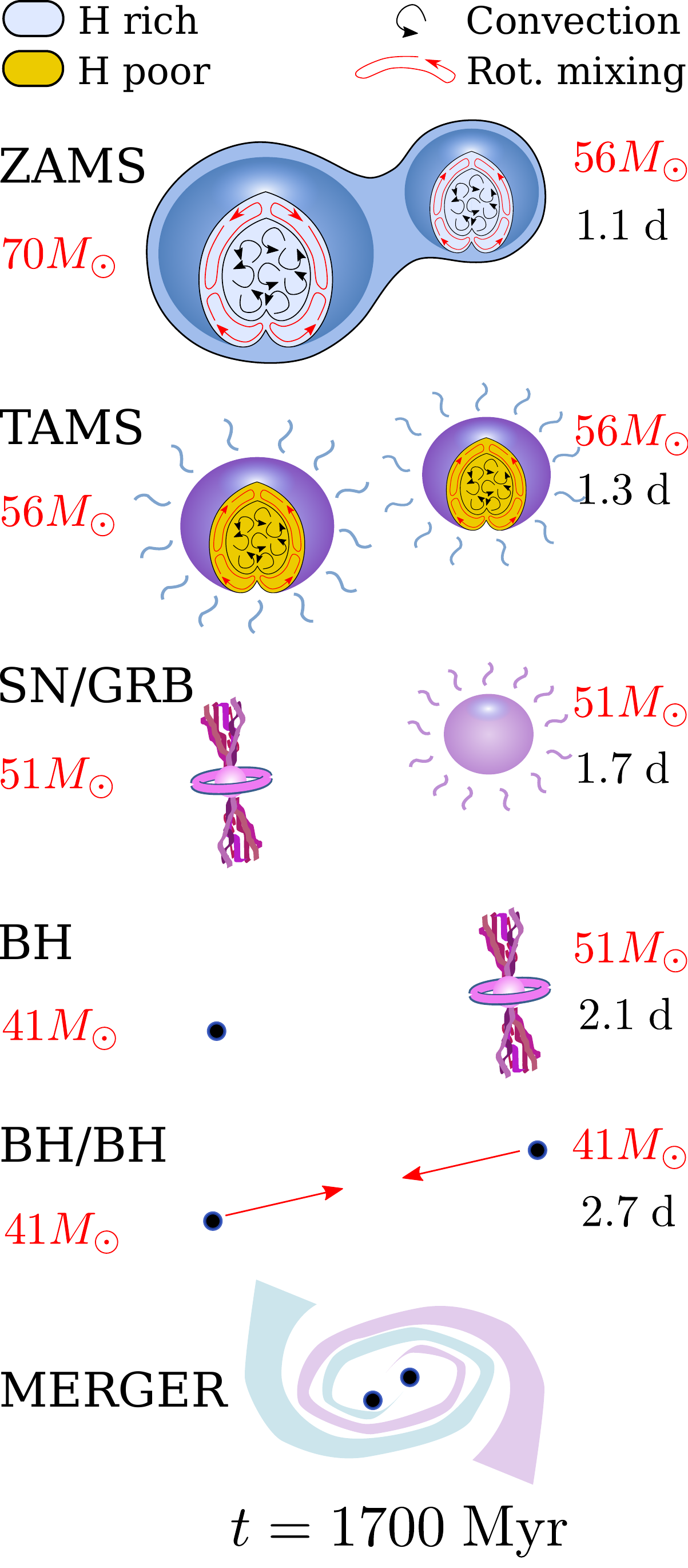}
\end{center}
   \caption{The CHE channel for the formation of
      double-BHs, including the occurrence of an overcontact phase as in
      \citet{Marchant+2016}.
      Numbers correspond to a system with
      $Z=Z_\odot/50$, initial masses
      $M_1=70M_\odot$ and $M_2=56M_\odot$, and a very short initial period at
      the zero-age main sequence (ZAMS). This
   system evolves early on into a contact configuration, where mass is
transferred back and forth until a mass ratio of unity is achieved. Efficient
rotational mixing distributes the helium rich material from the core throughout
the entire envelope, resulting in an almost pure helium star at the terminal-age
main sequence (TAMS). Depending on the final masses of each component, the system may
then proceed to form a compact BH binary that can merge within a Hubble time, or
explode as a PISNe. The models from \citet{Marchant+2016} had an
error in the computation of spin-orbit coupling which resulted in slightly wider orbits.
We have corrected for this, so the values differ slightly
from those in \citet{Marchant+2016}. We have also verified that the conclusions
of that work remain valid despite this issue.
}\label{fig:mob}
\end{figure}

The first observation from the twin LIGO detectors in Hanford and Livingston
of gravitational waves (GWs) from the inspiral and merger of
two $\sim 30M_\odot$ BHs (GW150914, \citealt{Abbott_GW150914_2016})
plays a particularly important role
in our understanding of ULX progenitor systems. Any formation channel that can
produce BHs above $30M_\odot$ is likely to be related to the formation of ULXs,
as the occurrence of RLOF would easily result in
very high luminosities. There are three main channels that can explain the
origin of GW150914, the classical field scenario involving
CE evolution \citep{TutukovYungelson1993,Belczynski+2016,Kruckow+2016}, the dynamical scenario in globular and
nuclear clusters \citep{PortegiesZwartMcMillan2000,Rodriguez+2016}, and the chemically homogeneous
evolution (CHE) channel
for field binaries \citep{MandeldeMink2016,Marchant+2016,deMinkMandel2016} which we
illustrate
in Figure \ref{fig:mob}. The ocurrence of CHE in binaries
was first proposed by \citet{deMink+2009} and has only recently been studied in more detail
\citep{Song+2016,MandeldeMink2016,Marchant+2016,deMinkMandel2016}.

Studying variations of channels for the production of GW sources
can then provide insight into
the origin of ULXs. For instance, in the CE scenario for merging binary BHs the primary is stripped
through stable mass transfer, collapses into a BH, and a CE phase happens when
the secondary expands to become a giant. However, considering the
possibility that stable mass transfer develops instead of a CE,
\citet{Pavlovskii+2017} showed that systems similar to the progenitor of
GW150914 could instead form a ULX with
a red supergiant as the donor. Recognizing different branches of binary BH
formation channels resulting not only in GW emission, but also in
electromagnetic waves, will play a fundamental role in constraining different
formation scenarios of GW sources.

In this paper we consider an alternative channel for the formation of
ULXs, which is a variation of the CHE channel for merging binary BHs. In this
scenario, the initial configuration is a very massive binary with a period
$\sim 1$ day and a mass ratio
far from unity. It is then expected that only the primary star undergoes efficient rotational mixing (as proposed by
\citealt{deMink+2009}), avoiding a binary interaction
before forming a massive BH. The less massive component in such a
system will evolve normally, eventually undergoing RLOF and
initiating mass transfer. As the resulting BH will usually be more massive than
the secondary, this results in a long-lived ULX phase, with mass transfer
proceeding
on a nuclear timescale. Such a channel of evolution
is strongly related to the formation of merging double BHs, PISNe and
LGRBs, and could be the source of the most luminous X-ray sources of stellar origin
that can be formed under any given conditions. In Section \ref{sect:methods} we
describe the setup of our stellar evolution models, and present our model for
ULX formation in Section \ref{sect:resultsULX}. We then discuss the consequences
of this channel for the luminosity distribution of ULXs in Section
\ref{sect:resultsLUM}, and their orbital parameters in Section
\ref{sect:orbit}.
In Section \ref{sect:resultsGW} we discuss the formation
of NS-BH and BH-BH binaries after a ULX phase, and the possibility to form
systems compact enough to merge in less than a Hubble time.
We give our concluding remarks in Section \ref{sect:conclusions}.

%%% Local Variables: 
%%% mode: latex
%%% TeX-master: "paper.tex"
%%% End: 

\section{Methods}
\label{sect:methods}
We extend the models computed by \citet{Marchant+2016} to lower 
initial mass ratios
to study the possibility of the primary evolving chemically
homogeneously and forming a BH, while the secondary evolves on a much longer
timescale, avoiding early interaction.
Our tool of choice for stellar modeling is version 8845 of
the \texttt{MESA} code
\citep{Paxton+2011,Paxton+2013,Paxton+2015}
\footnote{The inlist
   files and sources to reproduce our models are provided at
   \url{https://github.com/orlox/mesa_input_data/tree/master/2016_ULX}, together
with most of the data used for this paper.}.
We model about a $120\,000$ binary systems for metallicities in the range
$\log~Z=-2$ to $-6$ in steps of
$0.5$ dex, primary masses between
$\log M_1/M_\odot=1.5-2.5$ ($30M_\odot-300M_\odot$) in steps of $0.05$ dex, mass ratios
$q=M_2/M_1=0.05-0.6$ (which results in a range of secondary masses
between $1.5M_\odot-180M_\odot$)
in steps of $0.05$ and initial orbital periods between
$0.5$ and $3$ days, in steps of $0.05$ days. At higher mass ratios we expect the
formation of binary-BHs, as discussed in detail by \citet{Marchant+2016}.

\subsection{Stellar evolution}
For a given metallicity, the initial helium mass fraction is determined
by assuming that it linearly increases with metallicity from the primordial value $Y=0.2477$
\citep{Peimbert+2007} at $Z=0$ to $Y=0.28$ at $Z=Z_\odot$.
The value of the solar metallicity is taken as $Z_\odot=0.017$
\citep{Grevesse+1996}.

We use the Ledoux criterion for convection,
which we model using
mixing-length theory \citep{Bohm-Vitense1958} with a mixing length parameter
$\alpha=1.5$. In regions that are stable according to the Ledoux criterion, but
unstable according to the Schwarzschild criterion, we include semiconvective
mixing as in \citet{Langer+1983} with an efficiency parameter $\alpha_{\rm sc}=1$.
Opacities are computed using CO-enhanced tables from the OPAL project
\citep{IglesiasRogers1996} with solar scaled abundances from \citet{Grevesse+1996}. As we
do not need to follow the detailed nucleosynthetic evolution of our models, we
use the simple nuclear networks \texttt{basic.net}, \texttt{co\_burn.net} and
\texttt{approx21.net} that are provided with \texttt{MESA} and are
switched during runtime as needed to account for the later burning phases.

Stellar winds are implemented as in \citet{Brott+2011}, with mass loss for hot
hydrogen-rich stars modeled as in \citet{Vink+2001}.
Between a surface hydrogen composition of $X=0.7$ to $0.4$ we interpolate
between the Vink rate and a tenth of the mass loss rate for hydrogen-poor stars of
\citet{Hamann+1995}. For temperatures below that of the bi-stability jump, the rate
is taken as the maximum between the Vink rate and that of
\citet{NieuwenhuijzendeJager1990},
though in practice the stars we model remain blue over most of their
lifetimes, so this plays a negligible role. We scale the strength of stellar winds by a factor
$(Z/Z_\odot)^{0.85}$, extending the metallicity dependence predicted by 
\citet{Vink+2001} for O/B stars to hydrogen-poor Wolf-Rayet and cool stars.
Note
that the metallicity dependence of winds is constrained by observations of
massive stars in the Galaxy, the LMC and the SMC \citep{Mokiem+2007b}, the latter
of which has a metallicity of about $0.2\,Z_\odot$. Any model at lower
metallicities is neccesarily extrapolating these results. Measurements of mass loss at
lower metallicities have been reported by \citet{Tramper+2011}, but are
currently under dispute \citep{Bouret+2015}.

Rotational mixing and angular momentum transport are treated as diffusive
processes following \citet{HegerLanger2000}, including the effects of Eddington-Sweet
circulations, the Goldreich-Schubert-Fricke instability, and secular and dynamical shear, with an
efficiency parameter $f_c=1/30$ \citep{ChaboyerZahn1992}, and a sensitivity to composition gradients
parametrized by $f_\mu=0.1$ \citep{Yoon+2006}. We also include transport of angular
momentum due to the Spruit-Tayler dynamo \citep{Spruit2002} following the
implementation by
\citet{Petrovic+2005b}. The effect of centrifugal forces is modeled as in
\citet{EndalSofia1976}.
For the primary star, if central and surface helium mass fractions differ by
more than $0.2$, we consider the system not to be homogeneously evolving and
terminate the simulation.

\subsection{Binary evolution}

Both components in the binary are
assumed to be tidally locked at the ZAMS, although for the lowest mast ratios
and shortest orbital periods the Darwin instability \citep{Darwin1879} could make the formation of
such systems impossible. To account for this, we consider the minimum orbital
separation $a_{\rm Darwin}=\sqrt{3(I_1+I_2)/\mu}$ below which a binary would become
unstable, where $I_1$ and $I_2$ are the moments of inertia of both components
and $\mu$ is the reduced mass. Systems that have an initial orbital separation
below $a_{\rm Darwin}$ are ignored in our analysis.
Our models do not include the impact of
tidal deformation on stellar structure, but account for tidal synchronization
following \citet{Hurley+2002} and \citet{Detmers+2008}, which follow the model for
dynamical tides with radiative damping of
\citet{Zahn1975,Zahn1977}. The angular momentum deposited into each
component is distributed throughout the entire star, as described in
\citet{Paxton+2015}.

The evolution of orbital angular momentum considers the effects of
mass loss, gravitational wave radiation and spin-orbit coupling, as described in
\citet{Paxton+2015}. In particular, changes due to mass loss are
computed by assuming that winds carry the specific orbital angular
momentum corresponding to each component.

For these low mass ratios and short
orbital periods we expect systems to quickly evolve to an overcontact
configuration and a merger if the primary experiences RLOF. CE ejection
is unlikely in this case due to the short orbital period and the strongly bound envelope.
If the primary undergoes CHE, but the secondary expands and initiates mass transfer
before BH formation, the orbit is expected to widen as mass is transferred from
the less massive to the more massive star. This slows down the
rotation of the primary as it remains tidally locked. At the same
time, the
primary accretes hydrogen-rich material at its surface. These two
effects lead to the interruption of CHE. Because of this, we terminate our simulations if there is
mass transfer from any component before BH formation.

After BH formation, if the secondary does not evolve chemically homogeneously,
it will eventually expand and undergo RLOF.
Low-mass helium stars are expected to undergo an additional phase of mass transfer
after helium depletion, which is called Case ABB or BB depending on
whether the first mass-transfer event occurred before or after core-hydrogen
depletion \citep{DelgadoThomas1981,Dewi+2002}.
To study the possibility of forming merging BH-NS or BH-BH systems after
interaction, we need to consider this additional phase of mass transfer,
as due to the large mass ratios involved (the primary is expected to
become a BH of more than $20M_\odot$), it can result in
extreme orbital widening. To take this into account, we consider the evolution
of the secondary star until core carbon depletion, or until mass transfer
reaches hydrogen depleted regions. In case the latter happens, the orbit should
widen significantly making the system irrelevant as a source of GWs.

\vspace{0.1in}
\subsection{BH and NS formation}
\label{sect:bhnsform}

If the primary star evolves to helium depletion, with a
final mass outside the range $60-130M_\odot$, we assume it collapses directly to
a BH without losing mass or receiving a kick, while inside that range we assume the star explodes as a PISN
leaving no remnant \citep{HegerWoosley2002}.
The initial spin of the BH is computed as $a_0=Jc/M^2G$, meaning
that we assume all the spin angular momentum $J$ contained in the star
previous to collapse is retained. Note that this ignores possible mass loss due to PPISNe or 
LGRBs. PPISNe are expected to result in strong
mass loss for helium stars with final masses above $\sim 40M_\odot$, and to
limit the remnant mass to $\sim47M_\odot$ \citep{Woosley2016}.
Taking into account this effect would produce a reduction of $\sim 25\%$ in the
maximum luminosity we predict from sources below the PISN gap. This is much
smaller than the variations coming from uncertainties in the accretion rates.
LGRBs are also expected to occur through the collapsar scenario
\citep{Woosley1993} when the pre-collapse star has a large amount of angular
momentum that would result in $a_0>1$. In that case direct collapse into a BH is impossible without
shedding excess angular momentum and mass. For simplicity, we assume
direct collapse without mass loss to a maximally spinning black hole ($a_0=1$) when this happens.

For the donor star, the ULX phase results in its hydrogen envelope being
stripped, and its final mass plays a large role in determining whether it will evolve to
become a NS or a BH, and whether the binary system is disrupted or not.
For single stars there may not be a well defined threshold in
the ZAMS mass below which NSs are formed, and above
which the star collapses to a BH. Instead, detailed 1D models predict so-called
``islands of explodability'', where a range of initial masses results in NSs
and SNe explosions, but with lower and upper boundaries where BHs would be formed
instead \citep{Sukhbold+2016}. Translating this into a criterion for final
core-masses of envelope stripped stars is not straightforward, as the evolution
of these differs from that of single stars \citep{Brown+2001}.
Nevertheless, to study the final fate of our systems, we assume a simple
threshold for final masses of envelope stripped stars, and to take into account
possible uncertainties we vary this threshold between
$8M_\odot$ and $12M_\odot$. This range is chosen considering
predictions from some massive star models for which stars with helium core masses
up to $10M_\odot$ at core-collapse are predicted to explode as a SNe and produce a NS
\citep{Sukhbold+2016}.
For stars below the threshold we assume a $1.4M_\odot$ NS is
formed. There is also a lower mass threshold below which white dwarfs would
be formed instead of NSs, but in our simulations such low mass helium stars
undergo case ABB/BB mass transfer and lose their entire hydrogen envelopes so they are
already excluded from further analysis because of this.

We consider the effect of a kick on the newly-formed compact object, following a
Maxwellian distribution with a 1D root-mean-square (rms) $\sigma=265~\rm
km~s^{-1}$ for NSs \citep{Hobbs+2005}.
The \citet{Hobbs+2005} distribution for NS kicks might
be observationally biased towards larger kick velocities, as the small migration
distances of HMXBs with respect to their birth location
appear to favor smaller kicks \citep{ColeiroChaty2013}.
Systems undergoing Case ABB mass transfer can also result in
ultra-stripped CO-cores that might produce electron-capture and iron-core-collapse
supernovae (SNe), often with small kicks \citep{Tauris+2016}.
The post-kick orbital period $P_{\rm orb}$ and eccentricity
$e$ are computed following \citet{Tauris+1999} with no impulse velocity imparted
to the BH formed by the primary. If after the kick a bound system remains, we
compute its merger time due to radiation of GWs following \citet{Peters1964}.

In case the secondary forms a BH instead, for purely illustrative purposes we consider
the possibility of it receiving a kick, with $10\%$ of its mass being
lost.
This could be the case if instead of direct collapse a proto-neutron star is
formed first, with a weak explosion that unbinds only a small fraction of the
envelope, while the rest falls back \citep{FryerKalogera2001}.
We assume much weaker kicks
with $\sigma=26.5~\rm km~s^{-1}$, which mostly results in larger kicks than a momentum kick, where the BH
kick velocity is assumed to follow the NS kick distribution, scaled by
$1.4M_\odot/M_{\rm BH}$. Still, the strength of BH kicks remains quite
uncertain, with some arguing for weak kicks and
direct collapse \citep{MirabelRodrigues2003, Mandel2016, Adams+2016}, while
others argue for the opposite \citep{Repetto+2012, Janka2013}.

Modeling the effects of a kick on the BH formed by the primary is
significantly more difficult, as it requires us to sample different kick
velocities and directions and run individual binary stellar evolution models for
each. Nevertheless, we expect orbital periods well below $10$ days when the primary
collapses (see Section \ref{subsect:q}) for which small kicks would have little
impact.

\subsection{The Eddington limit for accretion to a BH}
A BH accreting matter through a disk at a rate $\dot{M}_{\rm acc}$ is expected
to have a luminosity $L_{\rm acc}=\eta \dot{M}_{\rm acc}c^2$, where $\eta\simeq
0.06-0.42$ is dependent on the position of the innermost-stable-circular orbit
(ISCO) of the BH, which in turn varies with its spin. If this radiation is emitted
isotropically, there is a limit at which the force exerted by radiation exceeds
the gravitational pull of the BH, which is given by the Eddington luminosity,
\begin{eqnarray}
   L_{\rm Edd}&=&\frac{4\pi G M_{\rm BH}c}{\kappa}\\
   &=&1.47\times 10^{39} \left( \frac{M_{\rm BH}}{10M_\odot}\right)
   \left( \frac{1+X_{\rm s}}{1.7} \right)^{-1}\;\rm erg\;s^{-1},
\end{eqnarray}
where $X_{\rm s}$ is the surface hydrogen mass fraction of the donor, and we
have assumed electron scattering to be the main source of opacity. The
mass-accretion
rate at which this luminosity is reached, is given by
\begin{eqnarray}
   \dot{M}_{\rm Edd}=2.6\times10^{-7}\left( \frac{M_{\rm BH}}{10M_\odot}\right)
   \left(\frac{1+X_{\rm s}}{1.7} \right)^{-1}\left( \frac{\eta}{0.1}
   \right)^{-1}\;M_\odot\rm\;yr^{-1},
\end{eqnarray}
and we assume that the accretion rate $\dot{M}_{\rm acc}$ is limited to this
value, i.e. if the mass-transfer rate from the donor is $\dot{M}_{\rm mt}$, then
$\dot{M}_{\rm acc}=\min(\dot{M}_{\rm Edd},\dot{M}_{\rm mt})$, with the
non-accreted material being ejected from the system with the specific orbital angular
momentum of the BH.

Although there are many
indications that beamed emission can allow for much higher mass-transfer rates and
luminosities, in particular in the case of the NS ULX \citep{Bachetti+2014}, we
consider isotropic emission as a lower limit
for the luminosities these objects can achieve. As it will be shown in Section
\ref{sect:resultsULX}, the high BH
masses produced through CHE can easily account for
ULX luminosities without the need of super-Eddington accretion;
the highest luminosities observed can be reached by accreting at only 3 times the Eddington
rate. As the energy
released as radiation will not contribute to the BH mass, it increases as
\begin{eqnarray}
   \dot{M}_{\rm BH}= (1-\eta)\dot{M}_{\rm acc},
   \label{equ:mdotbh}
\end{eqnarray}
and the remaining contribution $\eta \dot{M}_{\rm acc}$ that is radiated away
takes as well the angular momentum corresponding to the specific orbital angular momentum of the BH.

Following \citet{Podsiadlowski+2003}, we consider the evolution of the BH
spin as it accretes material, which for a BH with zero initial spin and mass $M_{\rm BH,0}$,
results in \citep{Bardeen1970}
\begin{eqnarray}
   \eta &=& 1-\sqrt{1-\left( \frac{M_{\rm BH}}{3M_{\rm BH,0}} \right)^2},\\
   a&=&\sqrt{\frac{2}{3}}\frac{M_{\rm BH,0}}{M_{\rm BH}}\left( 4 -
   \sqrt{ 18\left( \frac{M_{\rm BH,0}}{M_{\rm BH}} \right)^{2}-2
   }
   \right),
\end{eqnarray}
so long as $M_{\rm BH}<\sqrt{6}M_{\rm BH,0}$. If the BH mass reaches
$\sqrt{6}M_{\rm BH,0}$, we assume $a=1$ and $\eta=0.42$, though in practice the
absorption of radiation from the disc can produce a torque that limits the BH
spin to $\simeq 0.998$ \citep{Thorne1974}, with a correspondingly lower
$\eta$. If the BH has a non-zero initial spin parameter $a_0$, then we can still
make use of these expressions by computing an effective initial BH mass
$M_{\rm BH,0}^{\rm eff}$, corresponding to a BH with zero spin that would reach
$a=a_0$ after accreting material up to $M_{\rm BH,0}$. This effective mass can
be easily computed from a simple relation between the radius of the ISCO
$r_{\rm ISCO}$ and the
mass of the BH as it accretes \citep{Bardeen1970,Bardeen+1972}. If
$z=r_{\rm ISCO}/M_{\rm BH}$, then in geometrized units we have
\begin{eqnarray}
   M_{\rm BH,0}^{\rm eff}=\sqrt{\frac{z(a_0)}{6}}M_{\rm BH,0},
\end{eqnarray}
which reduces to $M_{\rm BH,0}^{\rm eff}=M_{\rm BH,0}/\sqrt{6}$ for $a_0=1$ and
$M_{\rm BH,0}^{\rm eff}=M_{\rm BH,0}$ for $a_0=0$, as expected.
Although no black hole in our models increases its mass by a factor
of $\sqrt{6}$, several are formed that are maximally rotating or close to
$a_0=1$.

%%% Local Variables: 
%%% mode: latex
%%% TeX-master: "paper.tex"
%%% End: 

\section{Formation of ULXs through CHE}
\label{sect:resultsULX}

\begin{figure}%[!ht]
   \begin{center}
   \includegraphics[width=\columnwidth]{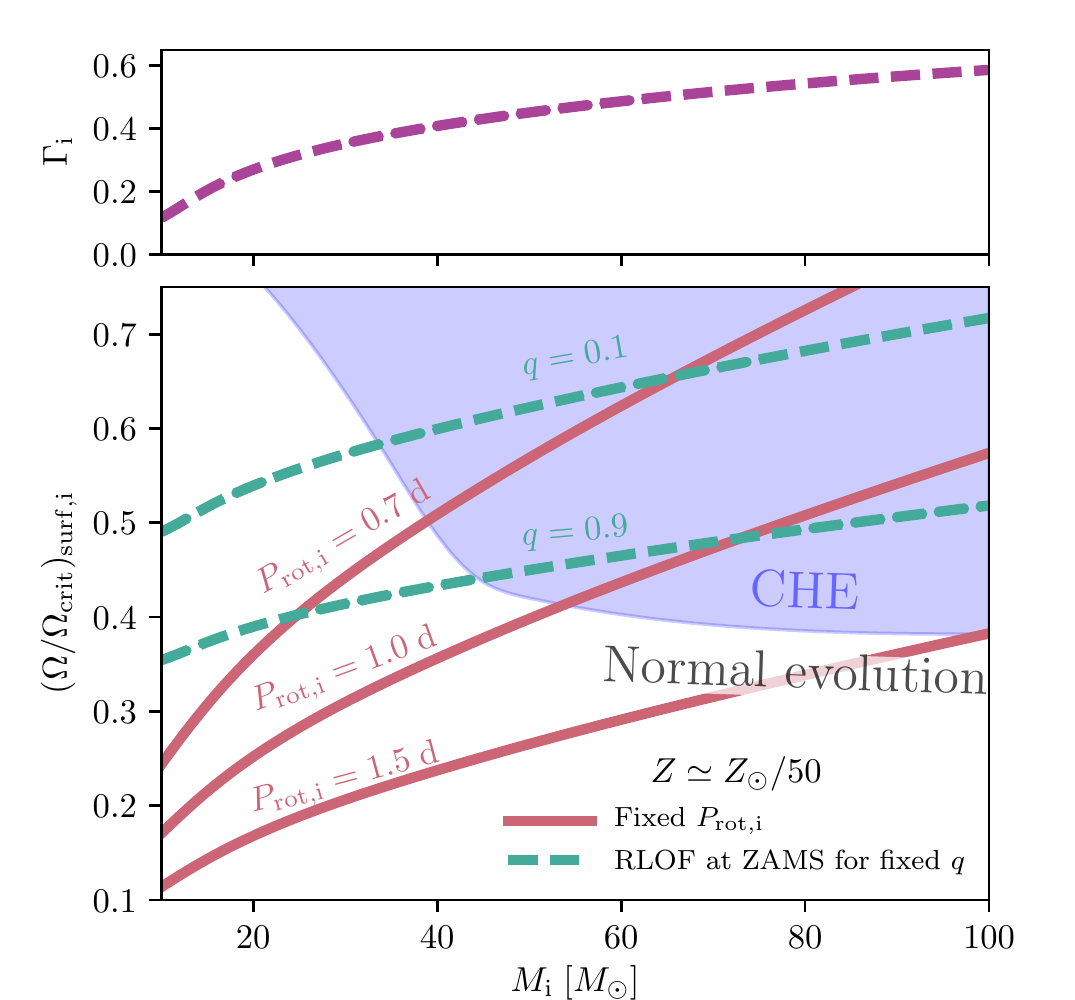}
   \end{center}
   \caption{(top) Eddington factor at the ZAMS for non-rotating stars at a
      metallicity $Z=10^{-3.5}\simeq Z_\odot/50$. (bottom) Initial conditions
      for the occurrence of CHE in single stars at the same metallicity, in terms of mass at the ZAMS and initial ratio of
   angular frequency $\Omega$ to its critical value $\Omega_{\rm crit}$ at the
   surface. The colored region indicates stars for
which the surface helium abundance at TAMS exceeds 0.8 and is a sharp
transition. Solid lines correspond to a fixed initial rotational period, while
dashed lines indicate the value of $\Omega/\Omega_{\rm crit}$ if the star is the
primary of a binary system at a fixed mass ratio $q$ which exactly fills its
Roche lobe at the ZAMS (see Equation \ref{equ_omegarlof}).}
\label{fig:chediagram}
\end{figure}

Our proposed model for ULX formation involves binary systems at low mass ratios,
where the more massive component undergoes CHE, while the secondary evolves
normally. This is in contrast to the CHE binary BH formation channel which requires
mass ratios closer to unity, for which both stars evolve chemically
homogeneously. Because of this it is important to understand under which
conditions one, both or neither of the components of a binary would experience
efficient rotational mixing. To illustrate this, Figure \ref{fig:chediagram}
shows the required initial rotation rates (in terms of the ratio of the angular
frequency to its critical value $\Omega/\Omega_{\rm crit}$) for which single
stars with a given ZAMS mass would undergo CHE, determined from a grid of single
star models computed with \texttt{MESA}.
The critical value of the angular frequency depends on the Eddington factor
$\Gamma$ at the surface of the star, and is given by \citep{Langer1997}
\begin{eqnarray}
   \Omega_{\rm crit}=\sqrt{\frac{GM}{R^3}(1-\Gamma)},\quad
   \Gamma\equiv \frac{L}{L_{\rm Edd}}=\frac{\kappa}{4\pi c G}\frac{L}{M},
\end{eqnarray}
where $\kappa$ is the opacity at the surface of the star.

Despite the reduction in stellar lifetimes with mass,
rotational mixing is expected to play a larger role
for the more massive stars, owing to the increasing importance of radiation
pressure which reduces the stability of the stratification in the radiative
envelope \citep{Yoon+2006}, and to the
larger mass of the convective cores relative to the total mass
\citep[see, eg.][]{Kohler+2015}.
This makes the
threshold $\Omega/\Omega_{\rm crit}$ for efficient mixing decrease with mass.
In contrast, $\Omega_{\rm crit}$ at the ZAMS decreases with mass, so at a
constant initial rotation period $\Omega/\Omega_{\rm crit}$ increases with mass,
as is shown by the solid lines in Figure \ref{fig:chediagram}. If we consider a
binary with tidally locked components, this means that
for mass ratios close to unity both stars can be inside the region for CHE,
while for lower mass ratios the less
massive component would evolve normally.

Another important point is that to form a ULX the binary has to avoid
RLOF before the primary forms a BH. This again is in contrast to the binary-BH
formation channel with CHE, where the detailed simulations of
\citet{Marchant+2016} showed that most systems need to be in contact to undergo
efficient rotational mixing. If the primary is tidally locked such
that its rotational frequency is $\Omega=\sqrt{G(M_1+M_2)/a^3}$,
the largest possible value that $\Omega/\Omega_{\rm crit}$
can have while avoiding mass transfer results when the primary is filling its Roche
lobe ($R_1=R_{\rm RL,1}=f(q)a$). In this case
$\Omega/\Omega_{\rm crit}$ is only a function of the mass ratio and $\Gamma$,
\begin{eqnarray}
   \left(\frac{\Omega}{\Omega_{\rm crit}}\right)_{\rm max}=\sqrt{\frac{(1+q)f(q)^3}{1-\Gamma}}
   \label{equ_omegarlof},
\end{eqnarray}
which is equal to $0.46(1-\Gamma)^{-1/2}$ and $0.33(1-\Gamma)^{-1/2}$ for
$q=0.1$ and $q=0.9$ respectively.
This is shown with dashed lines in Figure \ref{fig:chediagram} and it explains why
binaries with lower mass ratios can experience CHE while avoiding contact
\citep[cf. ][]{Yoon+2006,deMink+2009}.
For lower mass ratios,
binaries can have shorter orbital periods without undergoing RLOF, allowing for a larger range of
systems where the primaries fall into the CHE region.
The requirement of having a system that avoids RLOF at the ZAMS also limits the
minimum primary mass at which CHE evolution can happen in a binary.
In all of our binary simulations, the least massive primary that evolves chemically homogeneously
has an initial mass $M_1=45M_\odot$.

Although Figure \ref{fig:chediagram} is useful to illustrate the requirements
for CHE in a binary system, this boundary depends on how rotation rates change
due to mass loss through the full main sequence evolution, and this is different
for single and binary stars. In a tidally synchronized binary, changes in the
rotational period depend on how mass loss alters the orbital period, and this is
mass-ratio dependent. To assess whether a binary would undergo CHE we then need to
model each individual system in detail. In what follows, we describe in more detail how a ULX is formed through CHE,
what sets the lower and upper limits in mass ratio for the formation of ULXs, and discuss the effect of metallicity
on this channel.

\subsection{Mass-ratio dependence and sample case of ULX formation}
\label{subsect:q}

To exemplify the formation of a ULX via CHE, let
us consider the evolution of systems with metallicity $Z=10^{-3.5}\simeq
Z_\odot/50$ and primary mass $M_1=70M_\odot$
as the example shown in Figure \ref{fig:mob}, but for three different mass ratios
$q=0.05,0.2$ and $0.6$. For a
mass ratio $q=0.05$ (a secondary mass of $3.5M_\odot$) and an initial period of
$0.8~{\rm days}$, the primary is close to filling its Roche-lobe at the ZAMS,
with $R/R_{\rm RL}\simeq 0.92$. However, this binary would have an
initial orbital separation of $15R_\odot$, while the minimum separation at which
the system would avoid the Darwin instability is $a_{\rm Darwin}=21R_\odot$. Because of
this we do not expect this system to be formed, as it would have resulted in a
merger instead of a tidally synchronized binary which is our assumed initial
state.

\begin{figure}%[!ht]
   \begin{center}
   \includegraphics[width=\columnwidth]{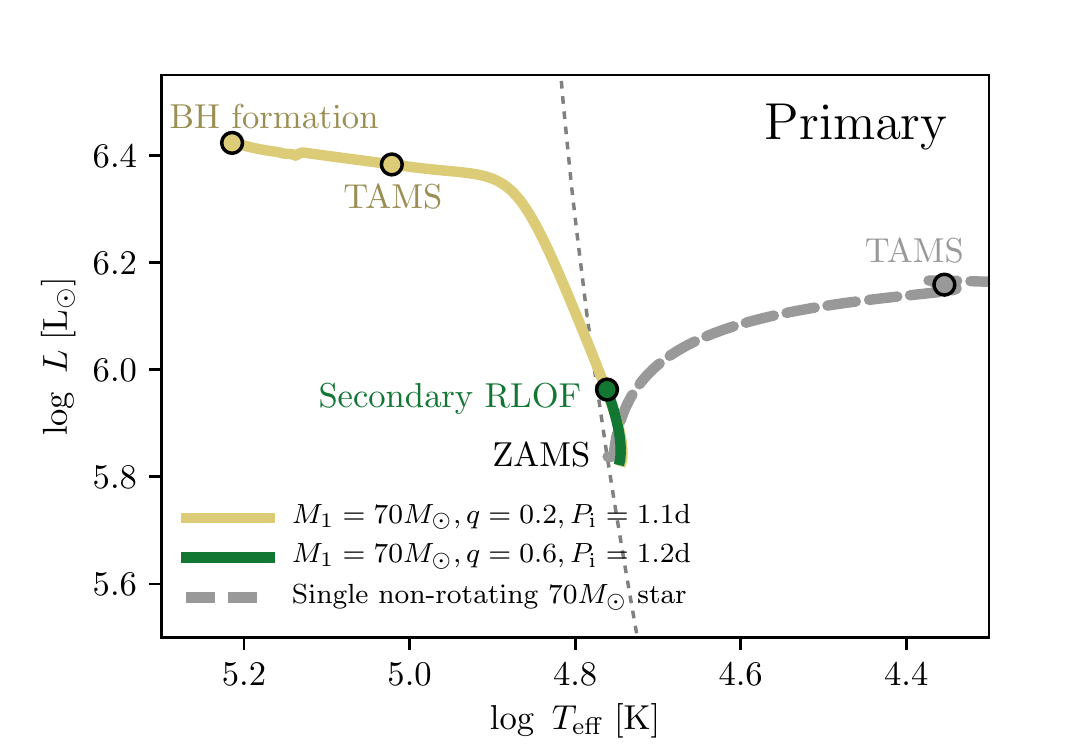}
   \includegraphics[width=\columnwidth]{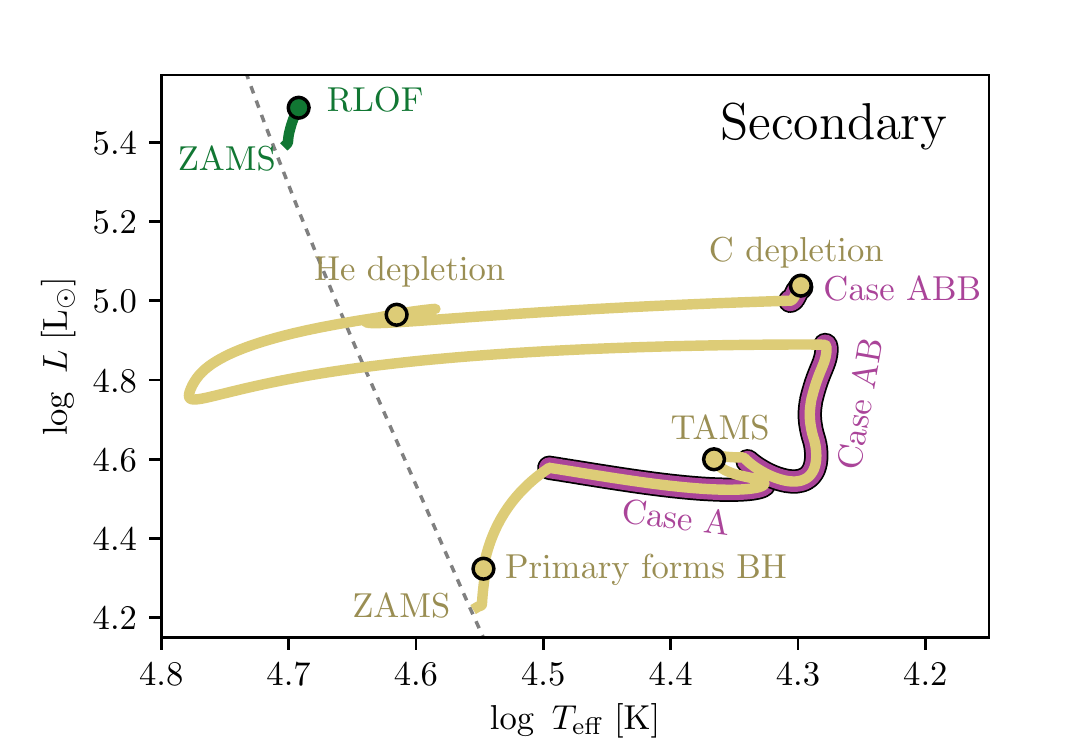}
   \end{center}
   \caption{Evolution in the Hertzsprung-Russell diagram of the primary (top) and secondary (bottom) stars in
   binary systems with $Z=10^{-3.5}\simeq Z_\odot/50$ consisting of a
   $70M_\odot$ primary with mass ratios
$q=0.2,0.6$ and initial orbital periods that are close to producing RLOF at
the ZAMS. The dotted line shows the location of the ZAMS for non-rotating stars,
and the track of a non-rotating $70M_\odot$ star is also shown for reference.
The system with initial mass ratio $q=0.6$ has the primary evolving chemically
homogeneously, but the secondary initiates mass transfer before a BH is formed.
The system with initial $q=0.2$ manages to form a BH and afterwards undergoes three
distinct phases of mass transfer. See Section \ref{subsect:q} for
details.}
\label{fig:HR}
\end{figure}

At larger initial mass ratios the initial configuration is not Darwin unstable,
and we show in Figure \ref{fig:HR} the evolution in the Hertzsprung-Russell diagram
for two systems with mass ratios $q=0.2$ and $0.6$. For an initial
mass ratio $q=0.6$ (a secondary mass of $42M_\odot$) and an initial period of
$1.2~{\rm days}$, after $1.6~{\rm Myrs}$ the orbital separation is still
$1.2~{\rm days}$, and the primary  experiences a significant amount of
mixing, with $Y_{\rm c}=0.44$ and $Y_{\rm s}=0.35$. However, the secondary does not evolve
homogeneously, and by this point it has expanded enough to undergo RLOF. Since
the secondary is the less massive component, mass transfer will widen the orbit
and transfer hydrogen-rich material on the surface of the primary.
The steep change in mean molecular weight at the base of the accreted material
prevents it from mixing inwards, so we terminate the simulation as we expect the
system to break away from CHE.

\begin{figure}%[!ht]
   \begin{center}
   \includegraphics[width=0.85\columnwidth]{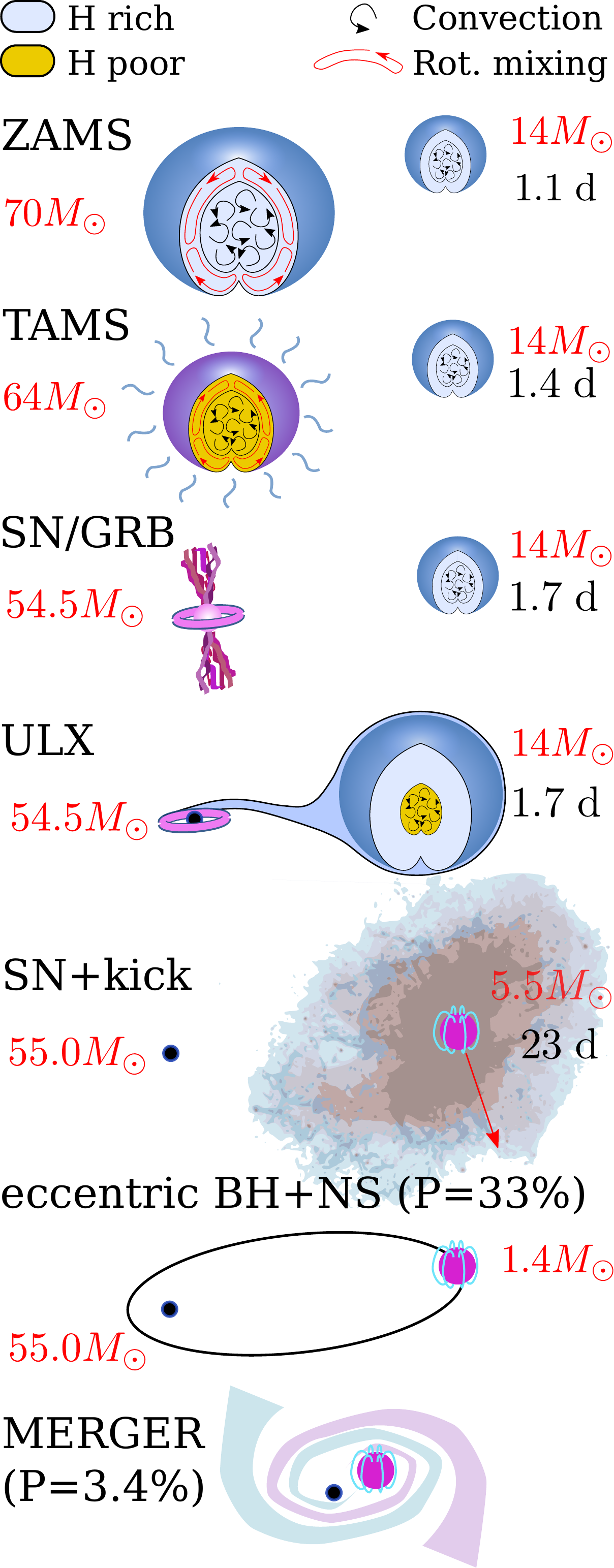}
   \end{center}
   \caption{Schematic evolution of a ULX model arising from CHE
   of the more massive component in a compact binary with
   unequal masses and $Z=10^{-3.5}\simeq Z_\odot/50$ (see Section
   \ref{subsect:q} for details). The phase of RLOF actually
   corresponds to three distinct mass-transfer phases. If at the moment of formation of the NS (or BH for the
most massive secondaries) there is a kick in a favorable direction, a compact
enough system can be formed such that a merger is possible in less than a Hubble
time. For this system in particular, assuming a Maxwellian kick distribution
with 1D root-mean-square $\sigma=265~{\rm
   km~s^{-1}}$, there is a $67\%$ chance that the binary is disrupted, and a
$3.4\%$ chance that it results in a NS+BH merger in less than a Hubble time.
For simplicity, mass loss at the moment of formation of the first BH is ignored.}
\label{fig:ulx}
\end{figure}

To form a ULX, a system with a mass ratio high enough to avoid the
Darwin instability, but small enough to avoid interacton before forming a BH is
needed. This is the case for the system shown in Figure \ref{fig:ulx},
depicting the evolution for an initial mass ratio $q=0.2$ and an
initial period of $1.1~{\rm days}$. The primary in this system evolves
chemically homogeneously, depleting central helium after $4.6~{\rm Myrs}$. At
this point the orbital period has slightly increased to $1.7~{\rm days}$, but
more importantly, the secondary has barely evolved, and its core hydrogen
mass fraction is $X_{\rm c}=0.62$.
At core helium depletion the primary is still rapidly rotating, with a dimensionless
spin angular momentum $a_0 = 1.25$ and a mass of $55M_\odot$. As discussed
in Section \ref{sect:bhnsform}, we ignore the possibility of a PPISN or a LGRB,
and assume the star collapses directly into a $55M_\odot$ BH with $a=1$. $12.6~{\rm Myrs}$ after the
formation of the system, and with $X_{\rm c}=0.24$, the secondary overflows its
Roche-lobe and
undergoes a phase of Case A mass transfer lasting $1.6~{\rm Myrs}$, and
reducing its mass from $14M_\odot$ to $8.6M_\odot$, while widening
the orbit from $1.7$ to $6.5$ days. The typical mass-transfer rate during
this phase is $~\dot{M}_{\rm mt}=10^{-5.7}~M_\odot~{\rm
yr^{-1}}$, which is only a factor of five above the Eddington rate of the BH.
The Eddington luminosity of the BH exceeds $8\times10^{39}~\rm
erg~s^{-1}$, so during mass transfer the system would be an ultra-luminous X-ray source.

After the secondary depletes its central hydrogen, it expands to undergo a
short-lived phase (lasting only $28000~{\rm yrs}$) of Case B mass transfer
which reduces its mass to $5.8M_\odot$, with mass-transfer rates as high as
$\dot{M}_{\rm mt}=10^{-3.4}~M_\odot~{\rm yr^{-1}}$.
At detachment the orbital period is
$20~{\rm days}$, and the star has a helium core of $3.6 M_\odot$, with a
significant hydrogen-rich envelope left. During core helium burning most
of the envelope is turned into pure helium, resulting in a $5.8M_\odot$ star with a
$5.1M_\odot$ hydrogen-depleted core. After helium depletion, the remaining envelope expands
and manages to initiate Case BB mass transfer; however carbon 
ignites during mass transfer and
is rapidly depleted after only $0.3M_\odot$ is transferred,
though this is already  enough to increase the orbital period to $23~{\rm
days}$. Note that the overall efficiency of all mass-transfer
phases is low, with the BH increasing its mass only by $0.5M_\odot$.
Assuming the $5.5M_\odot$ star explodes as a SN with a possibly strong kick oriented
in a random direction, there is a small chance ($3.4\%$) that the system remains
in a tight and very eccentric orbit that would allow a BH-NS merger within a Hubble
time (see Section \ref{sect:resultsGW}).

In general, considering our complete set of simulations, we find that
ULXs can be formed for initial mass ratios in the range $q\simeq0.1-0.45$. The
lower limit on mass ratios is a product of the Darwin instability, while the
upper limit results because secondaries initiate RLOF before BH formation,
interrupting CHE. For reference, the detailed outcomes of all our models are
shown in Appendix \ref{appendix:grids}.

\subsection{The impact of metallicity on the properties of ULXs}
\label{sect:Zdep}
\begin{figure}%[!ht]
   \begin{center}
   \includegraphics[width=\columnwidth]{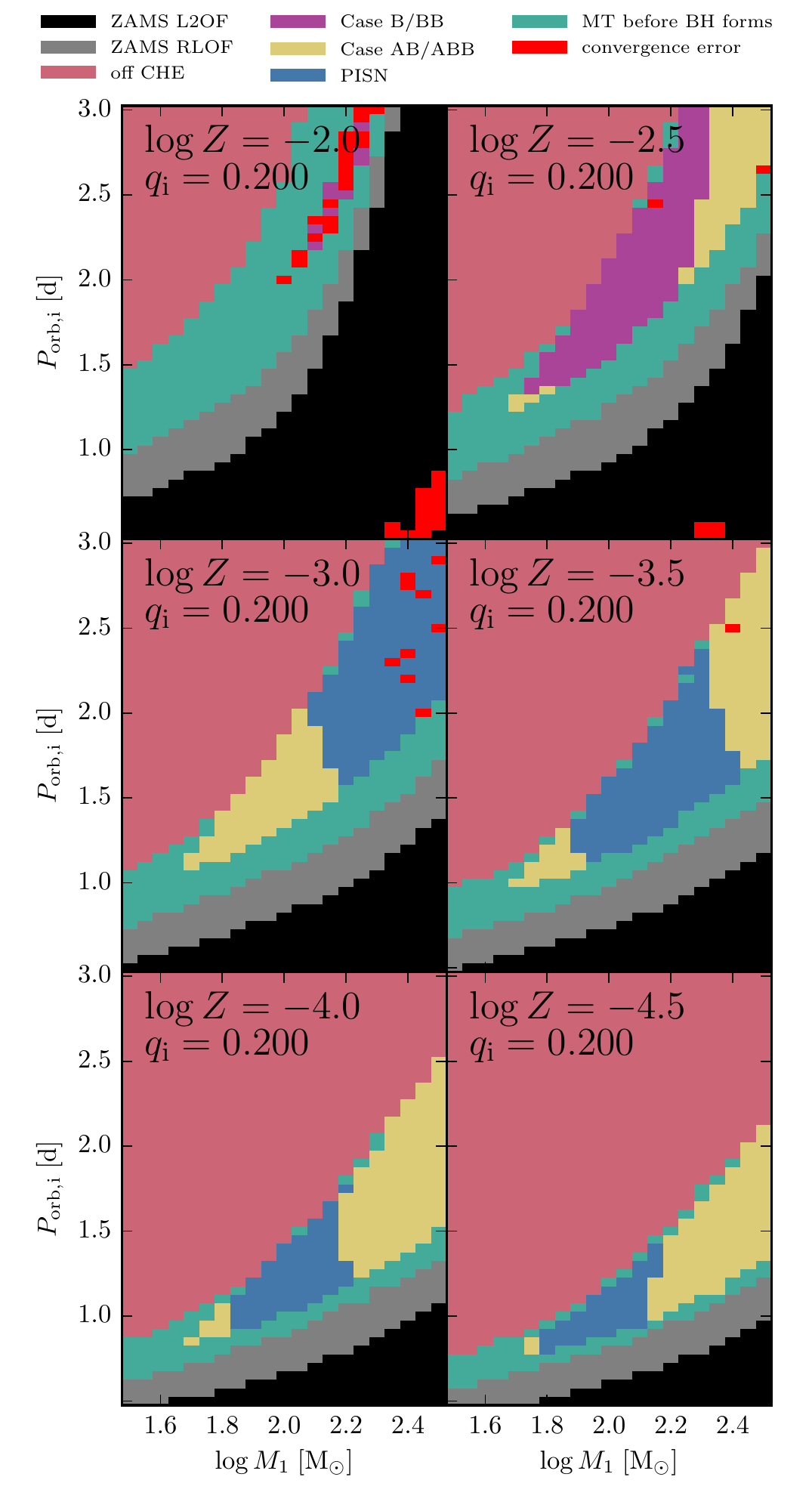}
   \end{center}
   \caption{Outcome of simulations for different metallicities and a fixed mass
   ratio $q$. Systems marked as Case B/BB or Case AB/ABB have primaries that
evolve chemically homogeneously and form BHs, to which the secondary then
transfers mass resulting in a ULX. Systems marked in blue have primaries that evolve chemically
homogeneously but have final masses resulting in PISNe. All other systems interact
before the formation of a BH and would not form a ULX. See
Appendix \ref{appendix:grids} for the outcome of all simulations. and a detailed
description of all outcomes.}
\label{fig:minisummary}
\end{figure}

Figure \ref{fig:minisummary} shows the outcome of simulations with $q=0.2$ for some of the
metallicities modeled. At a fixed mass ratio and metallicity, the initial
primary masses and orbital periods for which the primary can evolve chemically homogeneously are
very similar to those that can produce binary BHs from initial mass ratios closer to
unity \citep{Marchant+2016}, the main difference being that the period window for
contact-less evolution is much larger.
Although wind mass loss typically
disfavors CHE as it brakes the star's rotation, for the most massive primaries it
can help expose helium-enriched layers from their large convective cores,
significantly widening the window for this channel at the highest masses
\citep{Kohler+2015,Szecsi+2015}.

The properties of ULXs produced through CHE are strongly dependent on
metallicity. Metal-poor stars are more compact, making it possible for binaries
with the same component masses to have shorter initial orbital periods while
still avoiding RLOF at the ZAMS, as can be seen in Figure \ref{fig:minisummary}. Although
shorter initial orbital periods result in faster surface rotation velocities,
this does not translate into more systems undergoing CHE, as the relevant quantity
to consider mixing efficiency is not the absolute rotational velocity, but
rather its ratio to the critical velocity, which also increases as stars become
more compact at lower metallicity. The effect of metallicity-dependent
mass loss is more complex. For the highest metallicity modeled, $Z=0.01$, mass
loss results in significant orbital widening, which together with tidal coupling
significantly spins down the primary and results in very few systems evolving
chemically homogeneously\footnote{Most of our models with $Z=0.01$ that evolve
chemically homogeneously could not be modeled up to helium depletion due to
numerical issues arising from envelope inflation.
Still, only a small number of those models undergo this
channel of evolution, and only for very high primary masses, so at these high
metallicities the channel is negligible.}. In contrast, for extremely low metallicities,
reduced winds mean that the window for the channel does not widen too much at the
highest masses.

For systems undergoing CHE, mass loss
determines the occurrence of PISNe. At $\log~ Z=-2.5$, mass loss is strong enough
that systems with initial masses of $300M_\odot$ result in helium cores below
$60M_\odot$, avoiding explosion as a PISNe and producing BHs. At a metallicity
of $\log~ Z=-3$, the
most massive primaries have final masses well above $60M_\odot$, which we would
expect to explode as PISNe. At $\log~ Z=-3.5$ mass loss has reduced to the point
where we get primaries with final masses above $130M_\odot$, that could possibly
avoid the PISNe fate and instead produce very massive BHs. This would
translate into a gap in BH masses. At even lower metallicities, the
region where PISNe occur moves further down in terms of initial primary mass,
and as mass loss becomes negligible, the period window for CHE
becomes narrower.
This narrowing increases the minimum primary mass at which CHE occurs,
such that at an extremely low metallicity of $\log~
Z=-6$ only primaries above $70M_\odot$ undergo CHE. As mass loss is very weak,
these stars still fall into the mass range for PISNe, and there is no longer a gap
in BH masses; all the resulting BHs come from systems above the mass limit for PISNe.

\begin{figure}%[!ht]
   \begin{center}
   \includegraphics[width=\columnwidth]{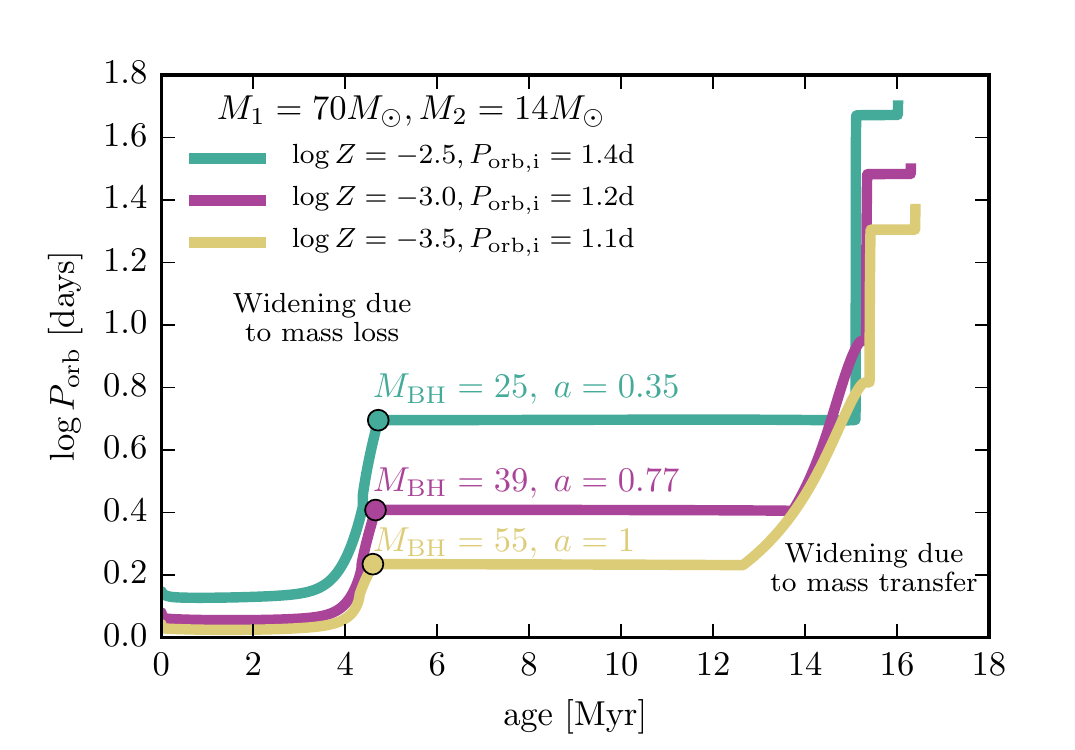}
   \end{center}
   \caption{Orbital separation as a function of time for three different systems
   with the same initial component masses and different metallicities. The
initial periods correspond to the shortest one in our simulations for which a
system with those component masses and metallicity evolves chemically
homogeneously and
undergoes a ULX phase. Circles mark the moment of BH formation, and the initial
mass and spin of the BH are shown.}
\label{fig:period}
\end{figure}

\begin{figure}%[!ht]
   \begin{center}
   \includegraphics[width=\columnwidth]{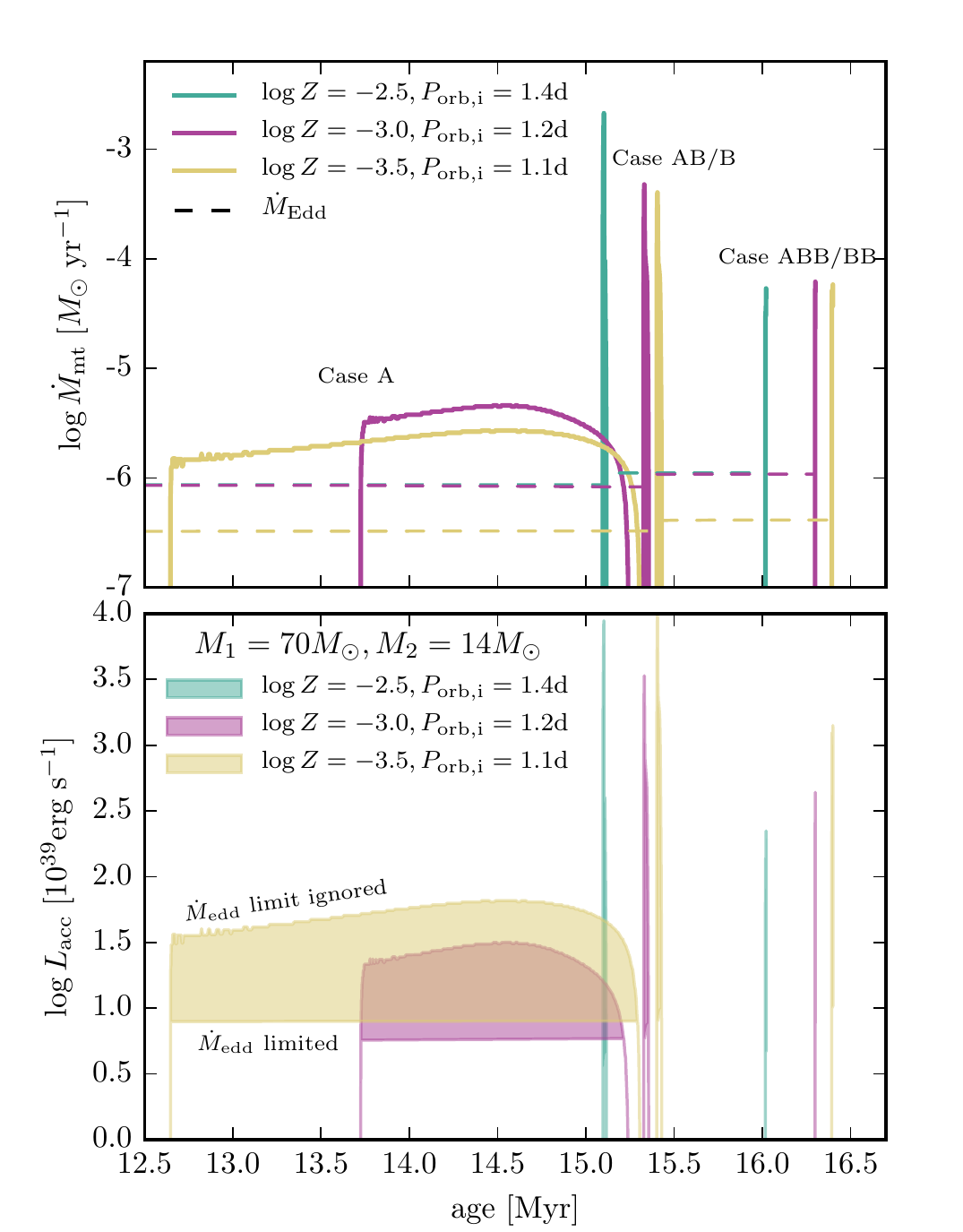}
   \end{center}
   \caption{Mass-transfer rates and accretion luminosities during mass-transfer
      phases for the three systems shown in Figure \ref{fig:period}. Accretion
   luminosities are shown as ranges going from the lower values given by strictly Eddington limited
accretion rates, to the potential luminosities that could be achieved if the Eddington
limit is ignored. Note that the Eddington accretion rates for the metallicities
$\log~ Z=-2.5$ and $-3$ overlap. This is due to the lower BH spin at
$\log~
Z=-2.5$ compensating for the higher BH mass at $\log~ Z=-3$.}
\label{fig:masstransfer}
\end{figure}

Mass loss of the primary star also affects the lifetime of a possible ULX phase.
Figures \ref{fig:period} and \ref{fig:masstransfer} show the
evolution of the orbital period and the mass-transfer phases for three of our
ULX models with initial primary masses of $70M_\odot$ and metallicities
$\log~ Z=-2.5,-3$ and $-3.5$. The highest metallicity model widens significantly
due to mass loss before the BH forms, resulting in the secondary initiating
RLOF only after core hydrogen depletion. This Case B mass-transfer
phase is short-lived, making it unlikely to detect ULXs during this
phase of evolution. In contrast, the two lower metallicity systems remain
compact enough after BH formation to undergo long-lived phases of nuclear-timescale
mass transfer,
with the duration of these increasing at lower metallicities as the orbit
widens less and mass transfer starts earlier while the secondary undergoes
core-hydrogen burning. The
resulting luminosities for these Case A systems are well above
$10^{39}~{\rm erg~s^{-1}}$ even when strictly limited to the Eddington rate.
During Case A, mass-transfer rates are not much higher than the Eddington rate,
which means that even if the Eddington limit is ignored, luminosities can only
increase by a factor of $\sim 5$. The situation is different during Case AB/B
and ABB/BB mass transfer, where mass-transfer rates can go many orders of
magnitude above $\dot{M}_{\rm Edd}$, resulting in potential luminosities going
above $10^{42}~{\rm erg~s^{-1}}$, which is the range for HLXs. However,
achieving these luminosities requires a complete disregard of the Eddington limit,
and even then, the short lifetimes involved would likely make these sources very rare. Note
that, with mass accretion limited to the Eddington rate, the BHs modeled
have only modest increases in their total masses and spins. The small increase
in $\dot{M}_{\rm Edd}$ that can be observed during Case AB/B mass transfer in
Figure \ref{fig:masstransfer} is only due to a decrease in the opacity of
accreted material, as helium rich layers from the secondary are exposed.

%%% Local Variables: 
%%% mode: latex
%%% TeX-master: "paper.tex"
%%% End: 

\section{Luminosity distribution function of ULXs}
\label{sect:resultsLUM}
To estimate the expected properties of observed ULX samples at a fixed
metallicity, we need to assume certain distribution functions {\rm describing
the population of binaries at zero age}. We follow the
choices made by \citet{Marchant+2016}, which consider a Salpeter distribution for
primary masses ($dN/d M_{1,\rm i}\propto M_{1,\rm i}^{-2.35}$), a flat distribution in
$\log~ P_{\rm orb}$ ranging from $0.5$ days to a year, a flat distribution in
mass ratio from zero to unity, and a binary fraction $f_b=0.5$ (i.e. out of
three massive stars two form part of a binary system). If we
assume the threshold mass for SNe is $8M_\odot$, and that the SN rate
is $10^{-2}~{\rm yr^{-1}}$ for a star-formation rate (SFR) of ${\rm
1 M_\odot~yr^{-1}}$, we can then compute expected distributions of
luminosities per $M_\odot~{\rm yr^{-1}}$ of SFR. 
This choice for the rate of SNe per $M_\odot~{\rm yr^{-1}}$ of SFR
is consistent with Milky Way values \citep{Diehl+2006,RobitailleWhitney2010}.
Note that the distributions we obtain depend linearly on this assumed ratio
between the supernova rate and the SFR, which is
uncertain to at least a factor of $2$. A detailed description
of how we derive formation rates and observable numbers of ULXs is provided in
Appendix \ref{appendix:rates}.

\begin{figure}%[!ht]
   \begin{center}
   \includegraphics[width=\columnwidth]{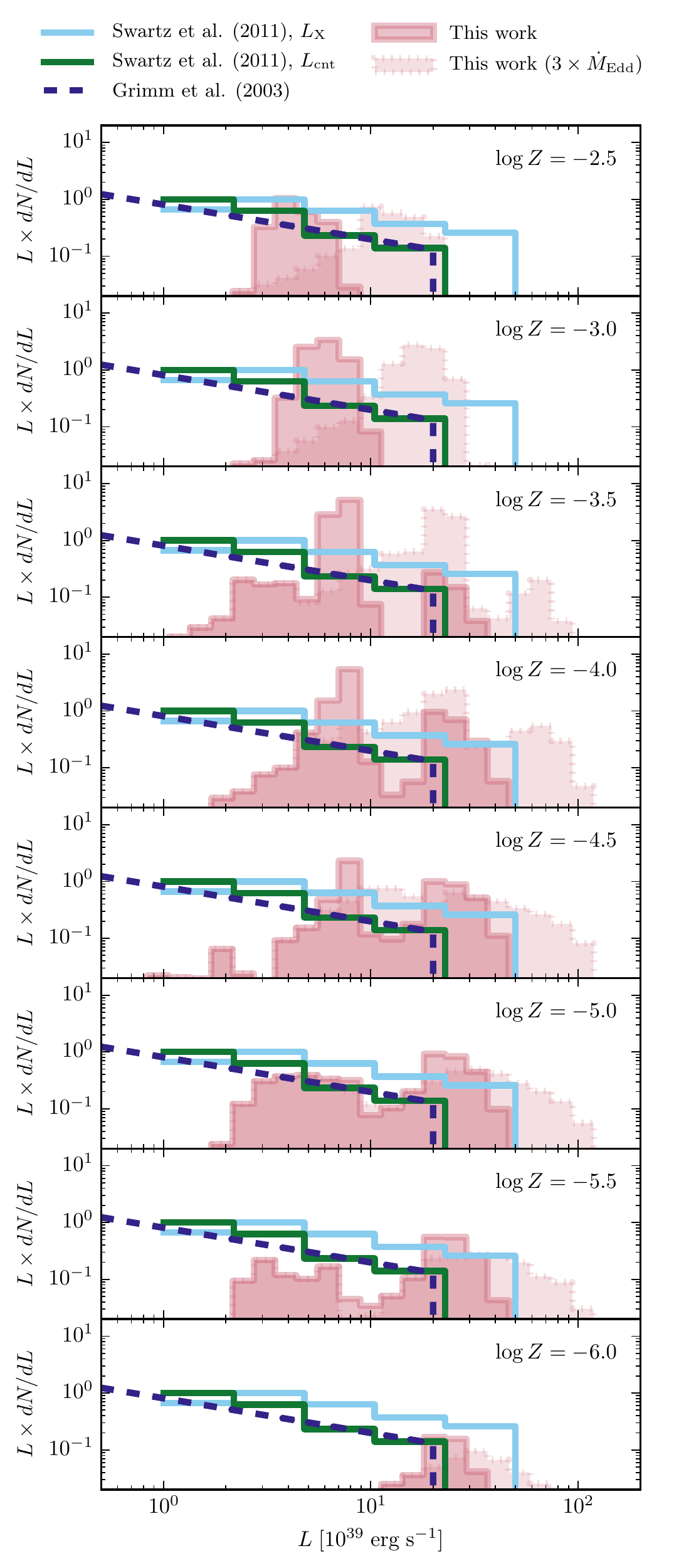}
   \end{center}
   \caption{Expected luminosity distribution function for ULXs formed
   through CHE, compared to the empirical distribution with a slope
   $\alpha=-1.6$ derived by \citet{Grimm+2003} and the sample of 117 ULXs from
   nearby galaxies described by \citet{Swartz+2011}, both of which cover
   metallicities $\log Z>-3.0$. For the sample of
   \citet{Swartz+2011} we include the distributions considering their estimates
   on source luminosities from spectral modelling $L_{\rm X}$, and
   that from number of counts $L_{\rm cnt}$. All distributions are normalized to
   a star-formation rate of ${1~M_\odot~\rm yr^{-1}}$.
}
   \label{fig:lumdist}
\end{figure}

\begin{figure}%[!ht]
   \begin{center}
   \includegraphics[width=\columnwidth]{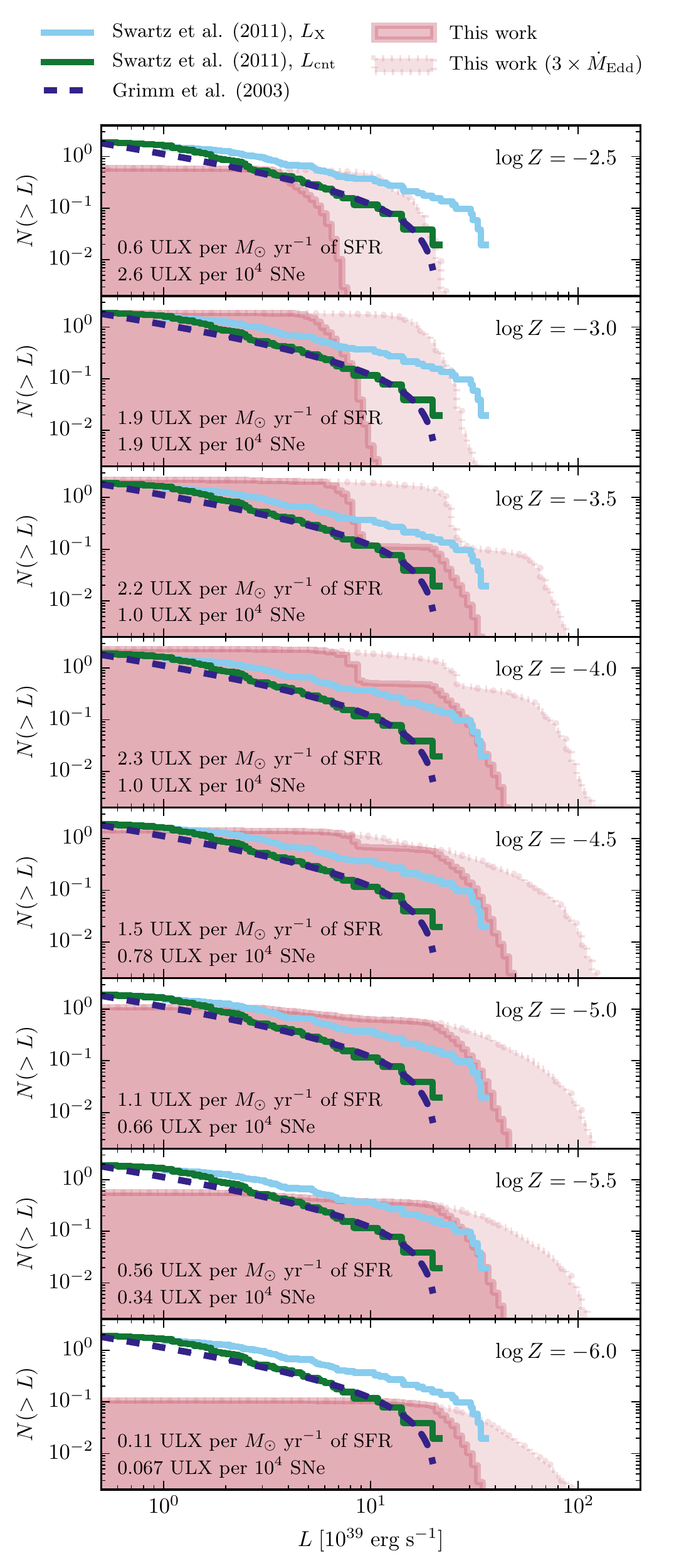}
   \end{center}
   \caption{Same as Figure \ref{fig:lumdist} but for the cumulative distribution
   function instead. Expected number of observable sources per $M_\odot~{\rm
      yr^{-1}}$ of SFR are also shown, as
      well as the expected formation rate of ULXs
      in terms of the SN rate. Although the number of
ULXs produced per SN increases with metallicity, at lower metallicities
phases with accretion are longer lived resulting in a peak in observable sources
per $M_\odot~{\rm yr^{-1}}$ of SFR at $\log~Z=-4$.
}
\label{fig:lumdistcum}
\end{figure}

Our predicted luminosity distribution function and cumulative distribution
function are shown in Figures
\ref{fig:lumdist} and \ref{fig:lumdistcum}, respectively. These take into account
the lifetime of the sources modeled, so they can be compared to observed samples
of ULXs.
To consider possible uncertainties on the efficiency
of mass transfer, we also include the distribution of luminosities if
accretion rates of three times the Eddington rate would be
possible. This is not done by running a full set of simulations with an adjusted
Eddington limit, but rather by considering the potential luminosities of our
models which are strictly limited to accrete at $\dot{M}_{\rm Edd}$.
The
resulting luminosities obtained in this way agree well with models computed
self-consistently with an adjusted $\dot{M}_{\rm Edd}$ (see Appendix
\ref{appendix:mdotedd} for details). For our
models we consider the full bolometric accretion luminosity $L_{\rm acc}$,
although a fraction of this would not be emitted in the bands detectable by
X-ray observatories.

To compare with observations, we also include the empirical
distribution described by \cite{Grimm+2003} for nearby (within $35~\rm
Mpc$) late-type starburst galaxies,
described by a power law with a slope of $\sim-1.6$, and a cutoff at a luminosity of
$2\times10^{40}~{\rm erg~s^{-1}}$. \citet{Grimm+2003} and
\citet{Gilfanov+2004} argue that the presence of such a cutoff is indicative of
a maximum possible mass for stellar BHs. We also
include the 117 ULXs described by \citet{Swartz+2011}, which represent complete
samples of ULXs for local galaxies of diverse types within $14.5~\rm Mpc$.
\citet{Swartz+2011} consider two
different methods to compute the luminosity of ULXs, one obtained
from photon
counts, and the other, for sources with $>130$ counts detected,
from spectral modelling. Although the luminosities from \citet{Grimm+2003} correspond to
the $2-10~\rm keV$ band, while the photon counts from \citet{Swartz+2011}
are corrected to give the luminosities in the $0.3-10~\rm keV$ range, the
two samples agree well with each other.

The galaxies considered by \citet{Grimm+2003} and \citet{Swartz+2011}
should be indicative of the formation of ULXs in the local
environment of our Galaxy and favor high metallicities, with no sources
below $\log Z<-3.0$. Moreover, as they do not
properly sample dwarf galaxies, we
expect an additional bias towards higher metallicities. In particular, there is
one ULX detected in the blue compact dwarf galaxy IZw18 \citep{Ott+2005}, which
has a very low metallicity of $Z_\odot/50$, and the study of ULXs in dwarf galaxies
provides hints of an increasing number of observable ULXs per $M_\odot\;\rm
yr^{-1}$ of SFR with decreasing galaxy mass \citep{Swartz+2008}. 

The CHE channel is
expected to produce the most massive BHs possible for a given
metallicity, as it transforms almost the whole star into a large helium core.
Since large initial masses are required to have efficient rotational mixing,
this results in the least
massive BH possible at $\log~ Z=-2.5$ to have a mass of $20M_\odot$, already falling into the ULX range
when accreting at the Eddington rate. As Figure \ref{fig:lumdist} shows,
there is a much less luminous tail of
objects that arises from brief moments at the beginning and end of mass-transfer
phases, when transfer rates are below the Eddington limit. At a metallicity of
$\log~ Z=-3$ we reach a luminosity cutoff due to the lower limit for PISNe at
about $10^{40}~\rm erg~ s^{-1}$, which, barring possible mass loss from
PPISNe,
means that accretion rates that are only a factor of a few above Eddington are
enough to explain the observed luminosity cutoff with our models. A population of lower mass BHs and
NS accretors is still required to explain the lower end of the luminosity
function of ULXs, extending to the HMXB regime. Such systems are likely to
originate from CE evolution, meaning that the BH accretor results
from an envelope stripped star \citep{Podsiadlowski+2003}, and thus should have lower masses than those
possible through CHE. Although the inclusion of a different channel should in
principle produce a break in the distribution function, a similar break that
should be visible due to differences between NS and BH accretors is not
observed \citep{Grimm+2003}, and the distribution function for our highest metallicity models
is not far off from that of the most luminous objects in the observed sample. It
might be possible then that the population of NS and BH accretors resulting from
CE evolution coexists with those produced by CHE and results in a
luminosity distribution function that can be described with a single slope.

At lower metallicities this should not be the case; Figure \ref{fig:lumdist} shows that a gap in the
luminosity function is expected. This is due to the formation of BHs above the
limit for PISNe. The gap is not completely deserted, as systems at the beginning
or end of mass-transfer phases to those very massive BHs accrete below
the Eddington rate, resulting in a wide range of luminosities for a short period
of time. This gap in the
distribution results in a clear feature in the cumulative
distribution, as shown in Figure \ref{fig:lumdistcum}. Observations in the local
universe would favor the observation of galaxies at the upper end of the
metallicities modelled, but deeper observations sampling lower metallicity
environments should show a significant digression from a population describable by a single
power law.

In terms of observable sources,
the CHE channel has a strong dependence on metallicity, with
almost no sources being produced at metallicities $Z\ge 0.01$, and rising to a
peak of $2.3$ ULXs per $M_\odot~{\rm yr^{-1}}$ of SFR
at a metallicity of $\log~ Z=-4$, though the rate is mostly flat in the range
$-4.5<\log~ Z<-3$. In a slightly counterintuitive way, at metallicities below
$\log~Z=-2.5$ the number of ULXs formed
per SNe monotonically decreases with metallicity, but it has to be taken into
account, as described in Section \ref{sect:Zdep}, that due to orbital widening
from wind mass loss, mass-transfer phases have shorter lifetimes at higher
metallicities. This compensates for the smaller number of sources produced per
SN at lower metallicities, resulting in the local maximum of observable sources at
$\log~ Z=-4$. Anyhow, with just a couple of systems formed every $10^4$ SNe, it
is clear that this evolution channel is only followed by a small fraction of massive stars.
Although there are important uncertainties in our calculations, in
particular in the choice of initial distribution functions at low metallicities,
this systematic behaviour with metallicity should be a generic feature, despite
uncertainties of at least a few in the rates we predict.

\subsection{Luminosity distribution at redshift $z=0$}
\label{subsect:lumdistlocal}

\begin{figure}%[!ht]
   \begin{center}
   \includegraphics[width=\columnwidth]{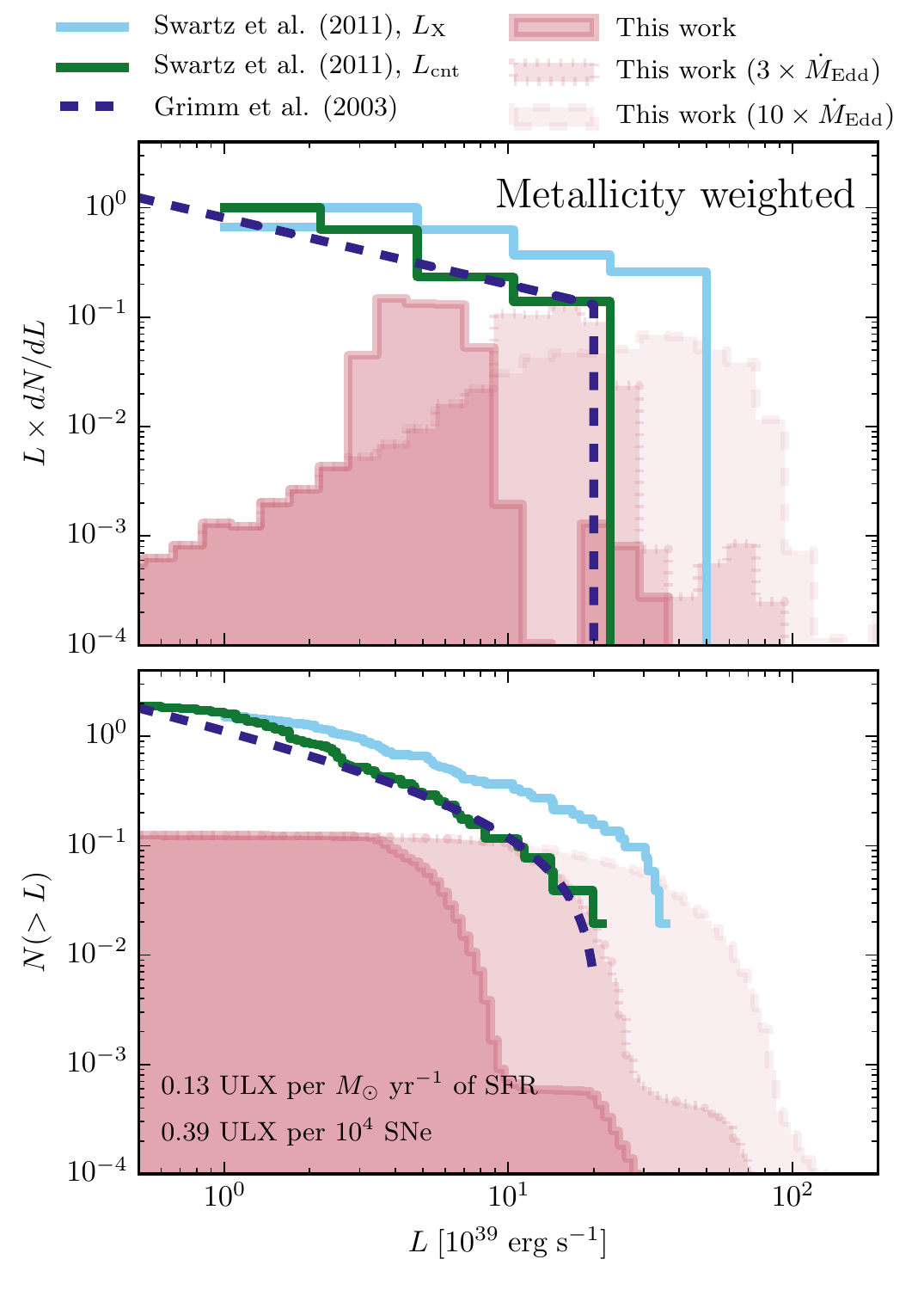}
   \end{center}
   \caption{Same as Figures \ref{fig:lumdist} and \ref{fig:lumdistcum}, but
   weighting our models according to the metallicity distribution of
   \citet{LangerNorman2006} at redshift $z=0$.
}
   \label{fig:lumweight}
\end{figure}

It is tempting
to relate the predicted rate of $\sim 2$ ULXs per $M_\odot~{\rm yr^{-1}}$ of SFR with the equivalent
observed number of sources by \citet{Swartz+2011}, but we need also consider the
contribution from the tail of the normal HMXB population that reaches up to ULX
luminosities. More importantly, both the \citet{Swartz+2011} and
\citet{Grimm+2003} sources sample the local universe. If we consider the metallicity
distribution of \citet{LangerNorman2006} evaluated at a redshift $z=0$, we would
only expect $\sim1\%$ of the star-formation in the local universe to happen at a
metallicity below $\log~ Z=-3$. We use this distribution to evaluate the
local luminosity distribution function of ULXs formed through CHE, which we show
in Figure \ref{fig:lumweight}. The metallicity weighting significantly reduces
the number of expected sources to $0.13$ ULXs per $M_\odot~{\rm yr^{-1}}$ of SFR,
but if we consider mass
accretion at three times the Eddington rate, the distribution we predict nicely
matches that of the brightest sources of \citet{Grimm+2003} and of
\citet{Swartz+2011} estimated by photon counts. For the majority of
ULXs which have lower luminosities, a different formation channel would be required.
The luminosities estimated by
\citet{Swartz+2011} through spectral modeling should better represent the
total luminosity of the source, which is what we consider for our models.
However, if accretion rates ten times larger than Eddington are allowed to try
to match these higher luminosities,
Figure \ref{fig:lumweight} shows that the gap produced by PISNe is lost.
This is
because not many systems transfer mass at those high rates (see Table
\ref{table:rates}), so the luminosity distribution does not simply shift to higher
luminosities.
If the increased luminosity is due to a relatively constant
beaming factor rather than
accretion above the Eddington rate, then we expect that luminosity gap would remain.
In any case, the spectrum of BHs $<60M_{\odot}$ and $>130M_\odot$
radiating at the same luminosity should differ significantly.

\begin{table*}
        \center
        \begin{tabular}{r|crlcc|crlc|c}
                %\hline  & & & & & \\
                \hline
                \noalign{\smallskip}
                $\log~ Z$             &
                \multicolumn{3}{c}{$\displaystyle\frac{n_{\rm ULX}}{\rm SFR}~{[M_\odot^{-1}~\rm yr]}$} &
                \%$>3 \dot{M}_{\rm Edd}$&
                \%$>10 \dot{M}_{\rm Edd}$&
                \multicolumn{3}{c}{$\displaystyle\frac{R_{\rm ULX}}{R_{\rm
                SN}}\times 10^{4}$}      &
                $\langle t_{\rm ULX}\rangle~\rm[Myr]$&
                $\displaystyle\left\langle \frac{L_{\rm X,gal}}{\rm
                SFR}\right\rangle~{\left[\frac{10^{39}~
                {\rm erg~s^{-1}}}{M_\odot~{\rm yr}^{-1}}\right]}$\\
                \noalign{\smallskip}
                \hline
                \noalign{\smallskip}
                -2.5 & 0.6  & (0.6,&\hspace{-0.15in}0) & 71 & 25 & 2.6    &
                (2.6,&\hspace{-0.15in}0) & 0.23 & 2.6\\
                -3.0 & 1.9  & (1.9,&\hspace{-0.15in}0) & 76 & 16 & 1.9    &
                (1.9,&\hspace{-0.15in}0) & 1 & 11\\
                -3.5 & 2.2  & (2.1,&\hspace{-0.15in}0.12) & 67 & 16 & 0.98   &
                (0.71,&\hspace{-0.15in}0.26) & 2.2 & 17\\
                -4.0 & 2.3  & (1.8,&\hspace{-0.15in}0.51) & 39 & 2.1 & 1.0    &
                (0.44,&\hspace{-0.15in}0.58) & 2.3 & 26\\
                -4.5 & 1.5  & (0.77,&\hspace{-0.15in}0.7) & 7.7 & 1.7 & 0.78   &
                (0.18,&\hspace{-0.15in}0.61) & 1.9 & 22\\
                -5.0 & 1.1  & (0.45,&\hspace{-0.15in}0.64) & 5.3 & 1.5 & 0.66
                & (0.13,&\hspace{-0.15in}0.53) & 1.7 & 17\\
                -5.5 & 0.56 & (0.18,&\hspace{-0.15in}0.39) & 5.2 & 1.4 & 0.34  &
                (0.044,&\hspace{-0.15in}0.29) & 1.6 & 9.9\\
                -6.0 & 0.11 & (0,&\hspace{-0.15in}0.11) & 2.6 & 0.85 & 0.067 &
                (0,&\hspace{-0.15in}0.067) & 1.6 & 2.3\\
                \noalign{\smallskip}
                \hline
                \noalign{\smallskip}
                local & 0.13 & (0.13,&\hspace{-0.15in}0.00062) & 70 & 21 &
                0.39 & (0.39,&\hspace{-0.15in}0.0011) & 0.33 & \\
                \noalign{\smallskip}
        \end{tabular}
        \caption{Rates and general properties of ULXs formed through CHE at different
        metallicities. Shown here are the expected number of observable ULXs
        $n_{\rm ULX}$ per
        $M_\odot~{\rm yr^{-1}}$ of SFR, the number of produced ULXs per SNe
        (i.e. the ratio between the formation rates of ULXs, $R_{\rm ULX}$, and
        SN, $R_{\rm SN}$), and the expected total
        X-ray luminosity of galaxies (from sources produced through CHE)
        per $M_\odot~{\rm yr^{-1}}$ of SFR. In parenthesis we indicate
  separately the number of objects with BHs below the pair-instability gap
  ($M_{\rm BH}<60M_\odot$) and above it ($M_{\rm BH}>130M_\odot$). Also included
  for the expected number of observable ULXs is the percentage of those systems
  that would accrete at 3 and 10 times their Eddington rates, while for the
  formation rates we include the average time that formed ULXs spend as such,
  $\langle t_{\rm ULX}\rangle$. The last column is computed under
  the assumption that the bolometric luminosity from accreting sources is
  released as X-rays, and that accretion is strictly limited to the Eddington
  rate. Local rates are estimated using the metallicity distribution of
  \citet{LangerNorman2006} at redshift $z=0$.
  A value for the locally weighted galactic $L_{\rm X,gal}$ is ignored, as the
  local environment contains many galaxies at higher metallicities where the
  total luminosity would be dominated by HMXBs instead of the ULXs described in
  this work. Values given in terms of SFR are computed assuming a SN rate of
     $0.01~\rm yr^{-1}$ per $1M_\odot~\rm yr^{-1}$ SFR.}
        \label{table:rates}
        %
        %\vspace{0.7cm}
\end{table*}

Locally, ULXs with BH masses $>130M_\odot$ would only represent a
small fraction of the total formed through CHE, around $\sim0.5\%$. Moreover,
since ULXs formed through this channel only represent the high-luminosity tail
of the luminosity distribution function,
they correspond to an even smaller fraction of the total. The upcoming {\it eROSITA}
X-ray observatory will perform a full-sky survey, which at a sensitivity limit
of around $2\times 10^{-14}~\rm erg~s^{-1}~cm^{-2}$ should detect sources
with luminosities of $10^{40}~\rm erg~s^{-1}$ up to a distance of $35~\rm Mpc$.
Considering the distribution of known sources, around
$\sim 100$ ULXs should be detected \citep{Prokopenko&Gilfanov2009}, so that
finding BHs above the PISN gap would appear unlikely. However, as shown in
Figure \ref{fig:lumweight}, if
BHs above the PISN gap can accrete at rates above a few times
$\dot{M}_{\rm Edd}$
their luminosities would approach $10^{41}~\rm erg~s^{-1}$,
with significantly larger detection volumes (for sources three
times more luminous than the cutoff luminosity the detection volume would be
$\sim 5$ times larger).
{\it eROSITA} could then potentially detect a few of these sources, likely in
metal-poor dwarf galaxies. On a longer timescale, the {\it Athena} X-ray
observatory will be capable of probing much deeper, and targeted observations to
dwarf galaxies with very low metallicities and high SFRs could test the
existence of these objects.

\subsection{$L_{\rm X, gal}-\rm SFR$ relation at low metallicities}
As we predict the luminosity distribution of X-ray sources to change
significantly at low metallicities, we also expect the relation between the
total X-ray luminosity of a galaxy $L_{\rm X,gal}$ and its SFR to be different
from that in our local environment. Locally, the X-ray luminosity of a galaxy
serves as a probe of its SFR \citep{Grimm+2003,Gilfanov+2004}, and
the presence of a luminosity cutoff results in a linear $L_{\rm
X,gal}-\rm SFR$ relationship for high enough SFR. \citet{Gilfanov+2004} argue that a
population of IMBHs would result in a break from the linear relationship at very
high SFR. As BHs formed above the limit for PISNe also form a distinct
population of very massive BHs, they could as well result in differences in
the dependence of $L_{\rm X,gal}$ with SFR. Moreover, if the $L_{\rm X,gal}-\rm SFR$
relationship changes
significantly at lower metallicities, it would need to be recalibrated to
serve as a probe of star-formation.

\begin{figure}%[!ht]
   \begin{center}
   \includegraphics[width=\columnwidth]{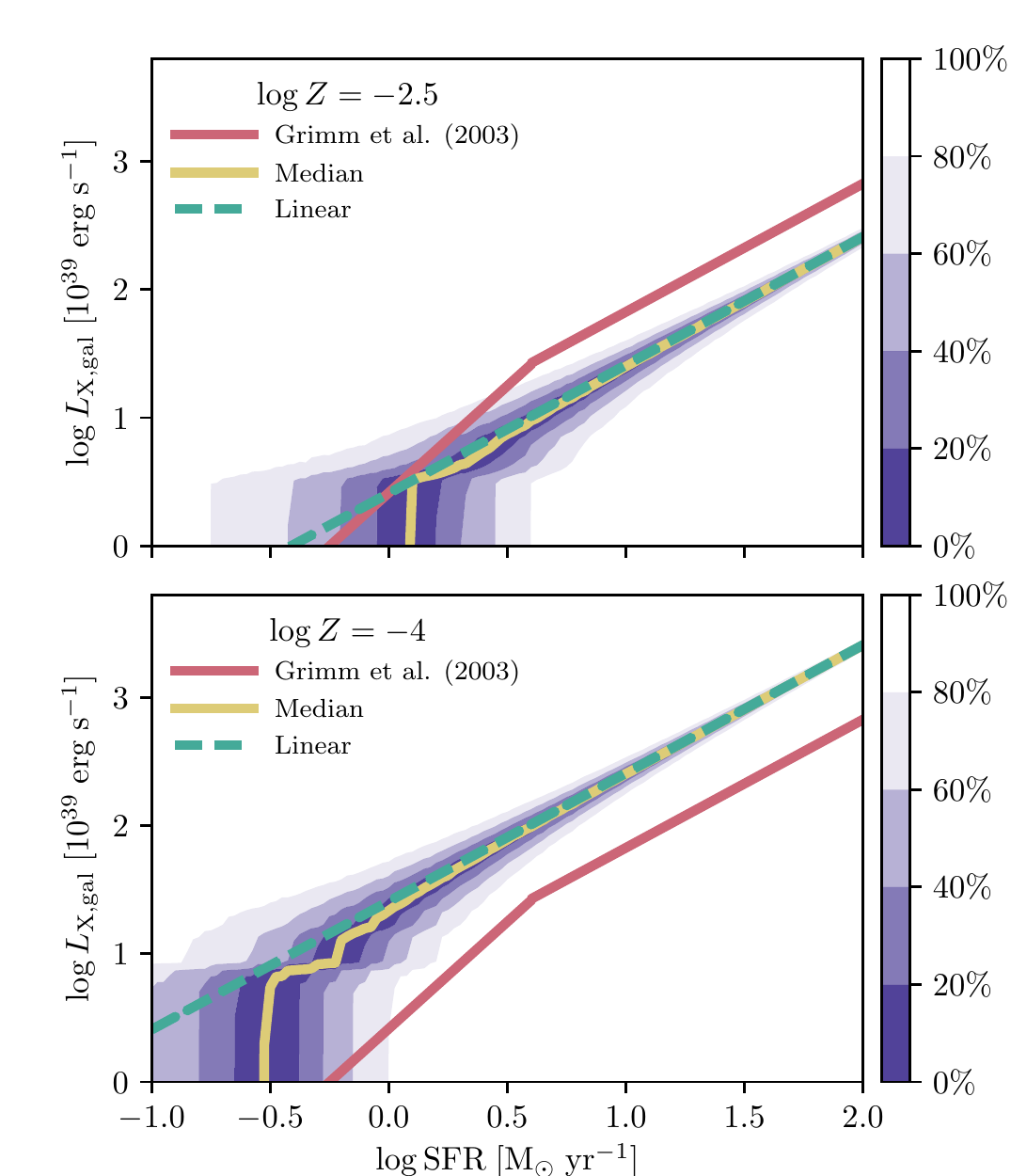}
   \end{center}
   \caption{$L_{\rm X,gal}-{\rm SFR}$ relationship arising from ULXs produced through
   CHE at two different metallicities, compared to the predicted
   {metallicity-independent} relationship
   from \citet{Grimm+2003} for the luminosity in the $2-10$ keV band. Systems at
   $\log~ Z=-4$ can also produce BHs above
the pair instability gap, resulting in the slightly different structure at low
SFR. For simplicity we assume all the luminosity from accretion is emmited as
X-rays. For different values of the SFR modelled we constructed 10000 synthetic galaxies, and the
color plor indicates the fraction of galaxies contained around the median
(i.e., the $20\%$ region is bounded by the fourth and sixth deciles).}
\label{fig:lxsfr}
\end{figure}

To assess the $L_{\rm X,gal}-{\rm SFR}$ relationship, using our ULX models
we construct multiple synthetic
galaxies for different SFRs (see Appendix \ref{appendix:synth} for
details), for which we show the distribution of
X-ray luminosities in Figure \ref{fig:lxsfr}. For galaxies with low star-formation rates,
that on average should have less than one ULX produced through CHE, low-number
statistics plays an important role, and this effect can clearly be seen at SFRs
less than $1~M_\odot~\rm yr^{-1}$. Unlike the relationship observed by \citet{Grimm+2003}, Figure
\ref{fig:lxsfr} shows a very steep increase in luminosity at low SFRs, but this
is just because our models do not include the contribution from HMXBs. Instead, the
luminosities jump from zero to above $10^{39}~{\rm erg~s^{-1}}$ for galaxies
that happen to have a single ULX.
Although we cannot properly assess the relationship at very low SFRs,
the switch from a non-linear to a linear relationship depends on the sampling of the most luminous
sources possible, and we expect ULXs formed through CHE to be the source of
these. The SFR at which \citet{Grimm+2003} observe a break in the
X-ray luminosity distribution of galaxies would then be
equivalent to the SFR at which most galaxies would sample a few of the ULXs
produced through CHE.

Figure \ref{fig:lxsfr} shows that the SFR at which the relationship becomes
linear, as well as the expected luminosities at high SFR, has an important
dependence on metallicity. This results both from the increased number of
sources expected per $M_\odot~{\rm yr^{-1}}$ of SFR at lower metallicities, and the formation of BHs above
the PISNe gap which can produce much higher luminosities. Table
\ref{table:rates} shows the ratio of $L_{\rm X,gal}$ to SFR that we expect in the linear
regime at different metallicities. This value can vary up to an order of
magnitude, which should be taken into account when using $L_{\rm X,gal}$ as a
measure of SFR. The presence of BHs above the PISNe gap does produce changes in
the $L_{\rm X,gal}-{\rm SFR}$ reationship, but this happens in the low-SFR regime, likely making
it hard to observe.

%%% Local Variables: 
%%% mode: latex
%%% TeX-master: "paper.tex"
%%% End: 

\section{Orbital parameters of ULXs formed through CHE}
\label{sect:orbit}

An additional tool to discriminate between different formation scenarios is the
detection of optical counterparts to ULXs, which can help identify the nature of
the donor star and the orbital parameters.
The largest sample of counterparts to date is given by
\citet{Gladstone+2013}, who detect potential counterparts in 22 out of 33 ULXs
studied. There are also two
ULXs for which dynamical estimates of the masses are available from measurements of
radial-velocity variations due to the orbital motion of the donor star, detected
as a WR: M101 ULX-1, with a BH mass likely
in the range $20M_\odot-30M_\odot$ \citep{Liu+2013}, and P13, with a BH mass
below $15M_\odot$ \citep{Motch+2014}. Both exclude the possibility of an IMBH as the
compact object and set constraints on the properties of the donor.
However, these dynamical mass estimates need to be considered with
care,
as is shown by the case of the HMXB IC10 X-1. Using
measurements of radial-velocity variations,
\citet{SilvermanFilippenko2008} concluded that this system contains a BH with a
mass in excess of $20M_\odot$, but \cite{Laycock2015} showed that the
radial-velocity variation detected
does not follow the stellar motion, but rather comes from a shadowed
region in the stellar wind. This means that the dynamical mass estimate is
incorrect, making the mass of the compact object much more uncertain and even
consistent with a NS accretor.

\begin{figure}%[!ht]
   \begin{center}
   \includegraphics[width=\columnwidth]{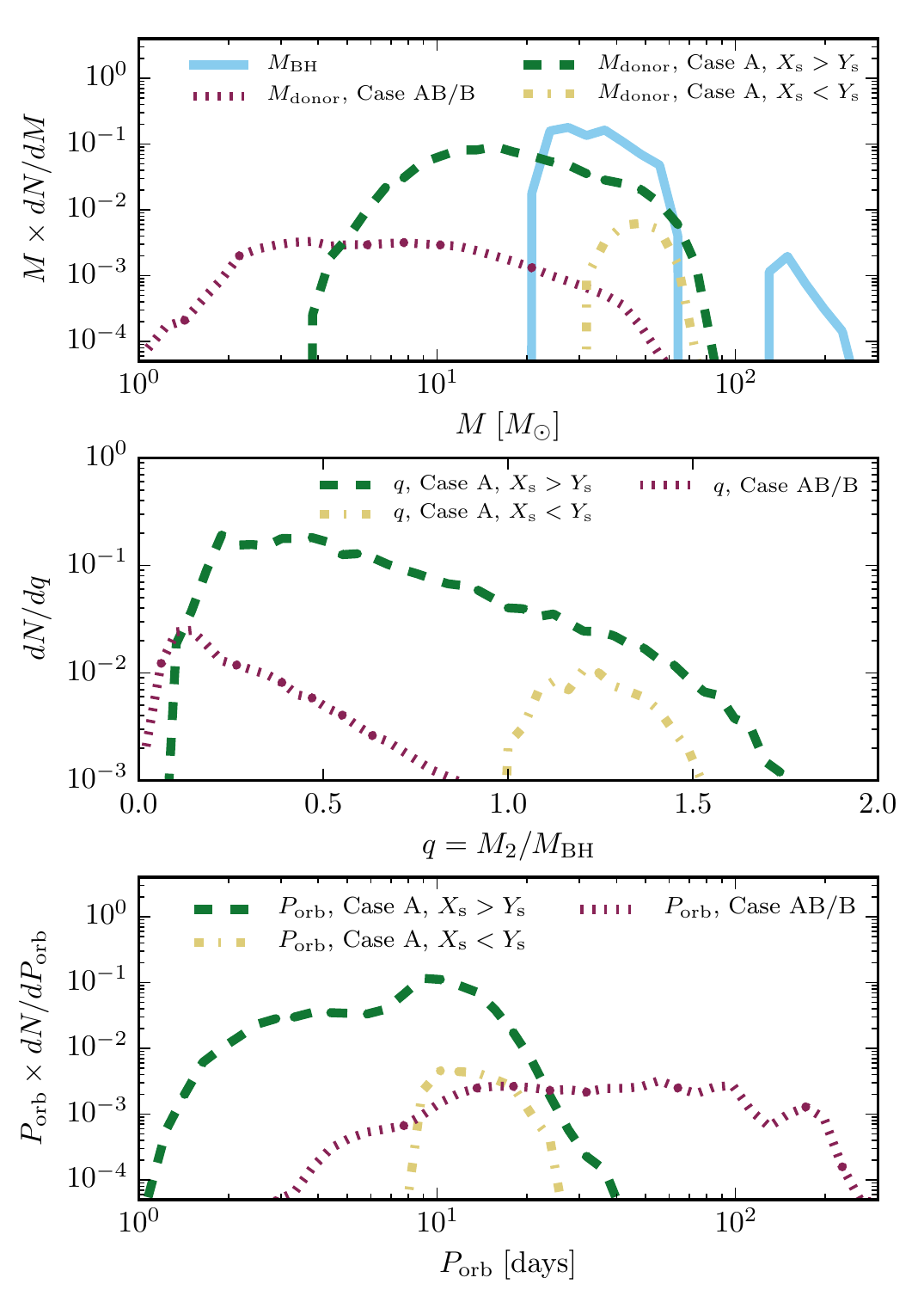}
   \end{center}
   \caption{(top) Distribution of BH and donor masses in ULX systems, weighting
      our models to the metallicity distribution of \citet{LangerNorman2006} at
      redshift $z=0$. (middle) Distribution of mass ratios using the same
      metallicity distribution. (bottom) Distribution of orbital periods using the same
metallicity distribution. Donor masses and orbital periods are separated between Case A and Case
B systems, and furthermore, Case A systems are separated between donors that are
hydrogen poor or hydrogen rich at their surfaces.}
\label{fig:massdist}
\end{figure}

It is beyond the scope of this paper to study in detail the optical properties of
our ULX models and compare them to observations, which needs modeling of the emission from the accretion disk,
but we can check the distribution of donor and BH masses,
together with orbital
periods, as is done in Figure \ref{fig:massdist}. This is done by
taking into account the lifetime of each phase, so it can be compared with the
observed distribution. As more
observations place better constraints on the orbital parameters of ULXs,
their origin can be better understood by comparing those to the predicted
distributions of different formation channels.
The bulk ($71\%$) of ULX models formed through
CHE should contain main-sequence (MS) hydrogen rich ($X_{\rm s}>Y_{\rm s}$) donors in the range $\sim
8M_\odot-30M_\odot$ with orbital periods below $20$ days. More massive MS donors
in the range $30M_\odot-70M_\odot$ are only a $16\%$ of the total, with
$3\%$ being hydrogen poor ($X_{\rm s}<Y_{\rm s}$). Although less numerous,
these massive optical
counterparts should be much easier to detect. Case AB/B systems correspond
to only $6\%$ of the total, have typical donor masses below $10M_\odot$, and typical
periods above $10$ days. As has already been mentioned, the predominance of
Case A sources is owed to these systems undergoing mass transfer on a
nuclear timescale, while mass transfer for post-MS donors operates on a much
shorter thermal timescale.

Independent of the formation channel, a common expectation for systems
undergoing RLOF is a preference for mass ratios $q=M_2/M_{\rm BH}$ below unity,
as this value plays an important role in the
lifetime of a mass-transfer phase \citep{Podsiadlowski+2003}. If the donor is
more massive than the BH, mass transfer typically results in a reduction of the
orbital separation leading to a short-lived thermal-timescale mass transfer,
similar to the situation for intermediate-mass X-ray binaries
\citep{Tauris+2000}. If instead the
donor is a MS star less massive than the BH at the onset of mass transfer, the orbital
separation increases as a result of mass transfer, leading to a much longer-lived
nuclear-timescale X-ray phase.

The mass-ratio distribution for ULXs produced through CHE at redshift $z=0$ is
shown in the second panel of Figure \ref{fig:massdist}. Although the initial
mass ratios of these systems are smaller than $q=0.5$, mass loss of the primary
before BH formation can lead to mass ratios above unity during the ULX phase;
but these are disfavored for the same reason discussed above.
The distribution favors mass ratios significantly below
unity, with $\sim 50\%$ of the sources having $q<0.5$. This preference for lower
mass ratios is stronger at lower metallicites, as the primary 
undergoes a smaller amount of mass loss before forming a BH and preserves its
initially low mass ratio.
Instead, for the CE channel, the primary which will form the BH expands and
initiates a CE phase,
and for very low secondary masses, a merger is expected rather than envelope
ejection \citep[see, e.g.][]{Kruckow+2016}. 
As a consequence, the distribution of mass ratios predicted from CE evolution
favors mass ratios below, but nor far, from unity \citep{Madhusudhan+2008}.
In the case of ULXs formed via dynamical
capture in clusters, much higher BH than donor masses are also expected, but the
orbital periods of these systems are well above $20$ days
\citep{MapelliZampieri2014}, which differentiates them from the bulk of systems
produced through CHE. The long orbital periods imply that such systems
would have post-MS donors with short lifetimes as active sources,
which reduces the likelihood of observing them.

The models of \citet{Madhusudhan+2008} have an
upper limit of $25M_\odot$ for the BH mass, which makes it
difficult to explain some of the brightest optical counterparts observed that
would require $\sim 50M_\odot$ donors. In consequence, they favor IMBHs as the
compact object in these sources, which would allow for long-lived mass transfer
phases due to the lower mass ratios. This could be avoided if CE could
produce higher mass BHs, but even at low
metallicities it is difficult to reach BH masses well above $30M_\odot$, as envelope
stripping significantly reduces the mass of the primary \citep{Linden+2010}.
ULXs formed through CHE can reach BH masses up to the
lower end of the PISN gap ($60M_\odot$), and even
if PPISNe would reduce this to $\sim 47M_\odot$ \citep{Woosley2016}, this easily
allows for stable (and long-lived) RLOF from very massive donors. For BHs formed above the PISN
gap, donor masses can be much higher but, at least in the local universe, ULXs
with these very massive BHs are expected to be uncommon (Section
\ref{subsect:lumdistlocal}).

For reference purposes, the distribution of several properties of our ULX
systems at different metallicities is provided in Appendix
\ref{appendix:properties}.

%%% Local Variables: 
%%% mode: latex
%%% TeX-master: "paper.tex"
%%% End: 

\section{NS-BH and BH-BH binaries after a ULX phase}
\label{sect:resultsGW}

After a ULX phase, the orbit widens significantly due to mass transfer, with
final orbital periods well in excess of $10$ days. Unless the secondary receives
a strong kick in a favorable direction, reducing its orbital period
and making the system very eccentric, a merger due to GW emission would not
happen. As an example, a binary with a $60M_\odot$ BH and a $1.4M_\odot$ NS at
a $10$ day orbital period would take more than $1000~\rm Gyrs$ to merge.
Instead, the same system with an eccentricity $e=0.9$ would merge in only
$3.5~\rm Gyrs$. This requires fine-tuning both the kick velocity and its
direction, making it an unlikely outcome which we study in this
section.

\begin{figure}
   \begin{center}
   \includegraphics[width=\columnwidth]{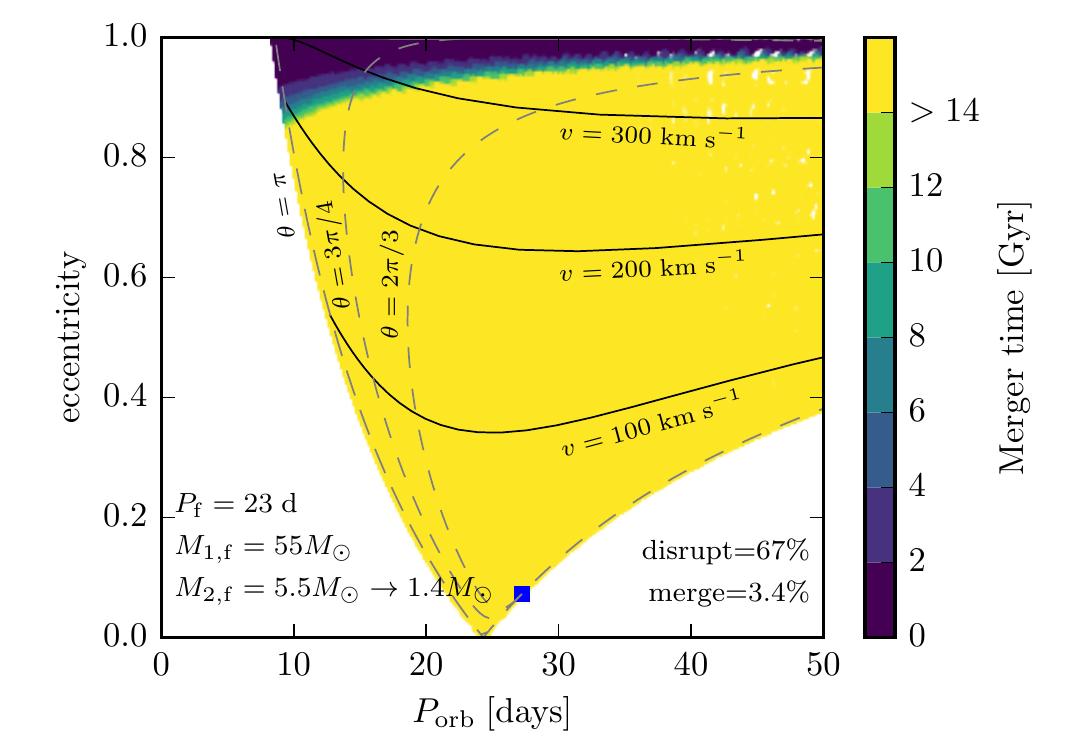}
   \includegraphics[width=\columnwidth]{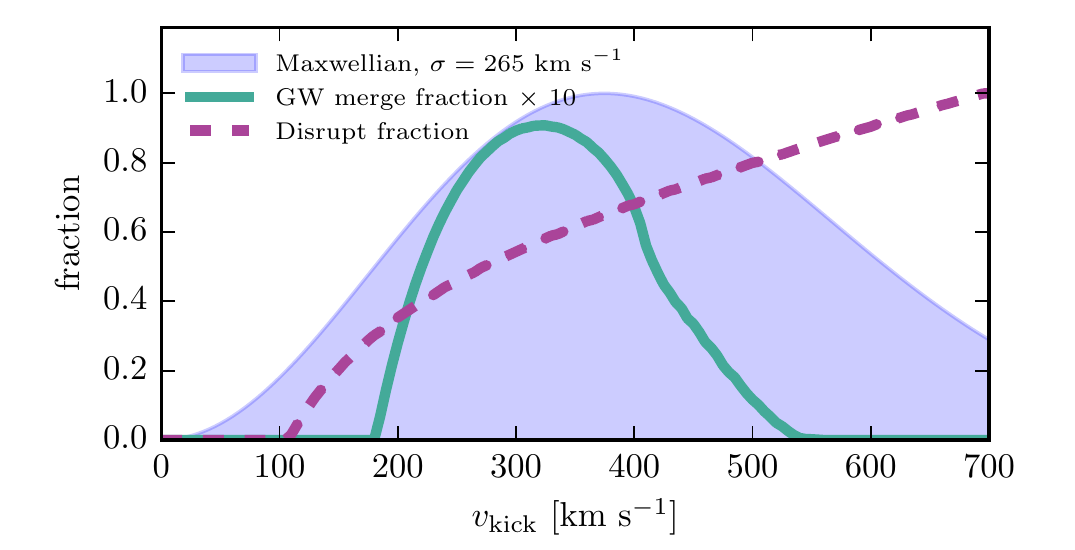}
   \end{center}
   \caption{(Top) Possible post-kick orbital properties for a NS-BH binary formed from a
      low-metallicity system ($\log~ Z=10^{-3.5}$) that passed through a ULX phase.
   The period and masses shown correspond to the pre-SN orbital parameters, and
   the kick distribution is taken to be a Maxwellian with a 1D rms
   of $\sigma=265\;{\rm km\;s^{-1}}$, and isotropic in space. After the kick, there
is a $67\%$ chance that the binary is disrupted, and a $3.4\%$ chance that the
resulting system is compact and eccentric enough to merge in a Hubble time.
Solid lines indicate final orbital parameters for fixed kick velocity $v$ and variable
angle $\theta$ formed between the kick velocity and the orbital velocity,
assuming the kick to be on the orbital plane. Dashed lines indicate the same,
except for a fixed $\theta$ and a variable $v$. The blue square at $P_{\rm
orb}\simeq 27$ days and $e\simeq 0.07$ corresponds to a symmetric SN (i.e. no
kick). (Bottom) Fraction of systems that would be disrupted or that would merge
in less than a Hubble time for an isotropic fixed kick velocity.}
\label{fig:nskick}
\end{figure}

Figure \ref{fig:nskick} shows the possible post-kick parameters when a NS is
formed in a system of metallicity $\log~Z=-3.5$ with initial masses
$M_1=70M_\odot$, $M_2=14M_\odot$ and an
initial orbital period of $1.1$ days (illustrated in Figure \ref{fig:ulx}). At core carbon
depletion of the secondary it consists of a $55M_\odot$ BH and a $5.5M_\odot$ star
with an orbital period of $23$ days. If there is no kick imparted on the NS,
then the result is a NS-BH binary at a separation of $P_{\rm orb}\simeq 27$ days
and a small eccentricity of $e\simeq 0.07$. Such a system would be too wide for
GW radiation to have an important effect, with an expected merger time well in
excess of $10000~\rm Gyr$. Still, detecting such a system while the NS is active as
a pulsar would be very interesting, but considering a typical pulsar lifetime of
$50~\rm Myr$, even if all ULXs would result in a NS-BH binary the expected
number of observable sources would be low. As an upper bound, consider systems at a
metallicity $\log~Z=-2.5$, for which we expect $2.6$ ULXs formed for every
$10^4$ SNe. For a galaxy with a SNe rate of $0.01~\rm
yr^{-1}$, this would mean $\sim 100$ NS-BH binaries with an active pulsar, which
accounting for beaming, should result in less than $\sim 30$ observable
pulsars.
These would be extragalactic sources, making them hard to observe in
radio, and as is shown in Figure \ref{fig:nskick}, we would expect an important
fraction to be disrupted from SNe kicks. However, the Square Kilometer Array
will be capable of detecting pulsars beyond the large Magellanic clouds, and the
discovery of a NS-BH binary is one of its key science goals \citep{Kramer2004}.

If the NS receives a kick of $\sim 200-500~\rm km~s^{-1}$ in a direction opposite to the orbital
velocity, then the orbit can become very eccentric, with a merger time from
GW radiation below a Hubble time. At lower kick velocities, despite the
direction of the kick, the system cannot be driven to a large eccentricity,
while for larger kicks the system would likely be disrupted. The system would circularize before entering
the LIGO band, but if it is observed earlier in the LISA band, it could still
retain a detectable eccentricity. As shown by \cite{Sesana2016}, GW150914 would
have been detectable by LISA, and eccentricity measurements for sources at
these high frequencies have been proposed as a way to distinguish
between different formation scenarios for merging binary-BHs
\citep{Nishizawa+2016,Breivik+2016}. This could
also play a role in understanding the origin of NS-BH mergers.

For the system shown in Figure
\ref{fig:nskick}, there is a $3.4\%$ probability that it would merge in
less than a Hubble time, and out of
those $44\%$ would have merger times under $1~\rm Gyr$. The
resulting inspiral would have a mass ratio $q=39$, and a BH with close to
maximal spin. The spin and the mass ratio have opposite effects on the
possibility of tidally disrupting the star and forming an accretion disk,
with larger spins favoring tidal disruption. The
simulations of \citet{Foucart2012} show that at high mass ratios and low
spins, the NS merges with the BH without being disrupted, producing no accretion
disk and no electromagnetic counterpart. For systems at a mass ratio
$q=10$ \citep[the largest considered by][]{Foucart2012}
spin parameters above $a=0.8$ are required to produce an
accretion disk, but even close to critical rotation, the disk might not be
massive enough to power a strong EM signal.
Owing to this, for the much higher mass ratios involved in our simulations, we would
not expect the production of counterparts to a GW signal even if the BH is
close to critical rotation.
In the absence of an electromagnetic counterpart, it would be difficult to
assess purely from a GW detection if the system contains a NS or if it is a BH-BH
binary, since there is a strong degeneracy between mass-ratios and spins in
parameter estimation \citep[see, e.g.][]{Hannam+2013}.

\begin{figure}
   \begin{center}
   \includegraphics[width=\columnwidth]{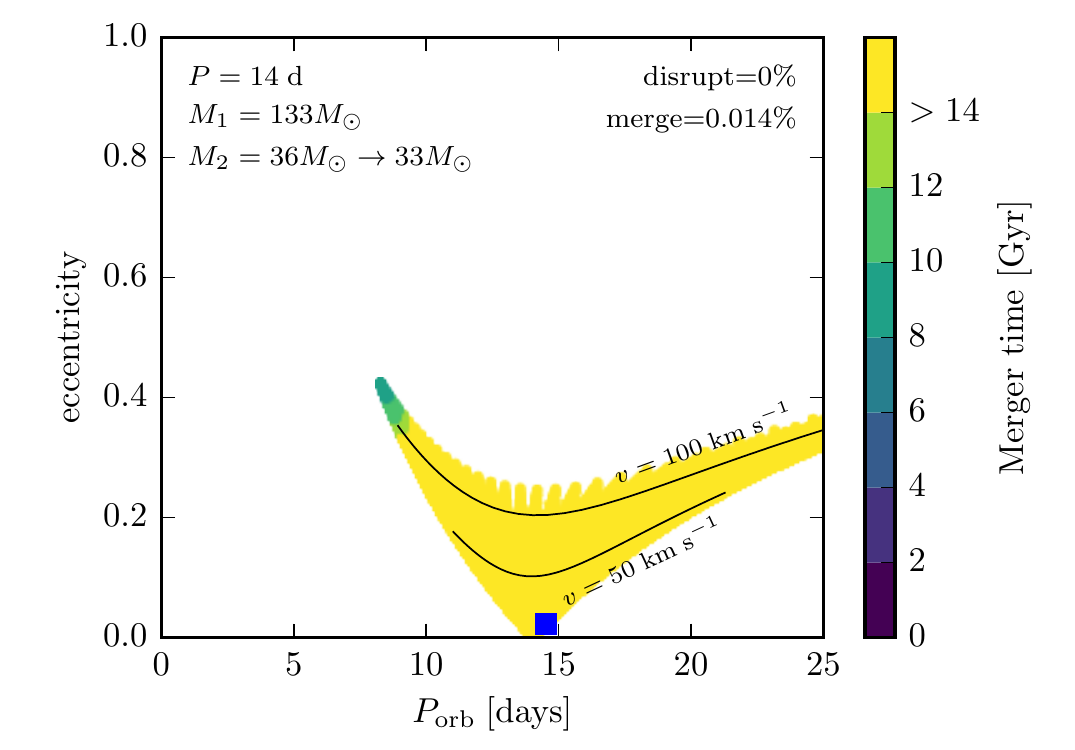}
   \includegraphics[width=\columnwidth]{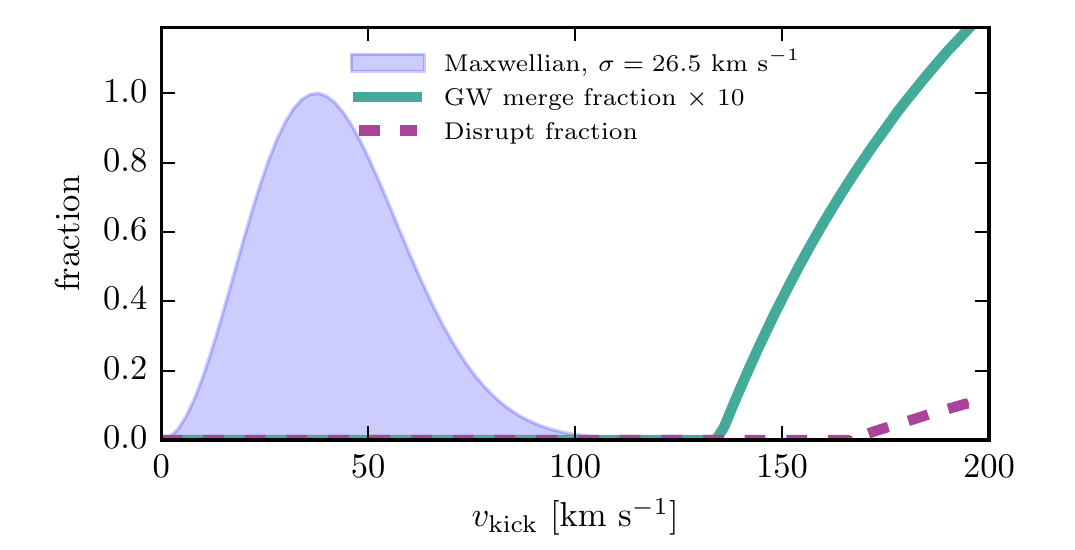}
   \caption{Same as Figure \ref{fig:nskick} but for a system where a BH would be
   expected to form from the secondary. We assume that $10\%$ of the stellar mass
   is lost during BH formation, and that the BH receives a kick with a 1D rms of
   $\sigma=26.5\;{\rm km\;s^{-1}}$. Kick velocities up to $v=135\rm
   km~s^{-1}$ are considered on the top panel, which corresponds to $99.999\%$ of the Maxwellian
distribution.}
   \label{fig:bhkick}
   \end{center}
\end{figure}

A case where the secondary would form a BH is depicted in Figure
\ref{fig:bhkick}. This system is the product of a binary with metallicity
$\log~Z=-3.5$, initial masses $M_1=250M_\odot$ and $M_2=63M_\odot$ with an
orbital period of $P_{\rm orb}=1.75$ days. The primary in this case forms a BH
above the PISNe gap with a mass of $133M_\odot$, while the secondary reaches
carbon depletion with a final mass of $36M_\odot$. Assuming a weak kick is
imparted to the BH as described before, the chances of the system being
disrupted are essentially zero, with $0.014\%$ of systems merging
in less than a Hubble time formed from kick velocities
at the tail of the Maxwellian distribution.

\subsection{Rate estimates for NS-BH and BH-BH mergers}

\begin{figure}
   \begin{center}
   \includegraphics[width=\columnwidth]{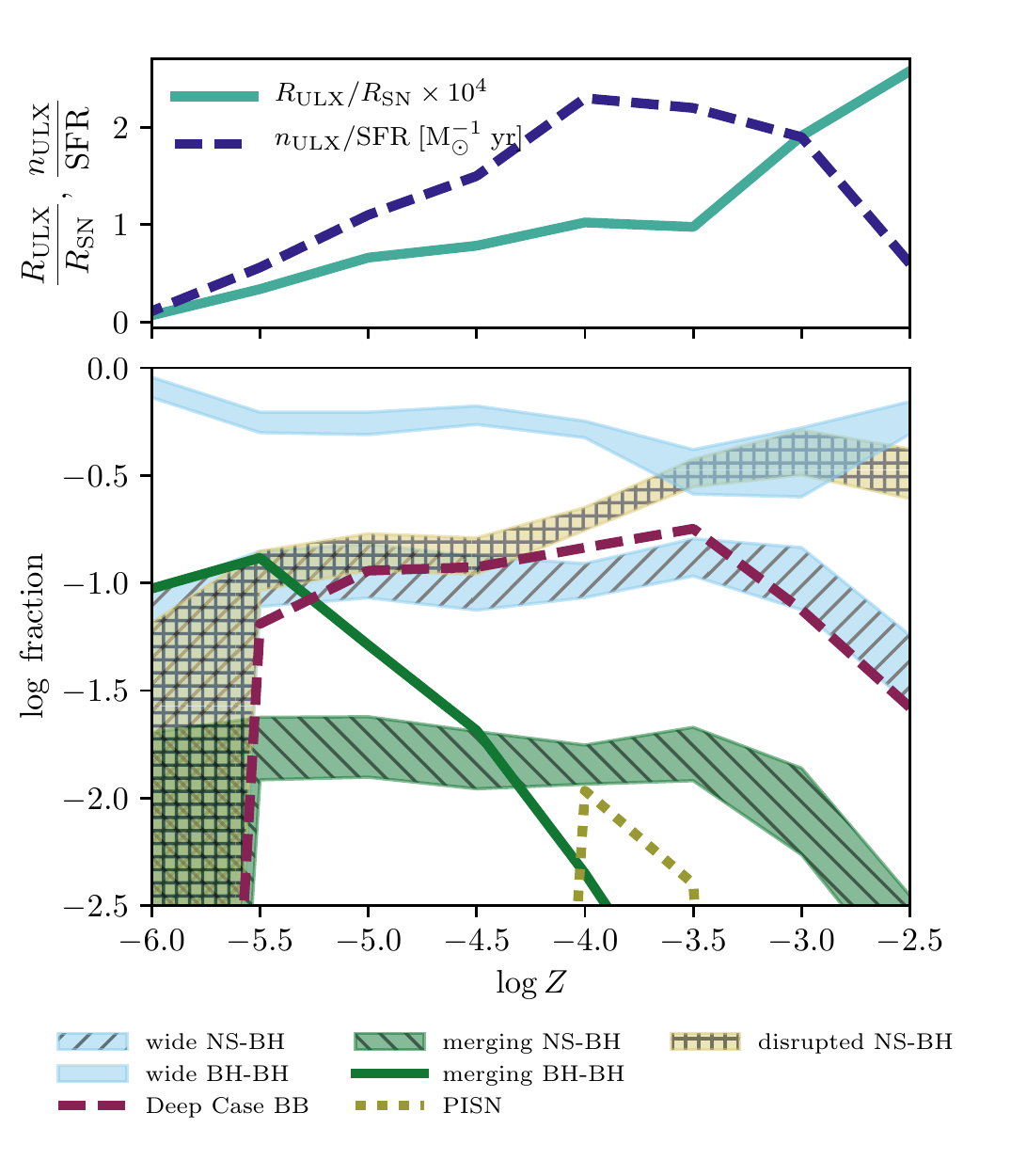}
   \caption{(Top) Production rate of ULXs in terms of the SN rate, and number
      of observable ULXs per $M_\odot~{\rm yr^{-1}}$ of SFR. (Bottom) Final outcomes after a
ULX phase, represented as the fraction of the total of ULXs.
Wide NS-BH and BH-BH systems correspond to binaries for which the
merger time from GWs is longer than $13.8~\rm Gyrs$, while merging NS-BH and
BH-BH systems are compact and/or eccentric enough to merge in less than that.
Disrupted NS-BH systems correspond to binaries that were disrupted due to the
kick imparted on the NS, while the number of disrupted BH-BH systems is
negligible. Systems marked as Deep Case BB widen significantly due to mass transfer
stripping hydrogen depleted layers of the donor, and we do not model
them until core-carbon depletion. A small fraction of systems marked as PISN
have secondaries massive enough to explode as PISNe, leaving only the BH
produced by the primary. The values for NS-BH binaries, and for wide BH-BH
binaries depend on the choice of threshold mass for BH formation.}
   \label{fig:kickoutcome}
   \end{center}
\end{figure}

Considering our full sample of ULX models, the different possible outcomes as a
function of metallicity are shown in Figure \ref{fig:kickoutcome}. These values
take into account the same distribution functions for the initial conditions as used
in Section \ref{sect:resultsLUM}, and also consider possible uncertainties on
the mass limit for BH formation by assuming a threshold of either $8M_\odot$ or
$12M_\odot$ for final masses above which BHs are formed. At all metallicities
considered the majority of systems would result in either
wide BH-BH binaries or NS-BH systems that are disrupted due to the kick to the
NS. The fraction that would result in a bound NS-BH binary is $\sim 10\%$,
further reducing the chances of detecting such a system with an active pulsar. A
similar fraction of systems undergo Case ABB/BB mass transfer driven by shell
helium burning before carbon ignition, resulting in layers of hydrogen depleted
material being stripped from the secondary. These stars are expected to lose
most of their helium envelopes, resulting in stripped CO cores and, owing to the
small final mass ratio, very wide binaries.

The most interesting possibility is the formation of NS-BH and BH-BH systems
compact enough to merge from the emission of GWs in less than a Hubble time.
For most of the metallicity range studied, $\sim2\%$ of the ULXs would become a
NS-BH binary compact enough to merge in a Hubble time, while for BH-BH binaries
it is only at metallicities below $\log~Z=-4$ that a non-negligible number
of sources could produce a merger. As the BH-BH rate is only relevant at
extremely low metallicities and is strongly dependent on the strength of BH
kicks which is not well understood, the numbers we provide for BH-BH mergers
should be considered speculative. Table \ref{table:merger} shows the
expected formation rates per SN for NS-BH and BH-BH compact enough to result in a
merger, including the values weighted with the metallicity distribution of
\citet{LangerNorman2006} at redshift $z=0$ to represent the local production rate.
\begin{table}
        \center
        \begin{tabular}{r|cc}
                %\hline  & & & & & \\
                \hline
                \noalign{\smallskip}
                $\log~ Z$             &
                %$n_{\rm NS-BH}/(10^6\times{\rm
                %SNe})$      &
                $\displaystyle \frac{R_{\rm NS-BH}}{R_{\rm SN}}\times 10^6$      &
                $\displaystyle \frac{R_{\rm BH-BH}}{R_{\rm SN}}\times 10^6$\\
                \noalign{\smallskip}
                \hline
                \noalign{\smallskip}
                $-2.5$ & $0.33-0.91$ & $0$\\
                $-3.0$ & $1.0-2.6$ & $0$  \\
                $-3.5$ & $1.2-2.1$ & $0.076$ \\
                $-4.0$ & $1.2-1.9$ & $0.045$ \\
                $-4.5$ & $0.87-1.6$ & $0.16$ \\
                $-5.0$ & $0.83-1.6$ & $3.4$  \\
                $-5.5$ & $0.41-0.80$ & $4.4$ \\
                $-6.0$ & $0-0.014$ & $0.63$  \\
                \noalign{\smallskip}
                \hline
                \noalign{\smallskip}
                local & 0.069-0.18 & 0.00029 \\
                \noalign{\smallskip}
        \end{tabular}
        \caption{Rate of production per SN of NS-BH and BH-BH systems formed
           after a ULX phase, which are compact enough to merge in less than a
        Hubble time. The NS-BH rate depends on the threshold mass at which a
     NS is formed, and the lower and upper bounds given correspond to
  assuming that for final core masses above $8M_\odot$ and $12M_\odot$
  respectively BHs are formed instead of NSs.}
        \label{table:merger}
        %
        %\vspace{0.7cm}
\end{table}
The expected local formation rate is below one per million SNe for NS-BH
mergers and below one per billion for BH-BH mergers, making
this a very unlikely outcome of the evolution of massive stars
\footnote{Note that
CHE evolution can result in a large number of detectable binary BH mergers, but
this requires both stars to evolve chemically homogeneously, as was shown by
\citet{MandeldeMink2016}, \citet{Marchant+2016} and \citet{deMinkMandel2016}}.

To put this in
the context of detectability by GW detectors, we can estimate a corresponding
volumetric rate for the production of these objects. Taking a volumetric SNe
rate of $1\times10^{5}~\rm Gpc^{-3}yr^{-1}$ (see, e.g.
\citealt{MadauDickinson2014})
and using our upper bound for the local production rate of NS-BH
binaries that would result in a merger, gives a very low rate of $0.018~\rm
Gpc^{-3}~yr^{-1}$, which owing to the large fraction of short delay times these
systems would have, is closely tracked by the rate of actual mergers. Even
assuming a metallicity of $\log~Z=-3.5$ at which we get the largest formation rate, this would
still give a very low upper boundary $<0.2~\rm Gpc^{-3}yr^{-1}$. These values
are comparable to the lower end of the estimates from CE models
\citep{Abadie_ratesummary_2010}. At its third science run the LIGO
detectors are expected to probe down to rates of $\sim 50~\rm
Gpc^{-3}~yr^{-1}$ \citep{Abbott_NSBHrates_2016} for the merger of a $1.4M_\odot$
NS with a $10M_\odot$ BH, which is well above our estimated rate.
The current generation of GW detectors is then unlikely to observe
any of these mergers, but third generation detectors like the Einstein Telescope
and the Big Bang Observatory, if operating at their expected sensitivities,
should detect several of these events per year. Although the contribution of the
CHE channel to the NS-BH merger rate might be sub-dominant, they would be
characterized by very heavy BHs, with masses well in excess of $20M_\odot$.

%%% Local Variables: 
%%% mode: latex
%%% TeX-master: "paper.tex"
%%% End: 

\section{Conclusions}
\label{sect:conclusions}
In this work we have studied a new formation channel for ULXs. 
We find ULXs to form from massive very compact binaries with large mass ratios,
where only the initially more massive star undergoes tidally induced chemically homogeneous
evolution (CHE), and evolves into a massive BH
without ever filling its Roche lobe. Thereafter, the less massive component expands
and undergoes mass transfer to the more massive BH, often on the nuclear time scale
(cf., Fig.\,4). We explore this channel by computing large grids of detailed binary 
evolution models (see Appendix\,\ref{appendix:grids}), which allows us to fully characterize the ensuing 
ULX population (Appendix\,\ref{appendix:properties}). We summarize our main conclusions as follows:
\begin{enumerate}
   \item At metallicities below $Z=0.01$, 
      in binaries with initial orbital periods of
      $1\dots 3\,$d and mass ratios of $q\simeq 0.1-0.4$, 
      primaries more massive than $45M_\odot$ may undergo CHE to form BHs of
      $20M_\odot$ or more. The secondary in these systems then expands and
      starts mass transfer to the BH. Assuming Eddington-limited 
      accretion leads to mega-year long phases with X-ray luminosities in excess of 
      $10^{40}\,$erg/s for many cases.
      This evolutionary path is expected to result in
      the most massive accreting stellar BHs possible at a given metallicity.
   \item The occurrence of PISNe, which leave no compact remnant, leads to
      a gap in BH masses in the range $\sim 60M_\odot-130M_\odot$. At
      metallicities higher than $\log~Z=-3$ no BHs above
      the gap are expected, resulting in a cutoff in BH masses that might be observed as
      a cutoff in ULX luminosities. 
      %This cutoff mass could be reduced
      %due to the ocurrence of PPISNe, but this produces a small difference compared to
      %the uncertainty in accretion rates. 
      At lower metallicities, very massive stars are expected to form BHs above the PISN gap, 
      potentially producing an observable gap in ULX luminosities (Fig.\,8).
   \item Locally, our new channel can account for the brightest observed ULXs, with
      $\sim 0.13$ sources per $M_\odot~\rm yr^{-1}$ of SFR. Observations of nearby
      galaxies give a rate of $2$ ULXs per $M_\odot~\rm yr^{-1}$ of SFR, so a
      different channel is required to explain the less luminous sources.
      The rate from our channel increases significantly in low
      metallicity environments, with a maximum of $2.3$ sources per
      $M_\odot~{\rm yr^{-1}}$ of SFR expected
      at a metallicity of $\log~Z=-4$.
   \item The metallicity dependence of both the number and the luminosity of the
      ULXs predicted through our channel, implies that the ratio of the total X-ray
      luminosity of galaxies and the SFR increases significantly at extremely low
      metallicities.
   \item The majority of our ULX binaries have orbital periods below $20\,$d
      and MS donors in the mass range $8M_\odot \dots 30M_\odot$, with a non-negligible
      number of donors up to $70M_\odot$, possibly explaining some bright
      optical counterparts to observed ULXs that are hard to explain with CE models. More
      than $90\%$ of our sources contain MS donors, transferring mass at
      rates below ten times $\dot{M}_{\rm Edd}$. ULXs formed through
      CHE are also expected to favor low mass ratios, with about $\sim 50\%$ of nearby
      sources having $q<0.5$.
   \item After a ULX phase, depending on the mass of the donor, a NS-BH or BH-BH
      binary could be produced. There is a small but finite probability to produce NS-BH
      systems which are compact enough to merge in less than a Hubble time, with a
      formation rate of $<0.2~\rm Gpc^{-3}~yr^{-1}$. The
      detection of such mergers in the near future is not likely, but they would be
      characterized by having large mass ratios, with BHs more massive than
      $20M_\odot$.
\end{enumerate}

\begin{figure}
   \begin{center}
   \includegraphics[scale=0.4]{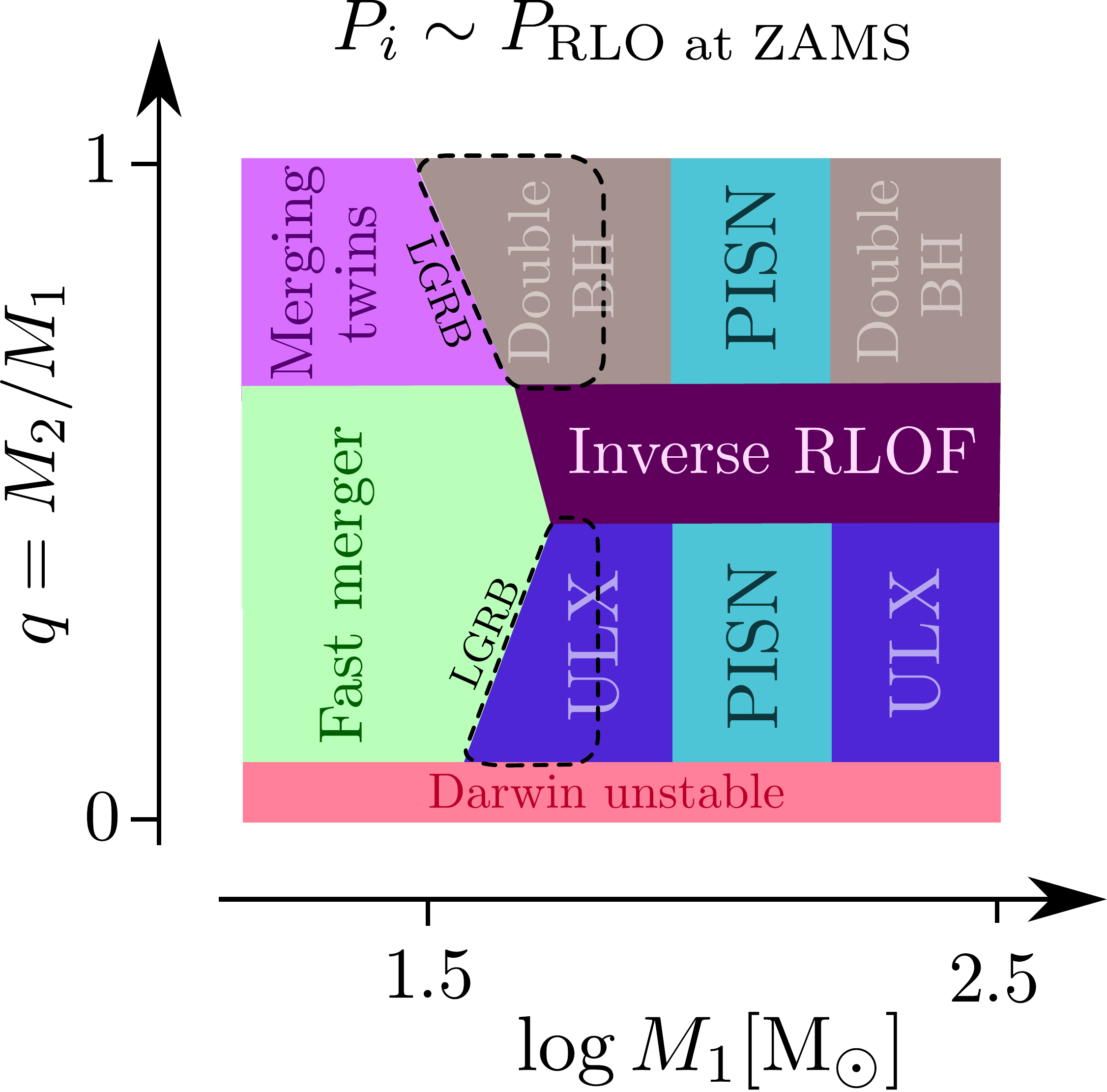}
   \end{center}
   \caption{Schematic representation of the evolutionary end stages of 
   massive low-metallicity binaries with the smallest possible initial orbital periods. 
%If rotational mixing is not strong enough to produce CHE, then
%this diagram would only be populated by fast mergers, twins merging and the
%Darwin unstable region. 
%ONLY EXPLAIN THE FIGURE HERE.  
See text for discussion.}
\label{fig:cartoon}
\end{figure}

Together with the results of \citep{Marchant+2016}, who investigated similar binary systems
to this study only focussing on mass ratios closer to one,
we find that the tightest low-metallicity massive binaries may produce a wealth of exciting phenomena
(Fig.\,\ref{fig:cartoon}).
Since the primaries above $\sim 30M_\odot$ evolve chemically homogeneously (Fig.\,2)
due to tidal synchronisation, they do not expand and may produce BHs with a
mass close to their initial mass. For mass ratios of $q \simgr 0.8$, the secondary may also 
follow CHE and form a second massive BH in the system, potentially leading to 
massive BH mergers. At lower mass ratios the secondaries follow ordinary evolution,
leading to ULXs. In both cases, the mass range of BH formation is interrupted
by the pair instability regime, leading to pair instability supernovae for primary masses
roughly in the range $60M_\odot \dots 130M_\odot$. Finally, as rapid rotation is required
for CHE, the BHs may form with high Kerr parameters, which may give rise to
LGRBs within the framework of the collapsar model. 

%The possibility of CHE in compact binary systems, opens up
%many interesting new branches for stellar evolution. Figure \ref{fig:cartoon}
%illustrates the many different outcomes that are possible if this channel
%exists. Normally we would expect binaries formed with very short
%initial orbital periods to interact before leaving the MS,
%resulting in a fast merger during thermal-timescale mass transfer. Some
%systems at mass ratios close to unity could possibly avoid a very early merger,
%remaining in an overcontact configuration with a mass ratio very close to one, as, 
%for example, VFTS 352, an overcontact system in the LMC composed of two
%$30M_\odot$ stars \citep{Almeida15}.
%Nonetheless, this particular system is expected to eventually
%merge. For higher initial masses, we expect rotational mixing to result in CHE,
%with the formation of compact binary BHs for mass ratios close to
%unity \citep{Marchant2016}, and in ULXs at low initial mass ratios,
%as has been shown in this work. At both ends, it may be possible to
%form maximally spinning BHs that could be a source of LGRBs
%through the collapsar model, while for certain mass ranges a significant
%number of PISNe could be produced. At intermediate mass ratios, the primary
%would evolve chemically homogeneously, but this would be interrupted due to RLOF
%from the secondary before the formation of a BH. These systems should quickly
%merge once the primary expands and fills its Roche lobe.

Many of these outcomes can be assessed observationally in the coming
years. We have argued that, at low metallicities, the distribution
of X-ray luminosities of ULXs could present a pronounced gap, which upcoming missions
such as {\it eROSITA} and {\it Athena} could possibly detect. A similar gap is
expected in the distribution of chirp masses of merging double BHs
\citep{Marchant+2016}, which may be detectable by aLIGO at its design
sensitivity. Current transient surveys such as the intermediate Palomar
Transient Factory \citep{Rau+2009}, the upcoming Zwicky Transient Facility
\citep{Bellm2014,Smith+2014} and the Large-Synoptic-Survey telescope LSST \citep{Tyson2002}
may provide strong constraints on the existence and rates of PISNe. All these
observations from very different instruments will provide strong
tests of our models, and in particular of CHE in the closest massive binary systems.

%%% Local Variables: 
%%% mode: latex
%%% TeX-master: "paper.tex"
%%% End: 

\begin{acknowledgements}
PM and NL are grateful to Bill Paxton for his continuous help in extending the
MESA code to contain all the physics required for this project over the last
years. PM would like to thank the Kavli Institute for theoretical physics of the
university of California Santa Barbara, together with all the participants of
the ``Astrophysics from LIGO’s First Black Holes'' workshop for helpful
discussion. PhP is grateful for a Humboldt Research award at the university of
Bonn. SdM acknowledges support by a Marie Sklodowska-Curie Action
(H2020 MSCA-IF-2014, project id 661502). IM acknowledges partial support from the STFC.
This research was supported in part by the National Science Foundation under
Grant No. NSF PHY11-25915.
We would also like to thank Richard Saxton and Luca Zampieri for helpful
discussion, and Martin Carrington for reporting an issue
with spin-orbit coupling in MESA. The authors would also like to thank the
anonymous referee for many helpful comments and suggestions.
\end{acknowledgements}

\bibliographystyle{aa}

\bibliography{references}

%\Online

\appendix
\section{Grids of binary models}
\label{appendix:grids}
\begin{figure*}
   \includegraphics[width=2\columnwidth]{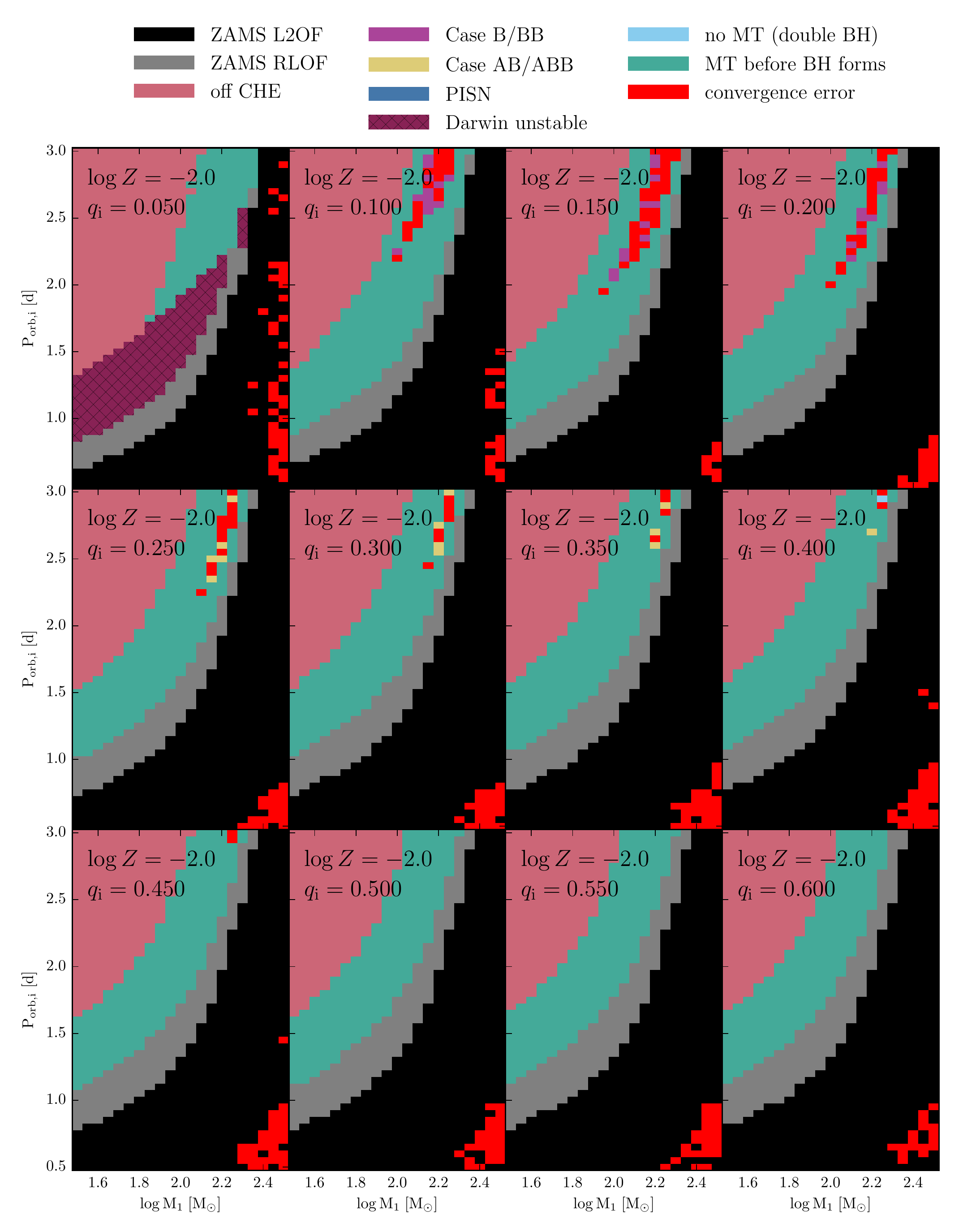}
   \caption{Grid of models for $\log Z=-2.0$. See text in Appendix
      \ref{appendix:grids} for an explanation.}
      \label{fig:grid1}
\end{figure*}

\begin{figure*}
   \includegraphics[width=2\columnwidth]{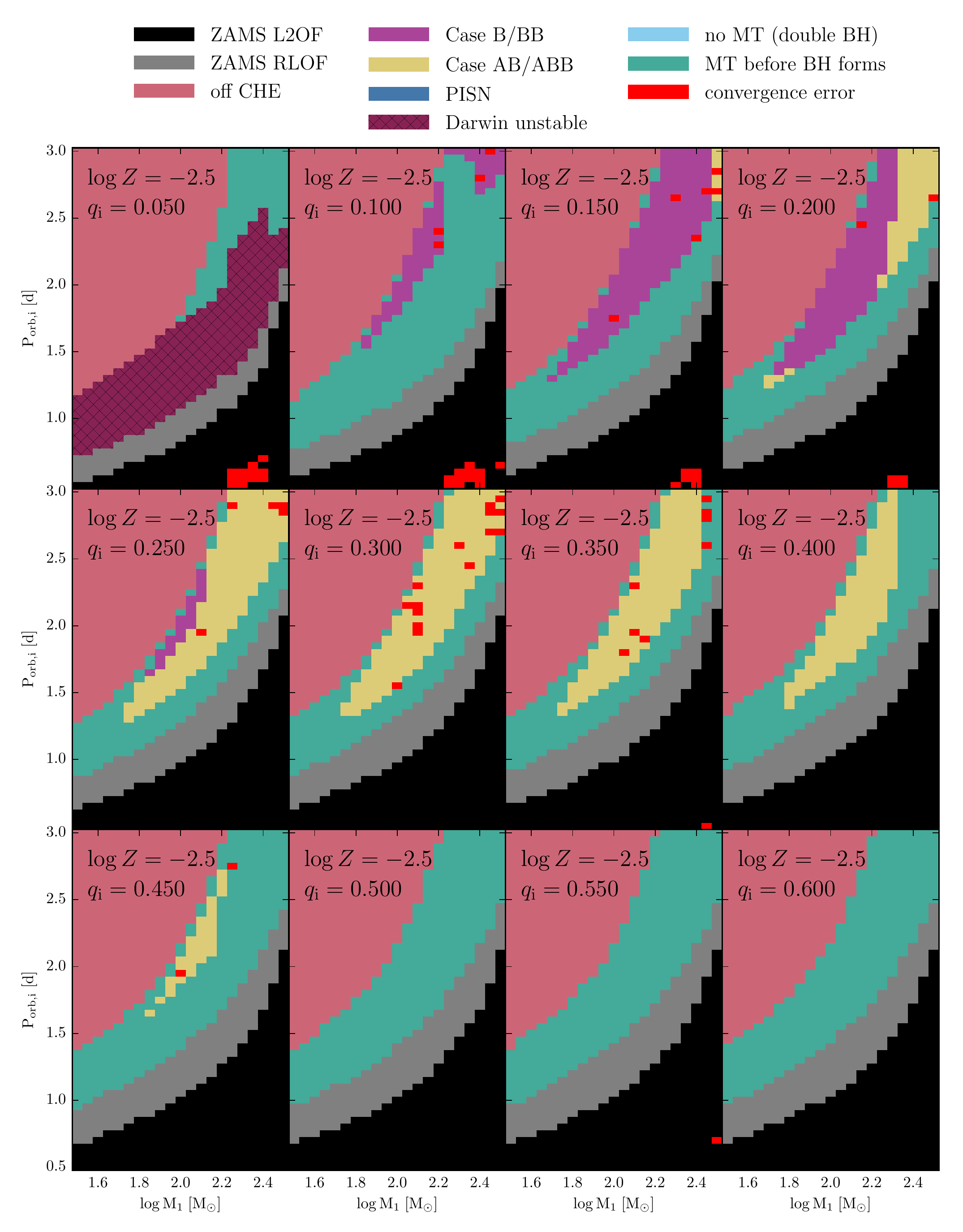}
   \caption{Grid of models for $\log Z=-2.5$. See text in Appendix
      \ref{appendix:grids} for an explanation.}
      \label{fig:grid2}
\end{figure*}

\begin{figure*}
   \includegraphics[width=2\columnwidth]{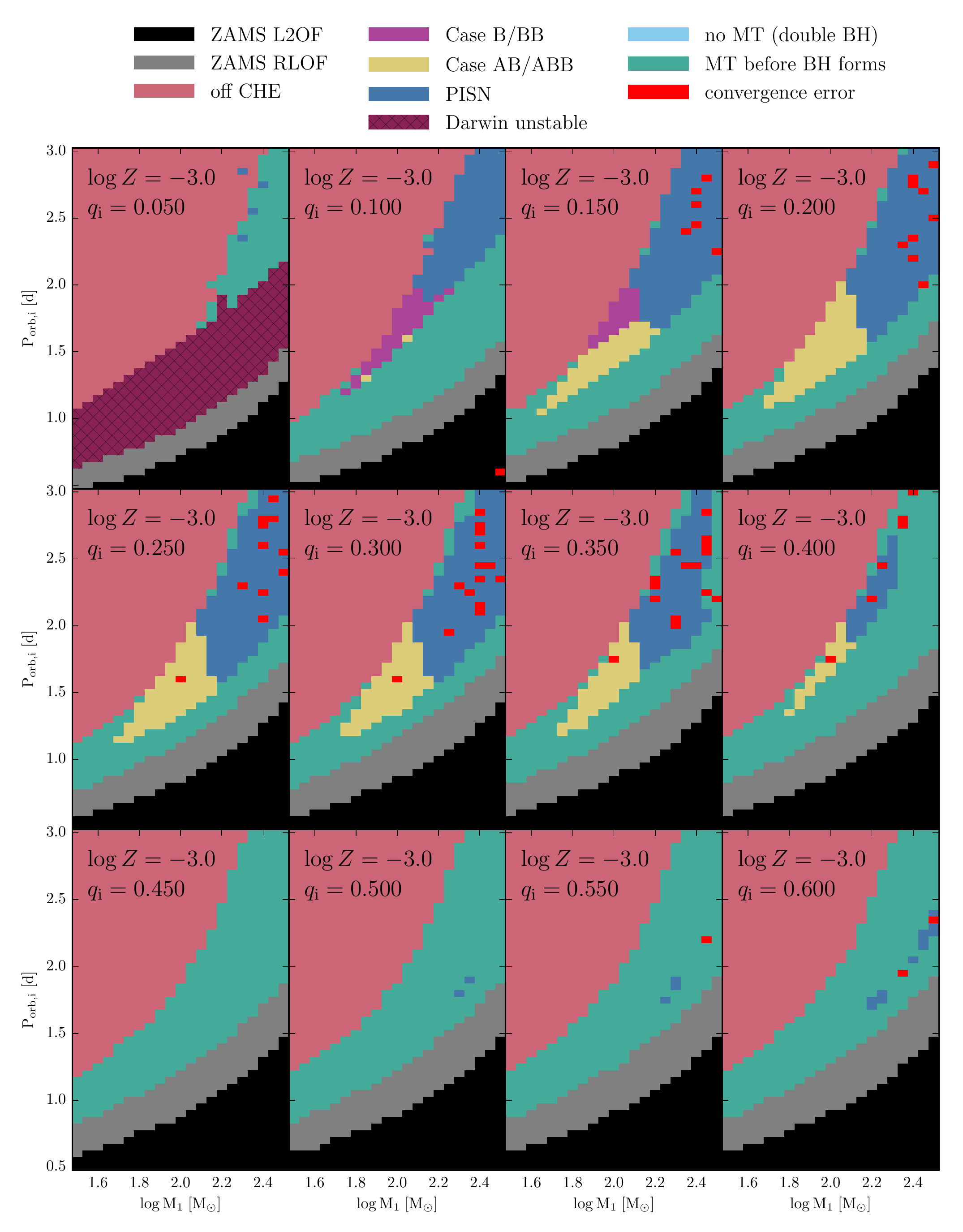}
   \caption{Grid of models for $\log Z=-3.0$. See text in Appendix
      \ref{appendix:grids} for an explanation.}
      \label{fig:grid3}
\end{figure*}

\begin{figure*}
   \includegraphics[width=2\columnwidth]{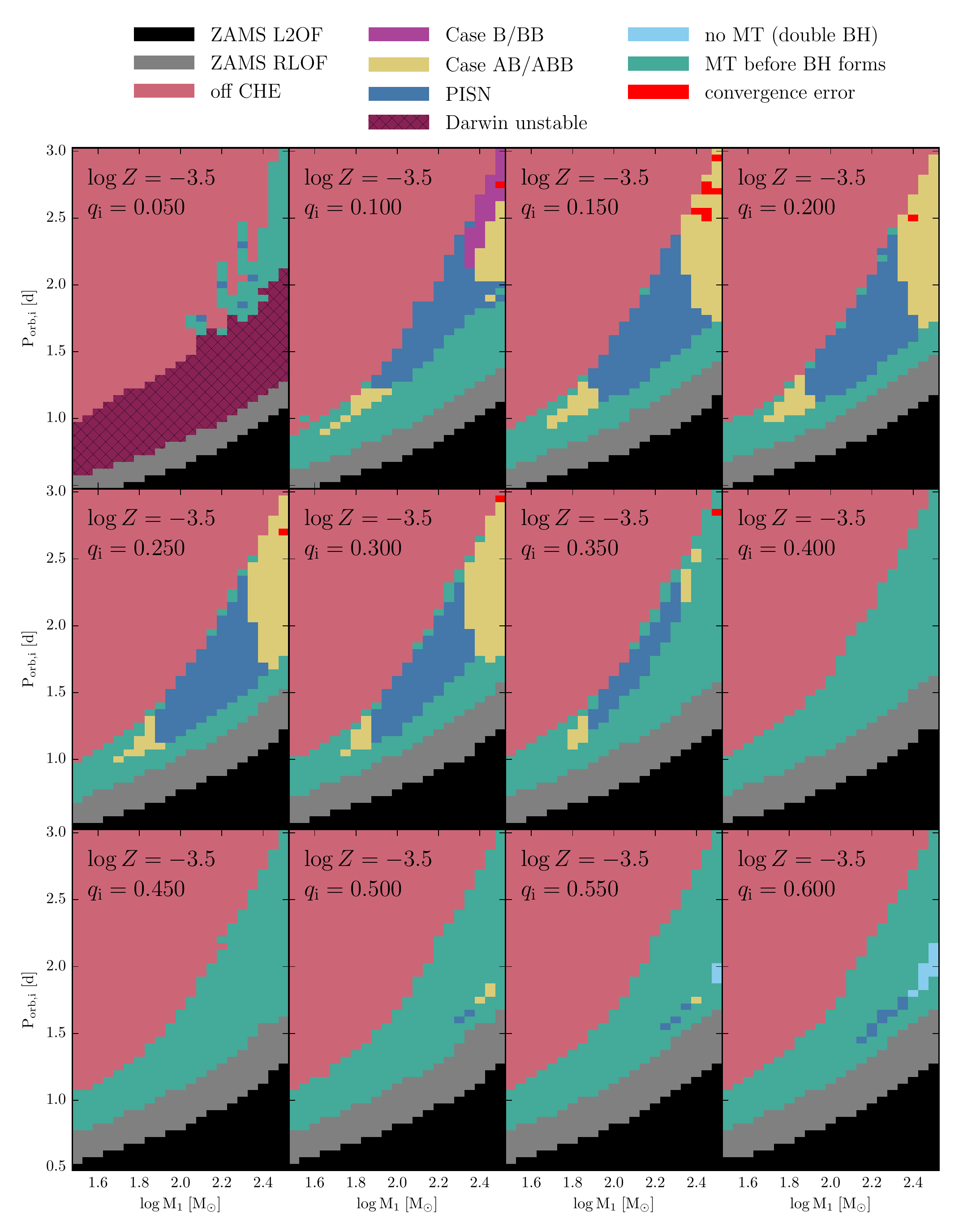}
   \caption{Grid of models for $\log Z=-3.5$. See text in Appendix
      \ref{appendix:grids} for an explanation.}
      \label{fig:grid4}
\end{figure*}

\begin{figure*}
   \includegraphics[width=2\columnwidth]{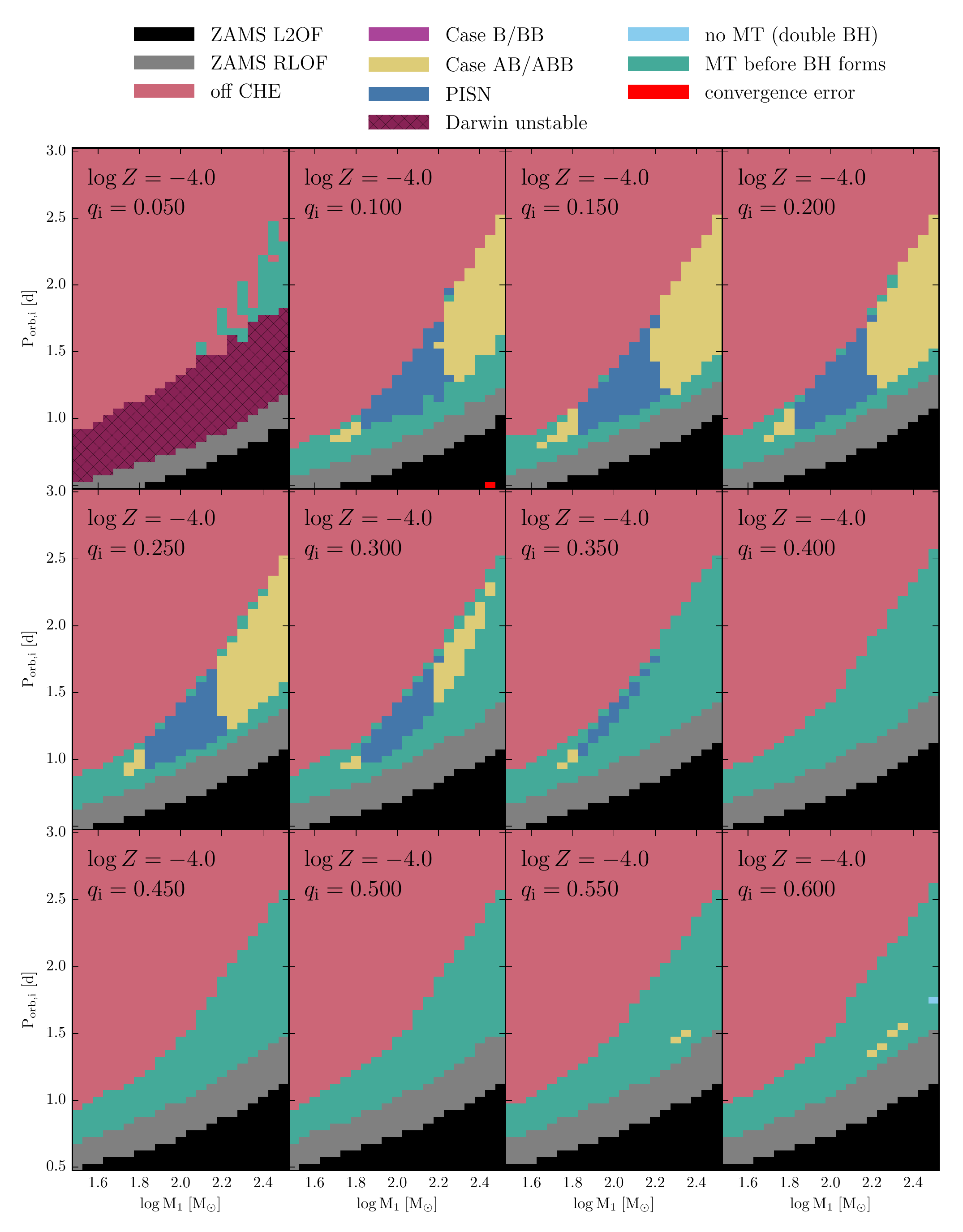}
   \caption{Grid of models for $\log Z=-4.0$. See text in Appendix
      \ref{appendix:grids} for an explanation.}
      \label{fig:grid5}
\end{figure*}

\begin{figure*}
   \includegraphics[width=2\columnwidth]{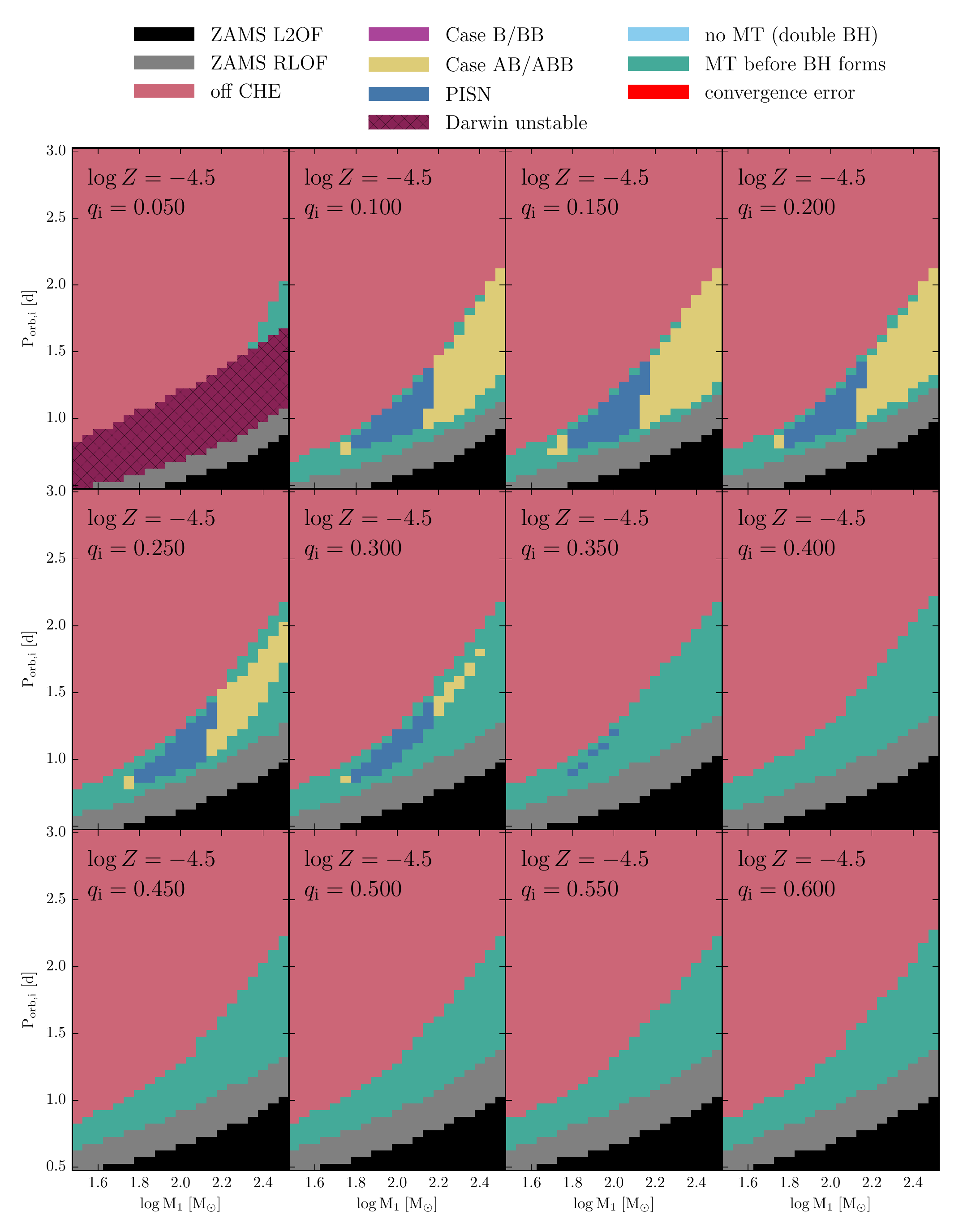}
   \caption{Grid of models for $\log Z=-4.5$. See text in Appendix
      \ref{appendix:grids} for an explanation.}
      \label{fig:grid6}
\end{figure*}

\begin{figure*}
   \includegraphics[width=2\columnwidth]{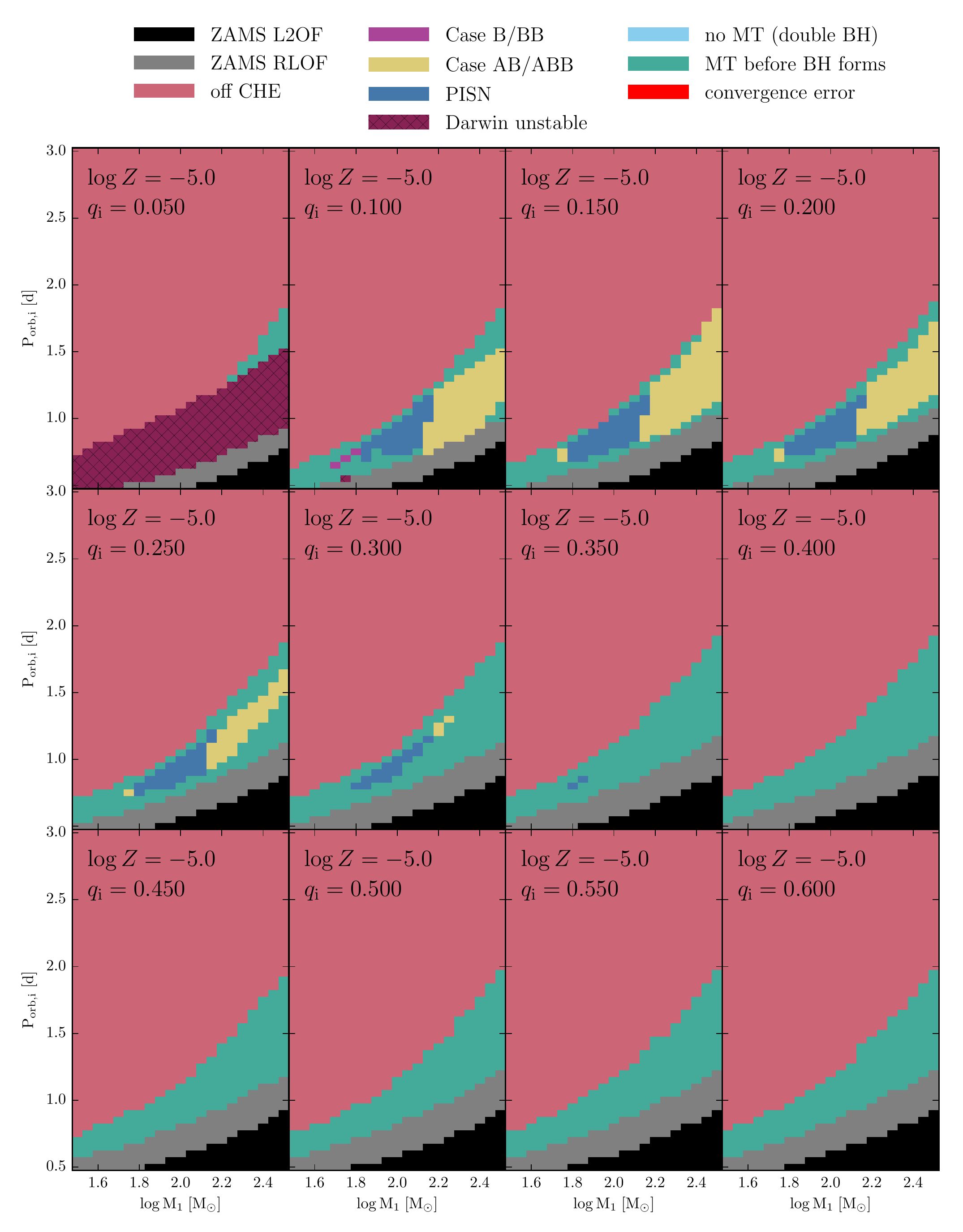}
   \caption{Grid of models for $\log Z=-5.0$. See text in Appendix
      \ref{appendix:grids} for an explanation.}
      \label{fig:grid7}
\end{figure*}

\begin{figure*}
   \includegraphics[width=2\columnwidth]{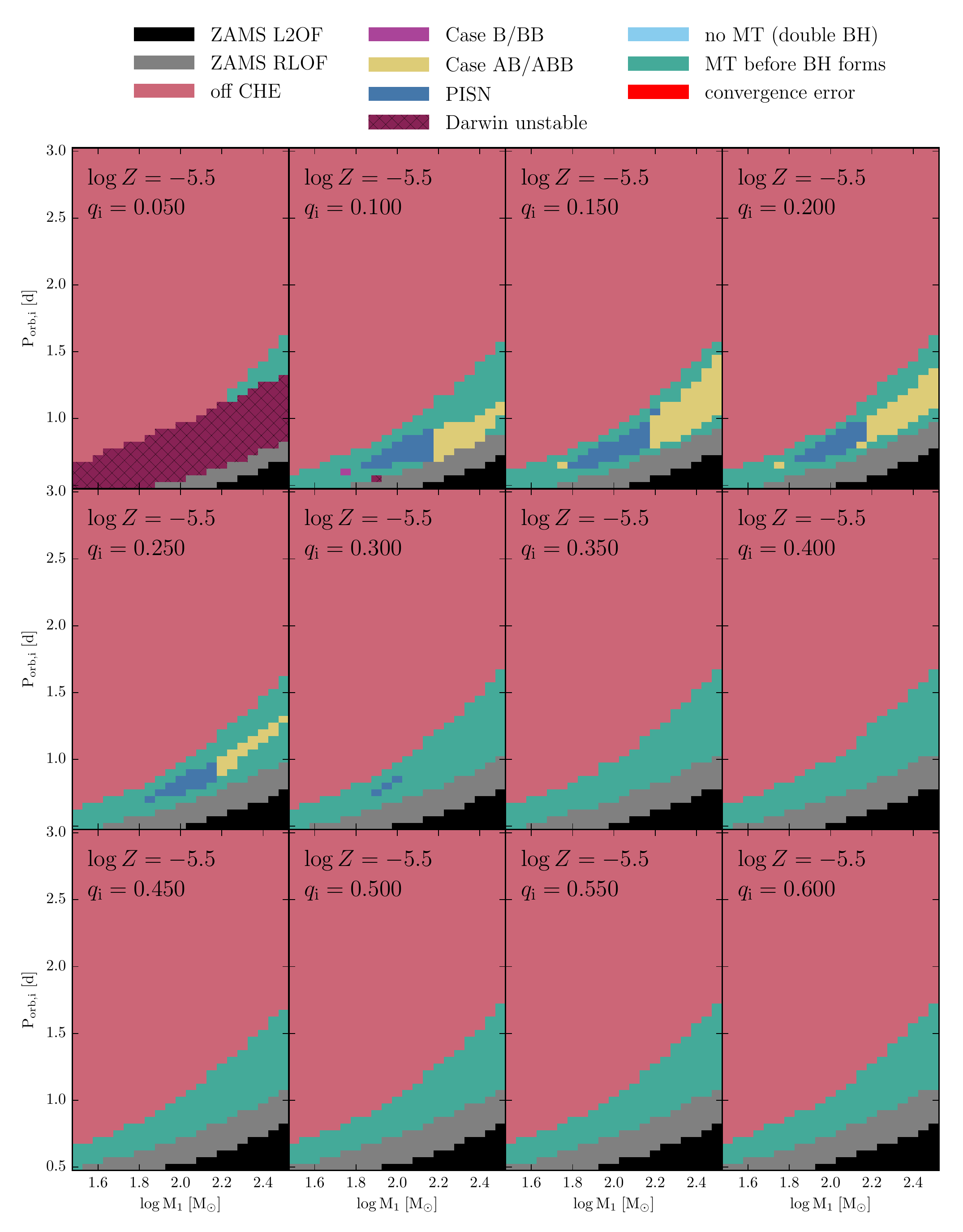}
   \caption{Grid of models for $\log Z=-5.5$. See text in Appendix
      \ref{appendix:grids} for an explanation.}
      \label{fig:grid8}
\end{figure*}

\begin{figure*}
   \includegraphics[width=2\columnwidth]{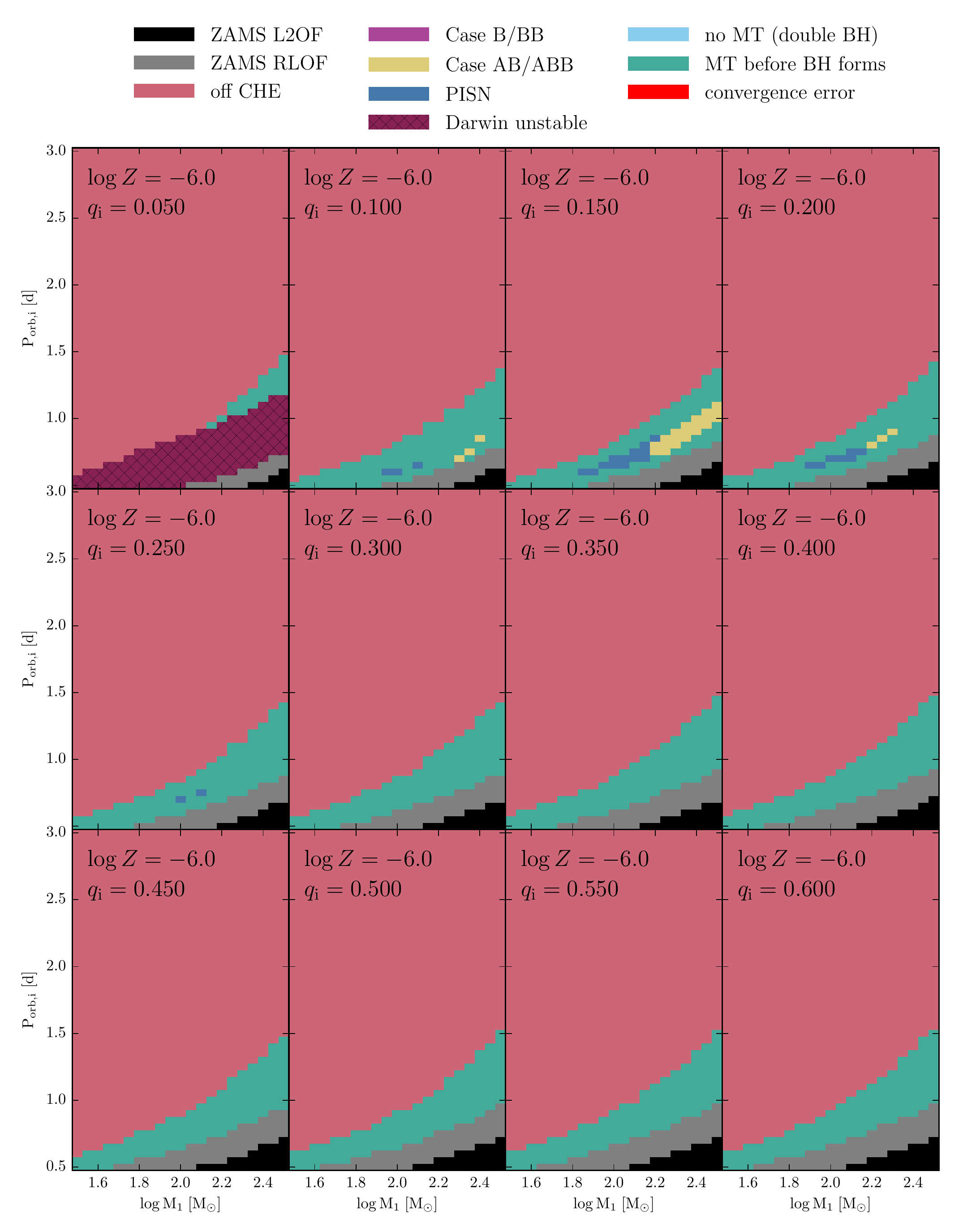}
   \caption{Grid of models for $\log Z=-6.0$. See text in Appendix
      \ref{appendix:grids} for an explanation.}
      \label{fig:grid9}
\end{figure*}

A summary of the outcomes of our simulations is presented in Figures
\ref{fig:grid1}-\ref{fig:grid9}. The meaning of the different labels in those
figures is as follows:
\begin{itemize}
   \item ZAMS L2OF: The initial orbital separation is so short that the system
      overflows the L2 Lagrangian point at ZAMS. Such a system should rapidly
      merge.
   \item ZAMS RLOF: System is undergoing RLOF at ZAMS. As shown in
      \citet{Marchant+2016}, these overcontact systems might survive interaction
      without merging, resulting in a binary with equal mass components.
      However, for low mass ratios we mostly expect the systems to evolve into
      deep contact and merge, and even for systems that avoid that, they would
      not follow the channel for ULX formation described in this work.
   \item off CHE: Primary reached a point where the difference between central
      and surface helium abundance is larger than $0.2$. We consider such
      systems are not evolving chemically homogeneous, and terminate these
      simulations.
   \item Case B/BB: The primary evolves chemically homogeneous and forms a BH.
      The secondary then initiates mass transfer and a possible ULX phase after
      depleting its central hydrogen. These systems are only expected to have a
      brief ULX phase. Some of these
      systems undergo an additional phase of case BB mass transfer after core
      helium depletion.
   \item Case AB/ABB: Similar to the previous one, only that mass transfer is
      initiated while the secondary is on the MS so it operates on a much longer
      nuclear timescale.
   \item PISN: The final mass of the primary at helium depletion is in the range
      $60<M_{1,\rm f}<130$, so we expect to result in a SN leaving the secondary
      as a single star.
   \item no MT (double BH): Both stars evolve chemically homogeneous, avoiding
      mass transfer and resulting in a compact binary BH. This is the path of
      evolution discussed in \citet{Marchant+2016}. For the mass ratios studied,
      only a handful of these systems are found.
   \item MT before BH forms: Mass transfer, either from the primary or the
      secondary, happens before BH formation. We expect such systems to either
      merge (in case the primary is the donor) or widen and
      interrupt CHE because of accretion of hydrogen-rich material (if the
      secondary is the donor), which would not result in the formation of a ULX.
   \item convergence error: Due to numerical problems the simulation was not
      completed.
   \item {Darwin unstable: At its initial state the system has an orbital
      separation smaller than $a_{Darwin}$, and thus is Darwin unstable. It
      would not be possible to form a
      synchronized binary with this orbital separation, as it would result in a
   merger instead. The moment of inertia is dependent in the initial rotation
rate, which results in some irregularities in the boundary between stable and
unstable models.}
\end{itemize}

\newpage
\clearpage

\section{Computation of the formation rate and number of observable
ULXs}
\label{appendix:rates}
To compute the number of ULXs formed per core-collapse SN, we first consider the binary
fraction $f_{\rm b}$ defined as the ratio between the number of binary systems
formed
to the total number of binaries and singles formed. Our
assumption that out of three massive stars two are formed in a binary implies
$f_{\rm b}=0.5$. We further assume that the lower mass threshold for a single star to result in
a SN is $M_{\rm SN}=8M_\odot$, and that binaries
containing primaries with masses $M_1>M_{\rm SN}$ produce two
SNe\footnote{In reality, very massive binaries resulting in BHs or binaries
   where the secondary is below $M_{\rm SN}$ produce less than two SNe events
per binary. This results in a systematic underestimate of the rate of massive
binary formation in terms of the SN rate (though we expect this effect to be
less than a factor of 2).}.
Using this, the formation rate of massive binaries $R_{\rm MB}$, defined as
the number of binaries
with primary masses above $M_{\rm SN}$ formed per unit time in a given galaxy
(or collection of galaxies), can be related to
the formation rate of SNe of that galaxy, $R_{\rm SN}$:
\begin{eqnarray}
   R_{\rm MB} = \frac{f_{\rm b}}{1+f_{\rm b}}R_{\rm SN}.
\end{eqnarray}
Taking into account the results of our binary models, we can define a function
$F_{\rm ULX}(M_{\rm 1,i},q_{\rm i},P_{\rm i},Z)$ which is either $1$ or $0$ depending on
whether a binary system with the corresponding initial parameters results in a
ULX or not. Considering a distribution of initial binary parameters given by
\begin{eqnarray}
   \frac{d N}{d M_{1,\rm i}}\propto f_{M}(M_{1, \rm i}),\quad
   \frac{d N}{d q}\propto f_{q}(q),\quad
   \frac{d N}{d P_{\rm i}}\propto f_{P}(P_{\rm i}),
\end{eqnarray}
the rate of formation of ULXs can be expressed in terms of the SN rate as
\begin{eqnarray}
   \begin{aligned}
      \frac{R_{\rm ULX}}{R_{\rm SN}}=\frac{f_{\rm b}}{1+f_{\rm
      b}}\times\qquad\qquad\qquad\qquad\qquad\qquad\qquad\qquad\\
   \qquad\qquad\frac{\displaystyle \int_{M_{\rm SN}}^{\infty}\int_{0}^{1}\int_{P_{\rm min}}^{P_{\rm max}}
   F_{\rm ULX}
   f_{M} f_{q} f_{P}\;d P_{\rm i} d q_{\rm i} d M_{1,\rm i}}
   {\displaystyle \int_{M_{\rm SN}}^{\infty}\int_{0}^{1}\int_{P_{\rm
   min}}^{P_{\rm max}}f_{M} f_{q} f_{P}\;d P_{\rm i} d q_{\rm i} d M_{1,\rm i}}.
\end{aligned}
\end{eqnarray}
To compute the number of observable ULXs in a galaxy, $n_{\rm ULX}$, we need to
take into account the duration of a ULX phase, which we define as $t_{\rm
ULX}(M_{\rm 1,i},q_{\rm i},P_{\rm i},Z)$. Furthermore, if we assume the
formation rate of SNe is proportional to the $\rm SFR$, $n_{\rm ULX}$ is then given by
\begin{eqnarray}
   \begin{aligned}
      \frac{n_{\rm ULX}}{\rm SFR}=\frac{f_{\rm b}}{1+f_{\rm
      b}}\frac{R_{\rm SN}}{\rm SFR}\times\qquad\qquad\qquad\qquad\qquad\qquad\qquad\\
   \qquad\qquad\frac{\displaystyle \int_{M_{\rm SN}}^{\infty}\int_{0}^{1}\int_{P_{\rm min}}^{P_{\rm max}}
   t_{\rm ULX}
   f_{M} f_{q} f_{P}\;d P_{\rm i} d q_{\rm i} d M_{1,\rm i}}
   {\displaystyle \int_{M_{\rm SN}}^{\infty}\int_{0}^{1}\int_{P_{\rm
   min}}^{P_{\rm max}}f_{M} f_{q} f_{P}\;d P_{\rm i} d q_{\rm i} d M_{1,\rm
   i}},
\end{aligned}
\end{eqnarray}
and for this work we have used $R_{\rm SN}/{\rm SFR}=0.01\;M_{\odot}^{-1}$. For a
different value of this ratio, all results presented in this paper can easily be scaled.
Finally, the average time that systems resulting in a ULX spend as active
sources can be computed as
\begin{eqnarray}
   \begin{aligned}
   \langle t_{\rm ULX}\rangle = \frac{n_{\rm ULX}}{R_{\rm
   ULX}}=\qquad\qquad\qquad\qquad\qquad\qquad\qquad\qquad\\
   \qquad\qquad\frac{\displaystyle \int_{M_{\rm SN}}^{\infty}\int_{0}^{1}\int_{P_{\rm min}}^{P_{\rm max}}
   t_{\rm ULX}
   f_{M} f_{q} f_{P}\;d P_{\rm i} d q_{\rm i} d M_{1,\rm i}}
   {\displaystyle \int_{M_{\rm SN}}^{\infty}\int_{0}^{1}\int_{P_{\rm
   min}}^{P_{\rm max}}F_{\rm ULX}f_{M} f_{q} f_{P}\;d P_{\rm i} d q_{\rm i} d M_{1,\rm
   i}}.
\end{aligned}
\end{eqnarray}
\section{Evolution of models accreting above $\dot{M}_{\rm
Edd}$}\label{appendix:mdotedd}
\begin{figure}
   \includegraphics[width=\columnwidth]{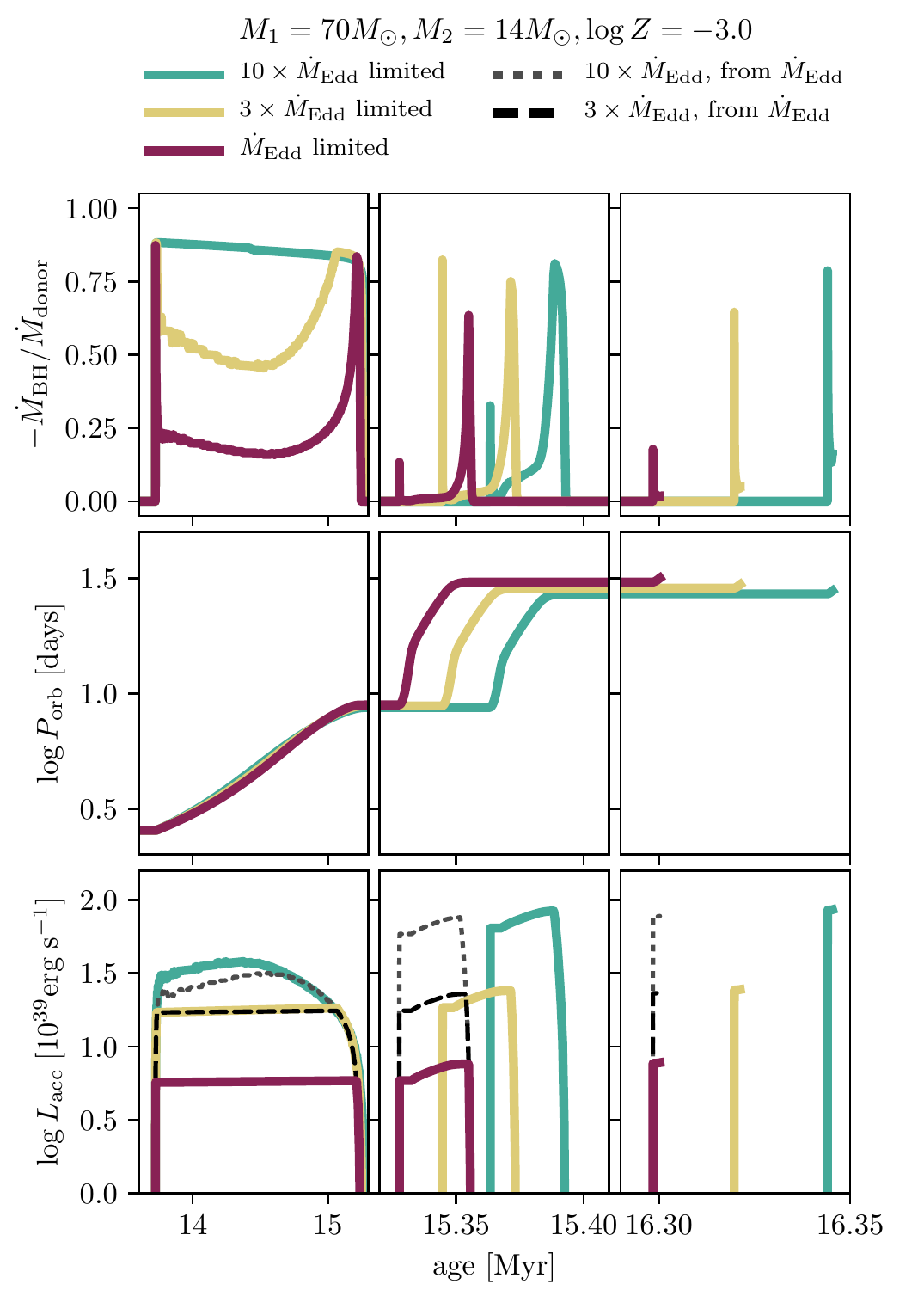}
   \caption{Zoom-in into the three mass transfer phases of the ULX model at
      metallicity $\log Z=-3.0$ shown in Figure \ref{fig:masstransfer},
   including simulations where the Eddington limit was increased by factors of
$3$ and $10$. Dashed and dotted lines show the estimated luminosity that
simulations with increased Eddington factors would have, using the potential
luminosity a source could reach in the simulation strictly limited to the
Eddington rate. (top) Mass transfer efficiency. (middle) Orbital periods.
(bottom) Accretion luminosity.}
      \label{fig:mdotedd_check2}
\end{figure}
\begin{figure}
   \includegraphics[width=\columnwidth]{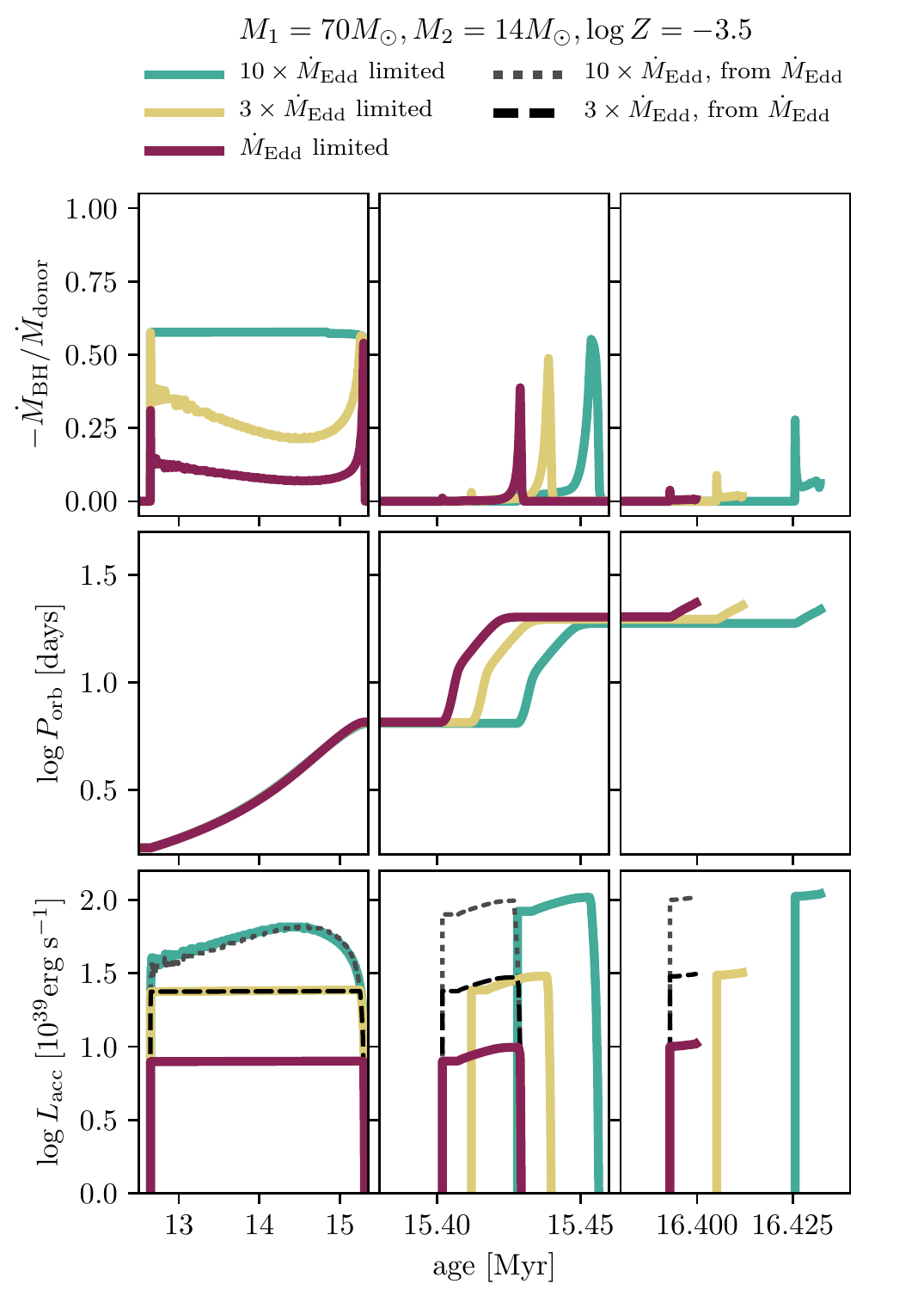}
   \caption{Same as figure \ref{fig:mdotedd_check2}, bur for a metallicity $\log
Z=-3.5$}
      \label{fig:mdotedd_check}
\end{figure}

In section \ref{sect:resultsLUM} we consider how the distribution of ULX
luminosities varies if the BH can accrete either at three or ten times
its Eddington rate. This is not done self-consistently, but rather by
post-processing our simulation grids that are strictly limited to the Eddington
rate, and considering the potential luminosity they could reach at different
phases. Figures \ref{fig:mdotedd_check2} and \ref{fig:mdotedd_check} show two
examples of this, comparing the estimated luminosities with simulations where
the Eddington limit is increased self-consistently.

The longer lasting X-ray phase,
which is expected to dominate in observed populations, is Case A mass transfer,
which is depicted on the left panels of Figures \ref{fig:mdotedd_check2} and
\ref{fig:mdotedd_check}. When accreting at 10 times the Eddinton rate, for both
metallicites shown, Case A mass transfer becomes as efficient as it can be, while
not becoming completely conservative due to wind mass loss from the donor and
the loss of a significant amount of accreted mass as radiation (see Equation
\ref{equ:mdotbh}). Despite the change in efficiency, the evolution of the
orbital periods remains more or less the same, as the final mass ratios are very
similar, and the evolution of orbital angular momentum is dominated by wind mass
loss of the donor and spin-orbit coupling, which are not modified by the
efficiency of accretion. For both metallicites, the non-self-consistent method
does a very good job in reproducing both the luminosities and lifetimes of ULX
phases.

For the post main-sequence mass-transfer phases shown in the middle and right
panels of Figures \ref{fig:mdotedd_check2} and \ref{fig:mdotedd_check},
mass-transfer rates are about two orders of magnitude above the Eddington limit, such
that all models considered accrete inefficiently, even when allowed to do so at
ten times the Eddington rate. The resulting orbital evolution is then almost
identical, with luminosities and lifetimes of ULXs in models with increased
Eddington rates being well reproduced from the model strictly limited to
$\dot{M}_{\rm Edd}$.

\section{Construction of synthetic galaxies at a fixed SFR}\label{appendix:synth}
To model the X-ray luminosities that individual galaxies would have for a given SFR,
we consider all individual timesteps for each binary model simulated at a
given metallicity, and account for the probability of each of these to be
observed. In the following, we consider a timestep of size $\Delta t$ at
a given point in time for a simulation with initial parameters $M_{1,\rm i}$,
$q_{\rm i}$, $P_{\rm i}$ and $Z$, during which the binary is predicted to be an X-ray
source with luminosity $L_{\rm X}$. The rate at which such a system would form
is independent of $\Delta t$, and can be computed in terms of the SFR as
\begin{eqnarray}
   \begin{aligned}
      \frac{R_{\rm \Delta t}}{\rm SFR}=\frac{f_{\rm b}}{1+f_{\rm
      b}}\frac{R_{\rm SN}}{\rm SFR}\times\qquad\qquad\qquad\qquad\qquad\qquad\qquad\\
   \qquad\qquad\frac{\displaystyle 
   f_{M} f_{q} f_{P}\;\Delta P_{\rm i} \Delta q_{\rm i} \Delta M_{1,\rm i}}
   {\displaystyle \int_{M_{\rm SN}}^{\infty}\int_{0}^{1}\int_{P_{\rm
   min}}^{P_{\rm max}}f_{M} f_{q} f_{P}\;d P_{\rm i} d q_{\rm i} d M_{1,\rm i}},
\end{aligned}
\end{eqnarray}
where $\Delta P_{\rm i}$, $\Delta q_{\rm i}$ and $\Delta M_{1,\rm i}$ are the
spacings in the parameter space of our simulation grids corresponding to the
particular model in question. The probability of observing $k$ such systems in a
galaxy can can then be determined using a Poisson distribution,
\begin{eqnarray}
   P(k)=\frac{\lambda^{k}e^{-\lambda}}{k!},\quad \lambda = R_{\Delta t}\Delta t.
\end{eqnarray}
A synthetic galaxy can then be constructed by sampling this probability
distribution for each timestep in all simulations at a given
metallicity, and adding up the individual contributions of $k L_{\rm X}$ to the
total X-ray luminosity of the galaxy, $L_{\rm X, gal}$.

The average ratio between galactic X-ray luminosities and SFRs can be
computed without the need to model a large population as
\begin{eqnarray}
   \begin{aligned}
      \left\langle\frac{L_{\rm X,\rm gal}}{\rm SFR}\right\rangle=\frac{f_{\rm b}}{1+f_{\rm
      b}}\frac{R_{\rm SN}}{\rm SFR}\times\qquad\qquad\qquad\qquad\qquad\qquad\qquad\\
   \qquad\quad\frac{\displaystyle \int_{M_{\rm
   SN}}^{\infty}\int_{0}^{1}\int_{P_{\rm min}}^{P_{\rm
   max}}\int_{0}^{t_{\rm f}}
   L_{\rm X}(t)
   f_{M} f_{q} f_{P}\;dt\,d P_{\rm i} d q_{\rm i} d M_{1,\rm i}}
   {\displaystyle \int_{M_{\rm SN}}^{\infty}\int_{0}^{1}\int_{P_{\rm
   min}}^{P_{\rm max}}f_{M} f_{q} f_{P}\;d P_{\rm i} d q_{\rm i} d M_{1,\rm
   i}},
\end{aligned}
\end{eqnarray}
where we integrate over the age $t$ of a binary with given initial parameters,
from the ZAMS to the endpoint of its evolution, $t_{\rm f}$.
We have verified that our synthetic galactic models satisfy this average, which
can also be seen from the models at high values of SFR in Figure
\ref{fig:lxsfr} following the expected linear trend.

\section{Properties of ULXs}\label{appendix:properties}
\begin{figure}
   \includegraphics[width=\columnwidth]{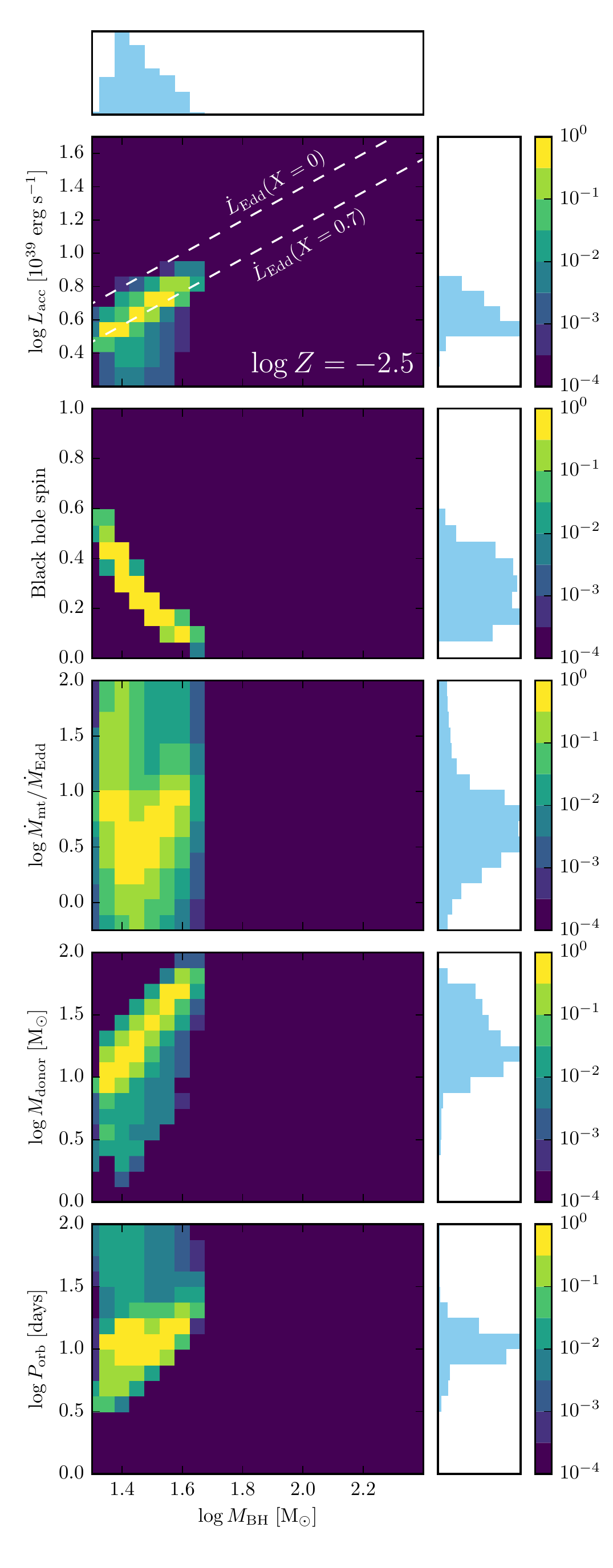}
   \caption{General properties of ULXs for $\log Z=-3.0$. See text in Appendix
      \ref{appendix:properties} for an explanation.}
      \label{fig:props1}
\end{figure}

\begin{figure}
   \includegraphics[width=\columnwidth]{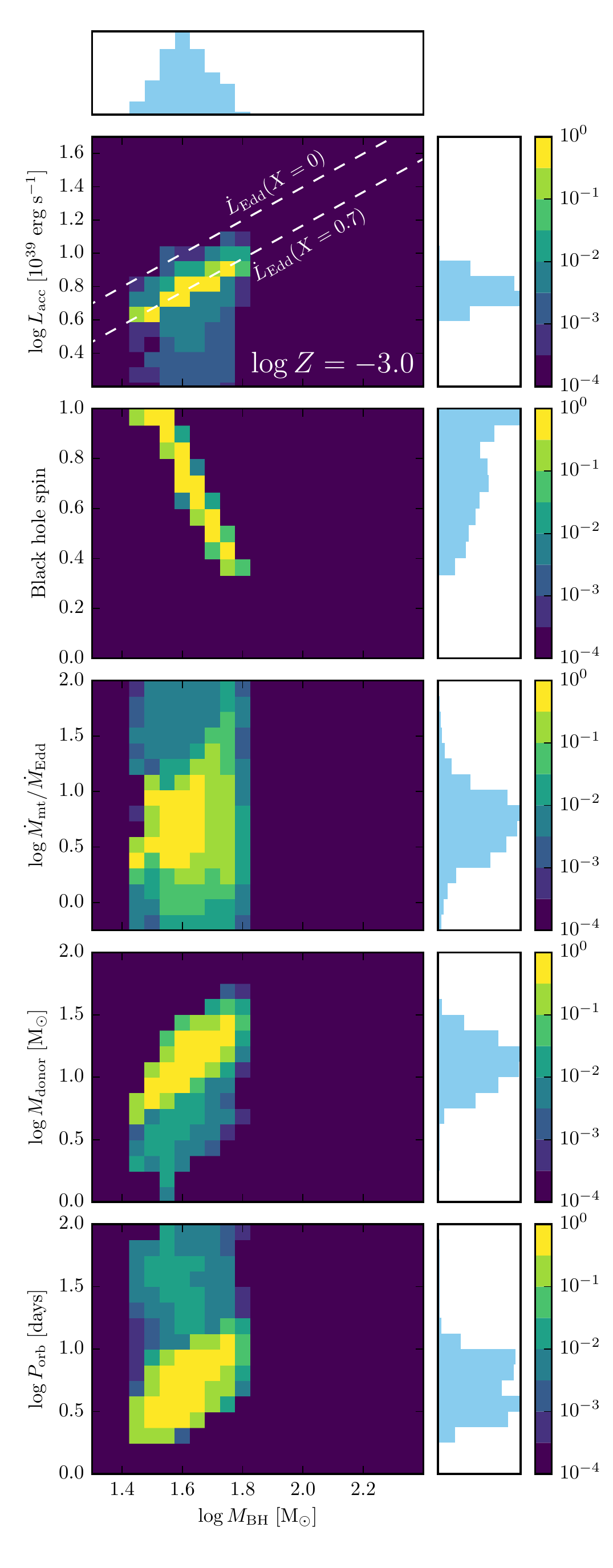}
   \caption{General properties of ULXs for $\log Z=-3.0$. See text in Appendix
      \ref{appendix:properties} for an explanation.}
      \label{fig:props2}
\end{figure}

\begin{figure}
   \includegraphics[width=\columnwidth]{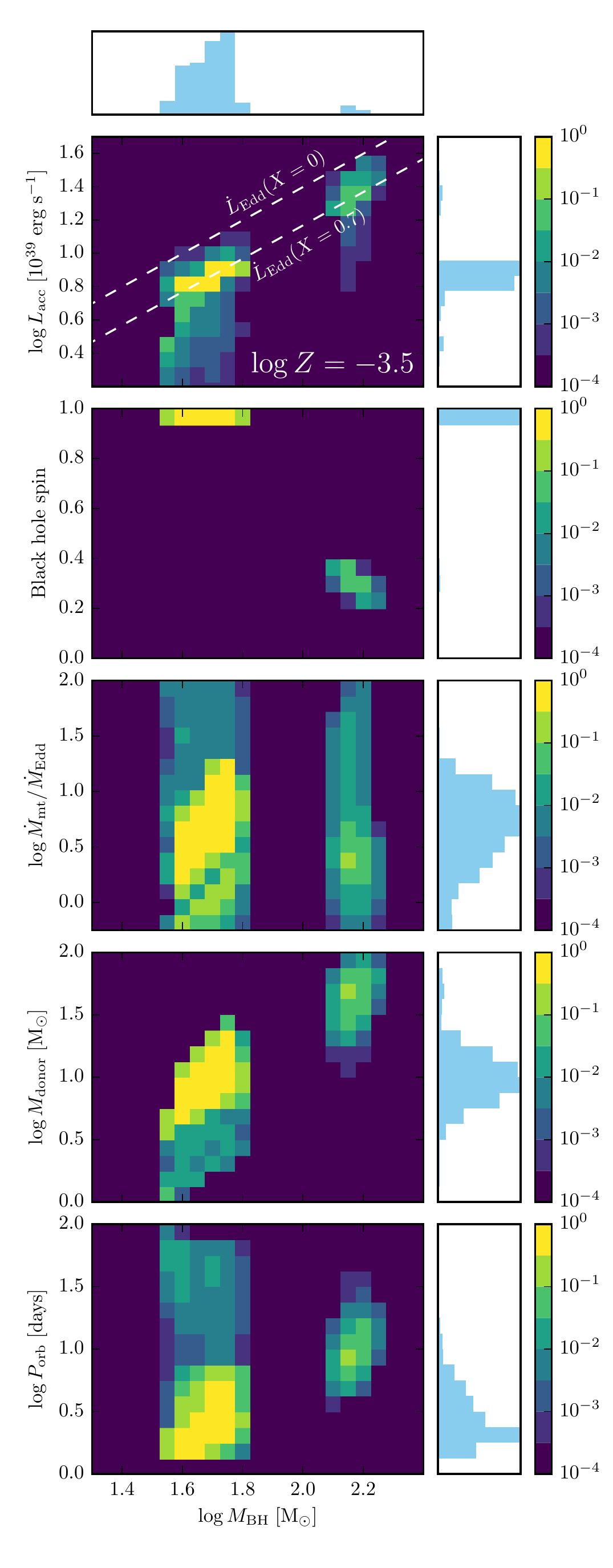}
   \caption{General properties of ULXs for $\log Z=-3.0$. See text in Appendix
      \ref{appendix:properties} for an explanation.}
      \label{fig:props3}
\end{figure}

\begin{figure}
   \includegraphics[width=\columnwidth]{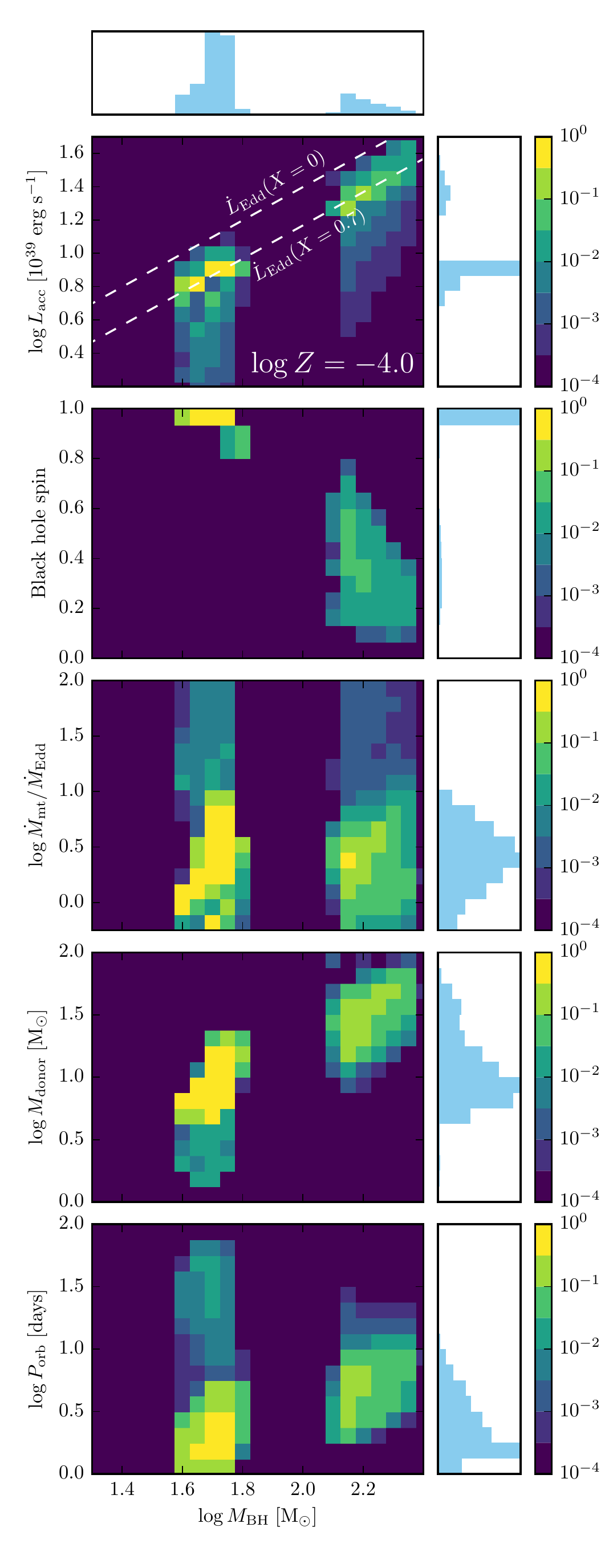}
   \caption{General properties of ULXs for $\log Z=-4.0$. See text in Appendix
      \ref{appendix:properties} for an explanation.}
      \label{fig:props4}
\end{figure}

\begin{figure}
   \includegraphics[width=\columnwidth]{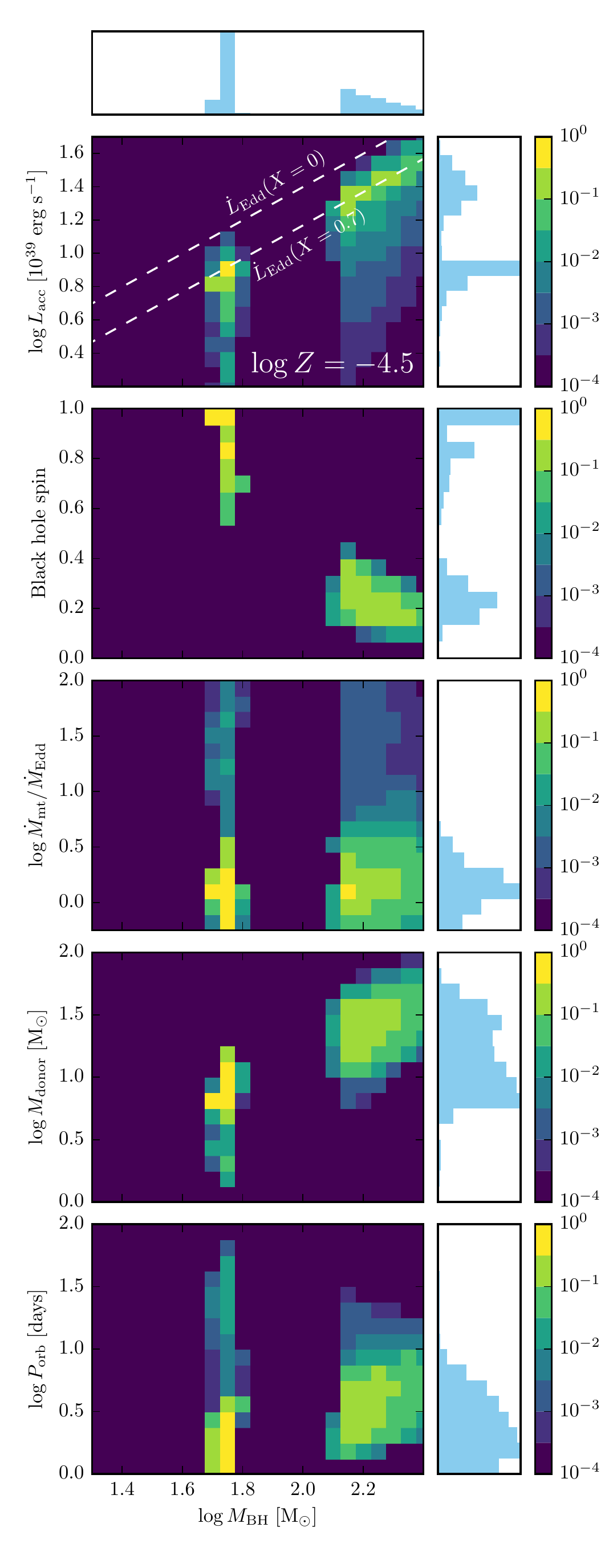}
   \caption{General properties of ULXs for $\log Z=-4.5$. See text in Appendix
      \ref{appendix:properties} for an explanation.}
      \label{fig:props5}
\end{figure}

\begin{figure}
   \includegraphics[width=\columnwidth]{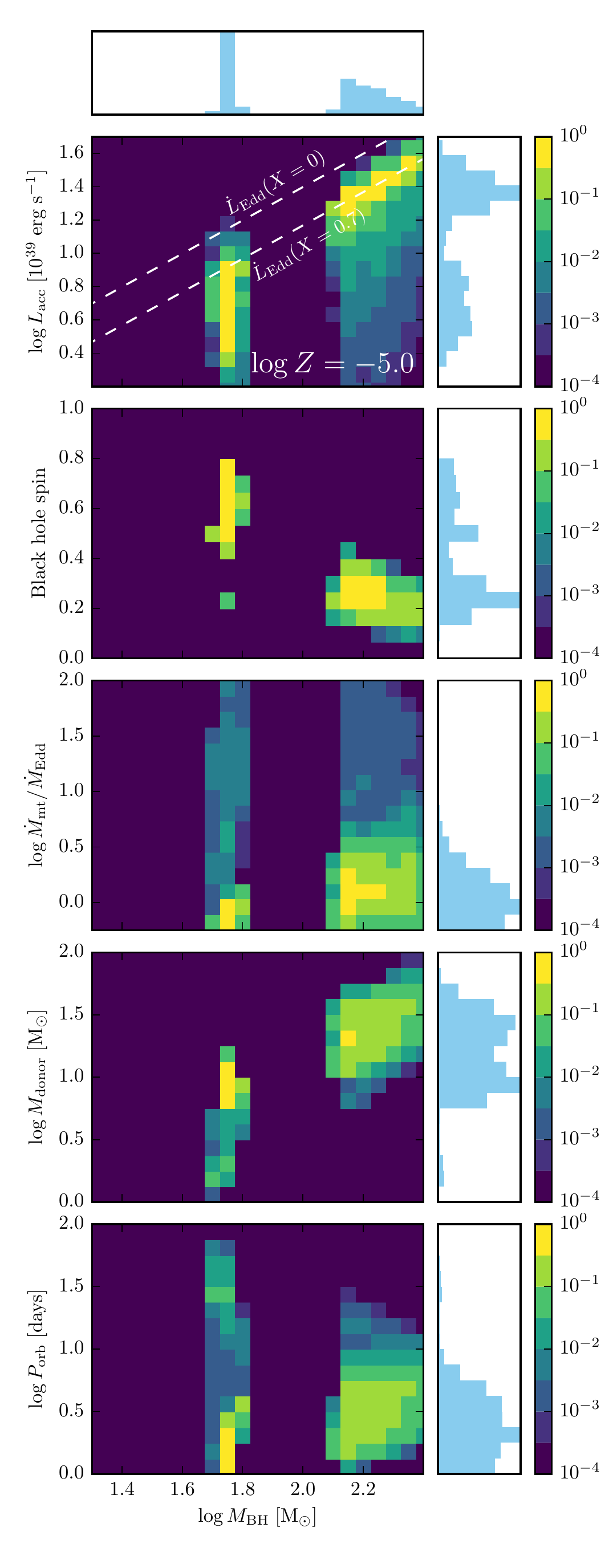}
   \caption{General properties of ULXs for $\log Z=-5.0$. See text in Appendix
      \ref{appendix:properties} for an explanation.}
      \label{fig:props6}
\end{figure}

\begin{figure}
   \includegraphics[width=\columnwidth]{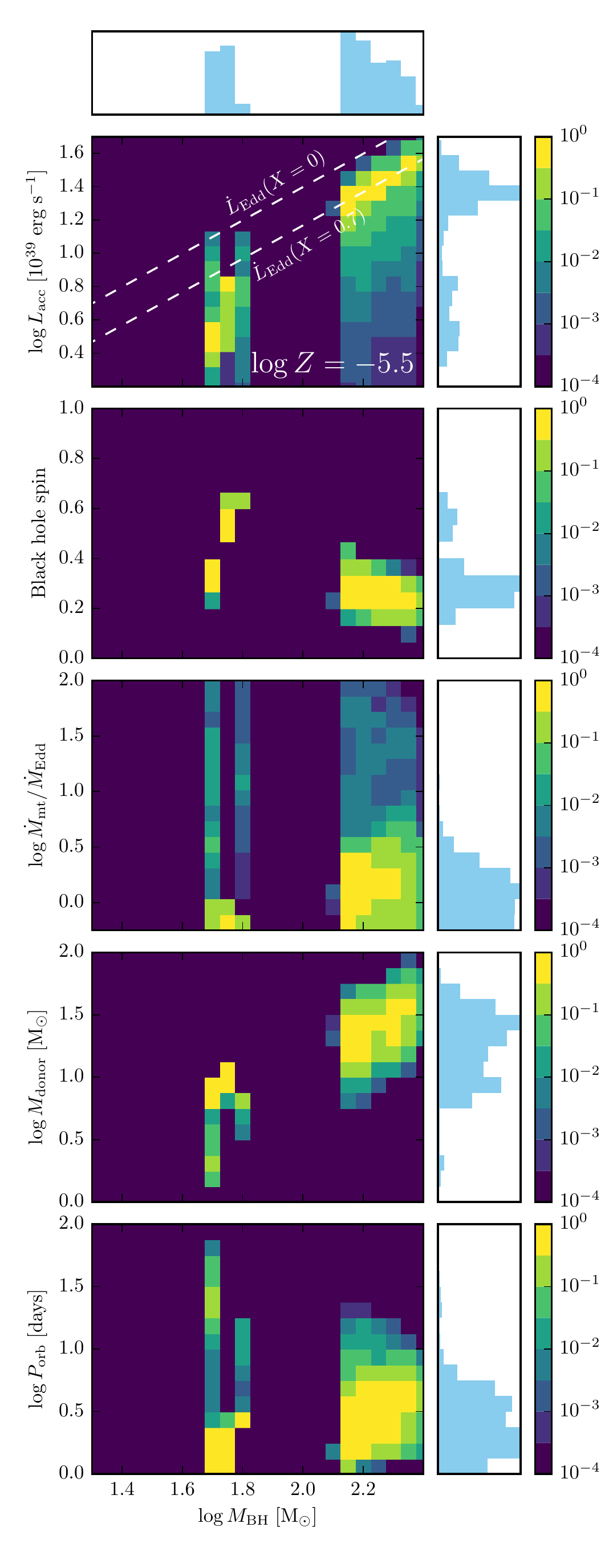}
   \caption{General properties of ULXs for $\log Z=-5.5$. See text in Appendix
      \ref{appendix:properties} for an explanation.}
      \label{fig:props7}
\end{figure}

\begin{figure}
   \includegraphics[width=\columnwidth]{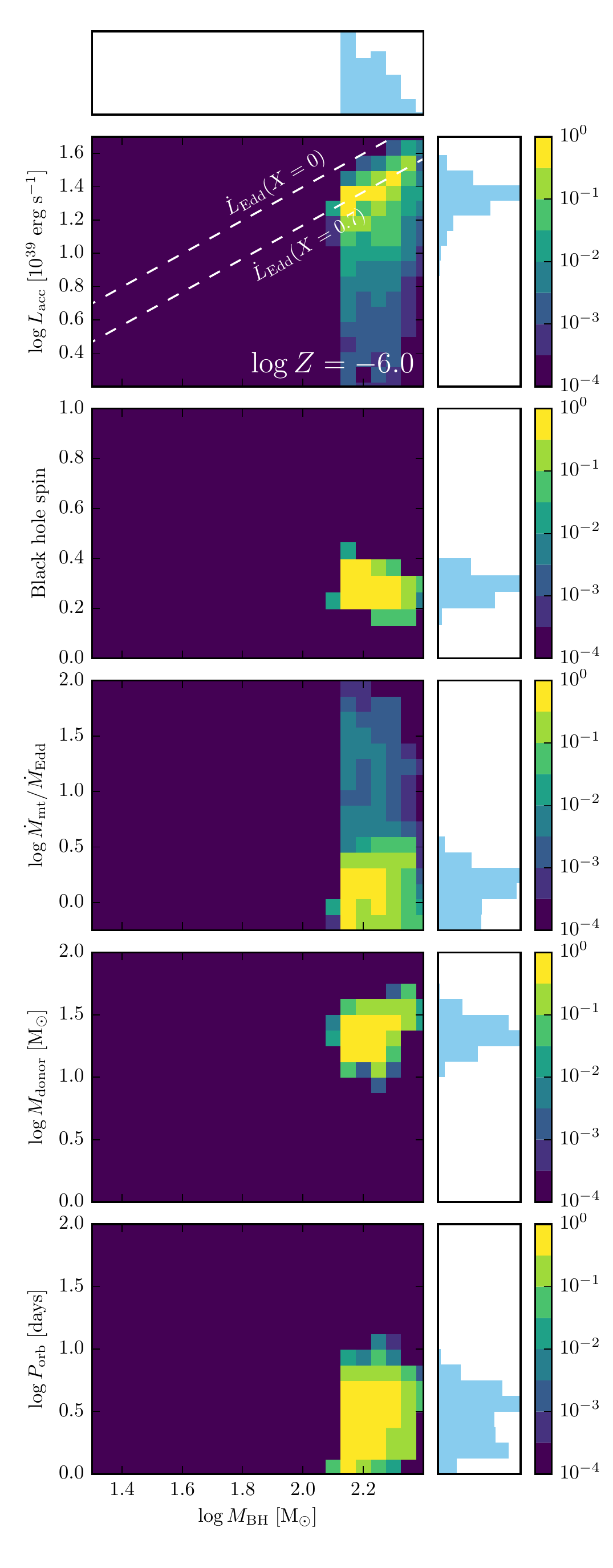}
   \caption{General properties of ULXs for $\log Z=-6.0$. See text in Appendix
      \ref{appendix:properties} for an explanation.}
      \label{fig:props8}
\end{figure}

Figures \ref{fig:props1}-\ref{fig:props8} show several properties of our ULX
models, including BH masses, accretion luminosities assuming mass transfer is limited to
the Eddington rate, BH spins, the ratio between mass transfer and the Eddington
rate $\dot{M}_{\rm mt}/\dot{M}_{\rm Edd}$, donor masses, and orbital periods.
Color plots indicate in logarithmic scale 2D density distributions of all
quantities against BH masses, while histograms are in a linear scale. For all
metallicities, we use the distributions described in Section
\ref{sect:resultsLUM} for the mass of the primary, the mass ratio, and the
orbital separation of the binary. These distributions take into
account the lifetimes of different phases, and so correspond to the observable
distributions at a fixed metallicity.


%%% Local Variables: 
%%% mode: latex
%%% TeX-master: "paper.tex"
%%% End: 

\end{document}